\documentclass[
 reprint,longbibliography,
superscriptaddress,
 amsmath,amssymb,
 aps,
pra,
]{revtex4-2} 
\usepackage{lipsum} 
\usepackage{amsmath}
\usepackage{comment}
\usepackage{graphicx}
\usepackage{dcolumn}
\usepackage{bm}
\usepackage{centernot}
\usepackage{comment}
\usepackage{circuitikz}
\usepackage[normalem]{ulem}
\usepackage{stackengine}
\usepackage{MnSymbol}
\usepackage{booktabs}
\usepackage{array}
\usepackage[table]{xcolor}
\usepackage{threeparttable}

\definecolor{DeepPurple}{HTML}{685599}
\definecolor{SoftPurple}{HTML}{EFE7F7}
\definecolor{VerySoftPurple}{HTML}{F8F4FC}

\newcommand{\purplerowlabel}[1]{\textcolor{DeepPurple}{\bfseries #1}}

\definecolor{DeepRed}{HTML}{A22B2B}
\definecolor{SoftRed}{HTML}{FDF6F7}
\definecolor{VerySoftRed}{HTML}{FDF6F7}

\newcolumntype{L}[1]{>{\raggedright\arraybackslash}p{#1}}
\newcolumntype{C}[1]{>{\centering\arraybackslash}p{#1}}

\newcommand{\rowlabel}[1]{\textcolor{DeepRed}{\bfseries #1}}

\stackMath
\newcommand{\tr}{{\rm tr}}

\usepackage[hidelinks]{hyperref}
 \hypersetup{
   colorlinks   = false
 }

\usepackage{mathrsfs}  

\makeatletter
\providecommand{\ilimits@}{\nolimits} 
\makeatother
\begin{document}

\newcommand\ak[1]{{\color{blue} Artemy: #1}}

\title{Wasserstein-2 gradient flows and the geometry of entropy production \\ in classical and quantum stochastic thermodynamics}

\author{Olga Movilla Miangolarra}\email[Corresponding author: ]{omovilla@ull.edu.es}
\affiliation{Departamento de Física, Universidad de La Laguna, La Laguna 38203, Spain}
\affiliation{Instituto Universitario de Estudios Avanzados (IUdEA), Universidad de La Laguna, La Laguna 38203, Spain}

\author{Ralph Sabbagh}
\affiliation{Department of Mechanical and Aerospace Engineering, University of California, Irvine, California 92697, USA}

\author{Artemy Kolchinsky}
\affiliation{ICREA-Complex Systems Lab, Pompeu Fabra University, 08003 Barcelona, Spain}
\affiliation{Universal Biology Institute, Graduate School of Science, The University of Tokyo, 7-3-1 Hongo, Bunkyo-ku, Tokyo 113-0033, Japan}

\begin{abstract}
The second law does more than set the direction of thermodynamic evolution: it endows nonequilibrium transformations with an underlying geometry. In this work, we provide a unified geometric description of entropy production in classical and quantum thermodynamics based on Wasserstein-2 structures arising from gradient flows of free energy. We review how relaxation to equilibrium,~in overdamped diffusions, discrete detailed-balanced Markov chains, and dissipative Lindblad dynamics, can be formulated as a gradient flow on the space of states.  The associated Wasserstein-2 distance bounds entropy production, yielding a finite-time refinement of the second law. We extend this framework beyond purely dissipative dynamics by introducing generalized Wasserstein-2 metrics that incorporate conservative (Hamiltonian) dynamics in both classical inertial systems and open quantum systems, yielding intrinsic distances that exactly characterize minimal entropy production under fixed dissipative mobilities. We establish equivalence bounds between purely dissipative and Hamiltonian–dissipative geometries, explicitly quantifying how inertial or coherent dynamics can reduce dissipation.  Finally, when restricted to equilibrium distributions, we recover the thermodynamic length of linear response—including the quantum thermodynamic length—thereby linking optimal transport, thermodynamic length, and counterdiabatic protocols within a single geometric framework.
Collectively, our results extend the Riemannian program of thermodynamics further from equilibrium and provide a geometric foundation for optimal protocols beyond the overdamped setting.
\end{abstract}

\maketitle

\section{Introduction}

Whether it be a living creature, the planet Earth, or the universe itself, the fate of complex systems is dictated by the second law of thermodynamics. When the inflow of free energy stops, when there is no more food to be eaten, no more photons to be absorbed, systems inevitably meet their boring, lifeless, equilibrium fate. This fact is ruthlessly enforced by the second law of thermodynamics, instilling in all evolution an arrow of time and giving free energy a price. Even as microscopic trajectories jitter and fluctuate, the macroscopic world is bound to relax, to spread, to dissipate. Yet the second law of thermodynamics can do more than that: it can also dictate \emph{how} systems relax to equilibrium, and what is possible when we are out of equilibrium.
It can tell us how much work can be harvested and at what cost, that precise control must be paid for in dissipation, and why computation typically leaves a thermal trace. From molecular assemblies to heat engines and information processors, the second law is the universal ledger that balances change with irreversibility. 

   \begin{figure*}[tb]
    \centering
\includegraphics[width=0.88\textwidth,trim= 0cm 0cm 0cm 0cm,clip]{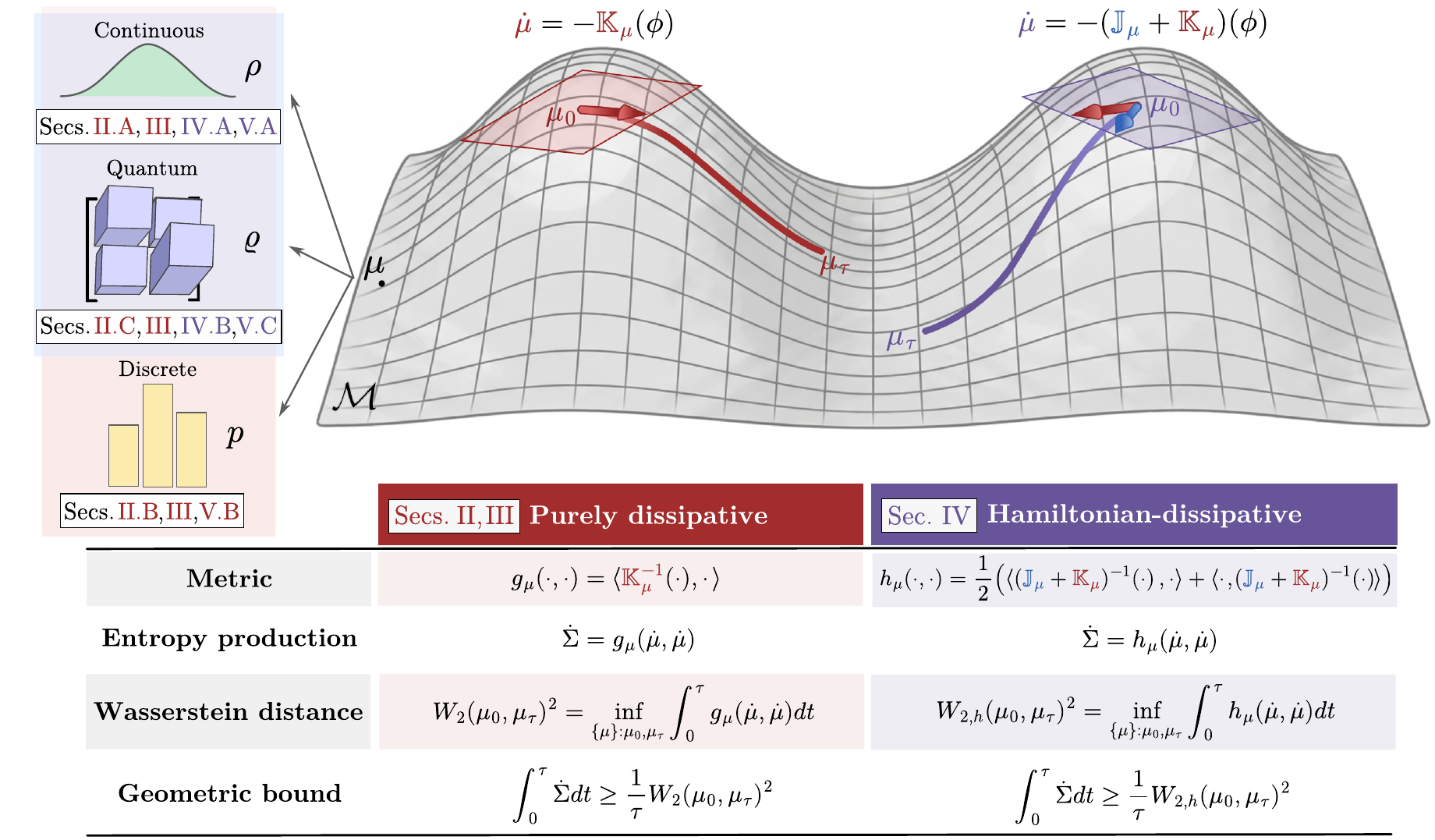}
    \caption{Summary of the Wasserstein-2 geometric picture of stochastic and quantum thermodynamics in purely dissipative (left) and mixed Hamiltonian-dissipative (right) systems.
    The Wasserstein-2 geometry arises from a gradient flow structure in purely dissipative systems (continuous, discrete, and quantum), and from a combination of gradient and Hamiltonian flows in mixed systems (continuous and quantum). In all cases, the distance squared bounds entropy production in finite-time transitions.
    }
    \label{fig:summary}
\end{figure*}

The approach to equilibrium together with the limitations of non-equilibrium transitions imposed by the second law can be beautifully captured by Riemannian geometry.
Riemannian frameworks that encode these principles have steadily evolved from equilibrium formulations to increasingly out-of-equilibrium descriptions. The earliest instances are traced to the independent works of Weinhold and Ruppeiner, which endow the space of equilibrium states—equivalently, the space of thermodynamic potentials—with Riemannian metrics that capture stability and fluctuation properties, thereby encoding the local equilibrium structure implied by the second law
~\cite{weinhold1975metric,ruppeiner1979thermodynamics}. This Riemannian viewpoint has been subsequently extended to the linear-response regime, where distances quantify dissipation during finite-time but slow transformations and geodesics identify dissipation-minimizing protocols~\cite{salamon_thermodynamic_1983,crooks_measuring_2007,sivak2012thermodynamic}. More recently, the approach has been generalized far from equilibrium, where system states are probability distributions and stochastic thermodynamics offers the natural language~\cite{aurell2011optimal,aurell2012refined,nakazato2021geometrical,dechant2019thermodynamic}. 


\nocite{mrugala1978geometrical,van2026geometric}
In the far-from-equilibrium regime, optimal transport theory has proven a fruitful candidate to capture the geometry of thermodynamics~\footnote{Optimal transport geometries capture the second law of thermodynamics by quantifying how much dissipation is required for a given finite-time transition.
There are other complementary  geometries that capture different aspects of thermodynamics. For instance, the first-law structure of equilibrium thermodynamics can be described using contact geometry~\cite{mrugala1978geometrical}, while, more recently, a geometry of physical maps, rather than thermodynamic states, has been used to formulate a third-law-type unattainability principle~\cite{van2026geometric}.}.
The optimal transport distance known as Wasserstein-2 distance \cite{benamou2000computational,villani2021topics} has a natural interpretation in the stochastic thermodynamics of overdamped Langevin systems. There, it provides a refinement of the second law by quantifying the minimum entropy production required to transition between two given states in a finite amount of time \cite{aurell2011optimal}. 
This result has been extended beyond the overdamped setting to
 underdamped regimes \cite{dechant2019thermodynamic,sabbagh2024wasserstein}, discrete stochastic systems~\cite{van2021geometrical,dechant2022minimum,van2023thermodynamic,yoshimura2023housekeeping,kolchinsky2026generalized},
 and quantum systems~\cite{van2021geometrical,van2023thermodynamic,yoshimura2025force}, among others~\cite{nagayama2025geometric}. 
These optimal-transport perspectives have found numerous applications, including a decomposition of entropy production for systems that do not satisfy detailed balance~\cite{maes2014nonequilibrium,dechant2022geometric2,yoshimura2023housekeeping,miangolarra2024minimal,kolchinsky2026generalized} and derivations of thermodynamic uncertainty relations~\cite{yoshimura2023housekeeping,nagayama2025geometric}. They have also led to speed limits and geometric bounds on energy extraction~\cite{movilla2021energy,miangolarra2022geometry}, as well as insights into 
the design of heat  engines~\cite{fu2021maximal,movilla2021energy,miangolarra2022geometry} and control protocols in information thermodynamics~\cite{taghvaei2021relation,nagase2024thermodynamically,oikawa2025experimentally,kamijima2025finite},
among other optimal thermodynamic control protocols~\cite{blaber2023optimal}. 

However, the aforementioned extensions of the overdamped result to other settings are not unique. Diverse approaches have been introduced, some focusing on extensions of the Wasserstein-1 optimal transport distance~\cite{dechant2022minimum,van2023thermodynamic,delvenne2024thermokinetic,nagayama2025infinite,kolchinsky2026generalized}, others on Wasserstein-2 distances ~\cite{van2021geometrical,yoshimura2023housekeeping,nagayama2025geometric,yoshimura2025force}.
Wasserstein-2 approaches arise by expressing detailed-balanced dynamics, which govern relaxation to equilibrium states, as a gradient flow over a Riemannian manifold of states. In this structure, Onsager-type operators that linearly map thermodynamic forces to fluxes play a central role, leading to a bound on entropy production in terms of a Wasserstein-2 distance. Unlike Wasserstein-1, Wasserstein-2 distances recover overdamped results in the appropriate limit, including linear-response metrics~\cite{sivak2012thermodynamic,scandi2019thermodynamic} when restricted to the equilibrium submanifold~\cite{zhong2024beyond,sawchuk2026thermodynamic}.

On top of these optimal transport geometric approaches, other optimal control problems that minimize and bound entropy production have proven of central relevance in stochastic thermodynamics. Foundational results minimize the work required to change the control parameters between two endpoints for overdamped~\cite{schmiedl2007optimal, Proesmans2020Landauer,loos2024universal}  and underdamped~\cite{gomez2008optimal,dago2023adiabatic,sanders2024optimal,sanders2025minimal} systems. 
Other entropy-minimizing protocols emerge in the discrete setting, where a natural optimization problem involves constraining the symmetric part of the transition rates~\cite{muratore2013heat,remlein2021optimality}. Similar optimal-control problems have been studied in open quantum systems, including protocols that minimize heat dissipation under Hamiltonian control~\cite{cavina2018optimal}.
 What these optimal-control approaches have in common is that they do not characterize minimal entropy production as a distance between endpoint states, possibly because of their endpoint constraints or other restrictions on the admissible dynamics. Here, we focus instead on a geometric formulation in which such a characterization arises.


In this work, we provide a unified perspective on the intimate relationship between Wasserstein-2 geometries arising from gradient-flow structures and entropy production in thermodynamic systems, see Figure \ref{fig:summary} for a schematic summary. 
The contributions of this work are twofold. First, we provide a unifying review of 
recent optimal-transport approaches to stochastic and quantum thermodynamics. 
Second, we derive new results that enable and extend this unified perspective. Among these results, we highlight: 
\begin{enumerate}
    \item A new Wasserstein-2 distance that captures minimal entropy production in inertial classical systems (see \eqref{eq:W2-minE-h}), leading to explicit optimal protocols in the linear Gaussian setting (see \eqref{eq:W2-m});
    \item  Metric equivalence bounds comparing dissipative and mixed Hamiltonian-dissipative metrics in classical and quantum systems (see \eqref{eq:hierarchy-c} and \eqref{eq:hierarchy-q}), thereby quantifying the advantages of Hamiltonian flows;
    \item The equality between the quantum Wasserstein-2 metric restricted to equilibrium states and the quantum linear-response thermodynamic metric~\cite{scandi2019thermodynamic} (see Section \ref{sec:q-LR}).
\end{enumerate}
Moreover, we show that the discrete and quantum Wasserstein-2 distances coincide when the endpoints commute (see \eqref{eq:q-d-W2}) and that Lindblad dynamics can be chosen so that any quantum trajectory is followed with arbitrarily small entropy production (see Appendix \ref{app:q-dechant} for the explicit construction).

The rest of the manuscript is organized as follows. First, in Section~\ref{sec:grad-flows}, we review gradient-flow structures in the context of optimal transport. In Section~\ref{sec:III}, we show that this leads to thermodynamic geometry, 
where distances reflect minimum entropy production needed to transition between states. 
In Section~\ref{sec:IV}, Hamiltonian flows are introduced in both classical and quantum settings, leading to new Wasserstein-2 distances that bound entropy production for mixed Hamiltonian-dissipative systems. Furthermore, the advantages of having Hamiltonian flows are quantified by providing equivalence bounds on Hamiltonian-dissipative distances in terms of the corresponding purely dissipative ones.  Finally, in Section~\ref{sec:LR}, we show that the proposed metrics reduce to the linear-response thermodynamic metrics~\cite{sivak2012thermodynamic,scandi2019thermodynamic} when restricted to the equilibrium submanifold for continuous, discrete, and quantum systems, both purely dissipative and mixed. Consequently, optimal-transport protocols may be understood as optimal counterdiabatic protocols that follow linear-response geodesics in finite time, even beyond the overdamped setting~\cite{zhong2024beyond}. In this way, the presented Wasserstein-2 approach further extends the reach of Riemannian thermodynamic geometry out of equilibrium.



\section{Gradient flows in optimal transport}
\label{sec:grad-flows}

Gradient flows in Riemannian manifolds require three key elements, namely, a smooth manifold of states, a Riemannian metric, and a free energy functional. Through these, dynamics that evolve the state in the direction of steepest descent of the free energy functional can be determined. 
Specifically, consider a smooth manifold $\mathcal M$, which may represent the set of states of a physical system. For example, $\mathcal{M}$ may represent the set of probability distributions over microstates for classical thermodynamic systems, or the set of possible density matrices in quantum systems (see Table I).

Together with the manifold $\mathcal M$,  consider a Riemannian metric $g$ such that $g_x:T_x\mathcal M\times T_x\mathcal M\to\mathbb R$ is given by
$g_x(u,v)=\llangle \mathbb G_x(u),v\rrangle$, where $x\in\mathcal M$, $u,v\in T_x\mathcal M$, and $\mathbb G_x:T_x\mathcal M\to T_x^*\mathcal M$ is an invertible, symmetric, positive-definite operator that takes vectors into covectors.
Here, $T_x\mathcal{M}$ and $T^*_x\mathcal{M}$ denote the tangent and cotangent spaces of $\mathcal{M}$ at the point $x$, respectively, and $\llangle\cdot,\cdot\rrangle$ denotes the natural duality pairing between covectors and vectors. In the context of non-equilibrium thermodynamics, 
 $\mathbb{G}$ maps fluxes (vectors, such as heat or mass flows) into thermodynamic forces (covectors, represented by temperatures or chemical potentials), incorporating phenomenological coefficients like transport properties (e.g., conductivity, viscosity, etc). The inverse map that transforms forces into fluxes, $\mathbb K_x=\mathbb G^{-1}_x:T_x^*\mathcal M\to T_x\mathcal M$, is known as the Onsager operator \cite{disser2015gradient}, \footnote{In the context of Riemannian geometry, the map $\mathbb{G}$ is simply the musical isomorphism induced by the metric $g$. That is, $\mathbb{G}=\flat: T\mathcal{M}\rightarrow T^*\mathcal{M} :  v\mapsto g(v,\cdot)$ is the flat map and its inverse, $\mathbb{K}=\sharp:T^*\mathcal{M}\rightarrow T\mathcal{M}: g(v,\cdot)\mapsto v$, is the sharp map, where $T\mathcal{M}$ and $T^*\mathcal{M}$ denote the tangent and cotangent bundles of $\mathcal{M}$, respectively. 
 }.



\begin{figure}
    \centering
    \includegraphics[width=0.8\linewidth, trim=0cm 0 1cm 0,clip]{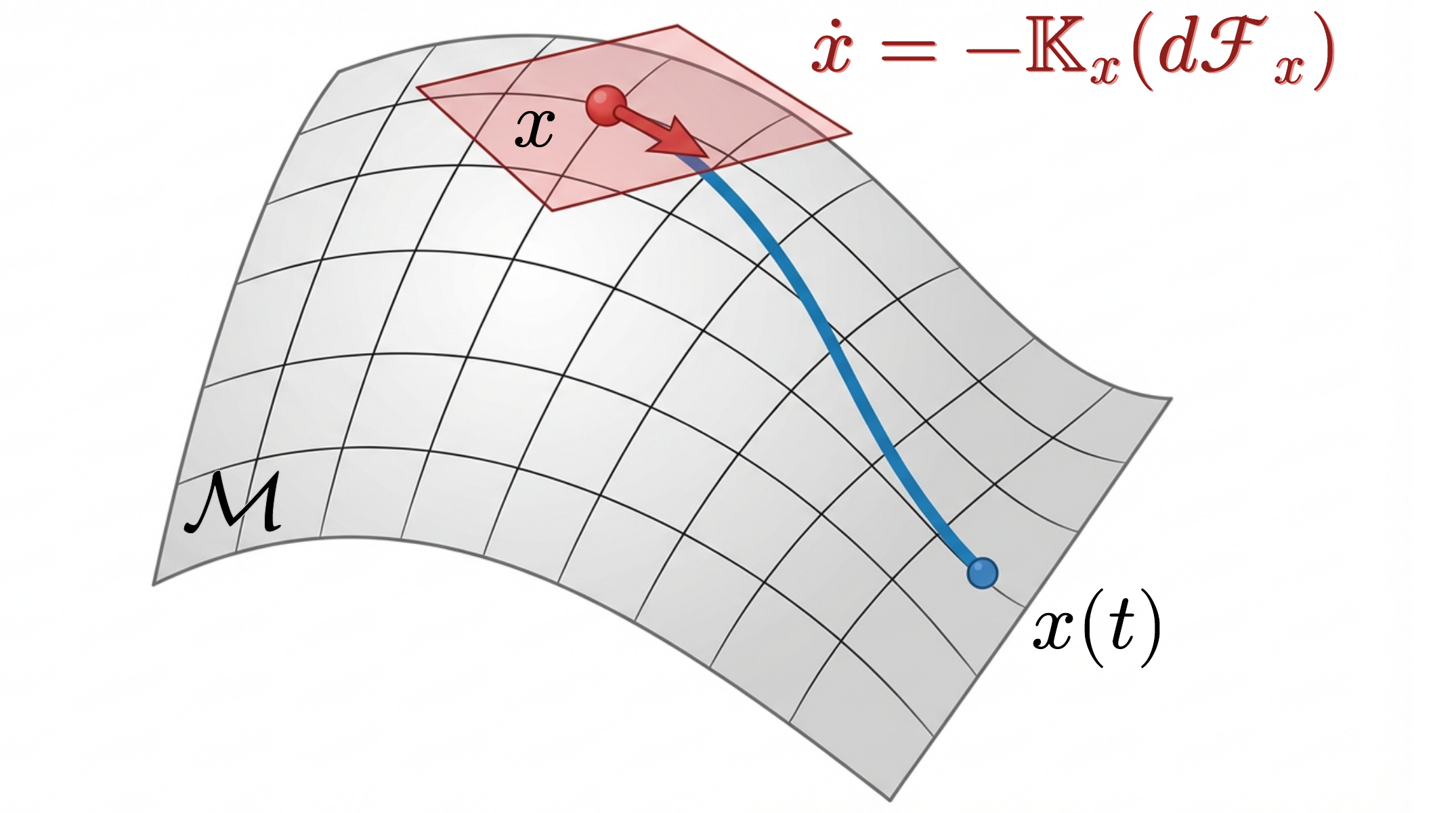}
    \caption{The state $x$ evolves on the Riemannian manifold $(\mathcal M, g)$ along the gradient flow of the free energy $\mathcal F$.}
    \label{fig:placeholder}
\end{figure}

The third ingredient is a smooth free energy function $\mathcal{F}:\mathcal{M}\rightarrow \mathbb{R}$. Let 
  $d\mathcal F$ denote its differential, which is a 1-form so that, when evaluated at  a point $x\in\mathcal{M}$, $d\mathcal F_x\in T_x^*\mathcal M$.
The gradient flow of $\mathcal F$ on $(\mathcal M, g)$ is given by (see Figure \ref{fig:placeholder} for an illustration) 
\begin{equation}
    \label{eq:grad-flow}
    \dot x(t)=-\mathbb K_{x(t)}(d\mathcal F_{x(t)}),
\end{equation}
where the right-hand side is the negative of the gradient of $\mathcal F$ in the metric $g$~\footnote{The gradient of $\mathcal F$ in the metric $g$ is defined, for all $v\in T_x\mathcal M,$ by
$$
g_x(\nabla_g\mathcal F_x,v)=\llangle d\mathcal F_x,v\rrangle.
$$
Therefore, $\llangle \mathbb K^{-1}_x\nabla_g\mathcal F_x,v\rrangle=\llangle d\mathcal F_x,v\rrangle$, implying that $\nabla_g\mathcal F_x=\mathbb K_x(d\mathcal F_x)$
}. 
 For a thermodynamic state $x$, this flow may be understood as a kinetic relation linking thermodynamic forces to fluxes. Moreover, it is precisely in the direction of steepest descent of $\mathcal F$ with respect to the metric $g$~\footnote{The steepest descent direction is given by the velocity $v\in T_x\mathcal M$ with norm $1$, i.e., $g_x(v,v)=1$, that minimizes
$$ \frac{d}{dt}\mathcal F(x+tv)|_{t=0}=\llangle d\mathcal F_x,v\rrangle
.$$ Up to a normalization factor, the optimal velocity is $v=-\mathbb K_x(d\mathcal F_x)$
}. Indeed, by the chain rule,
\begin{align}
   \frac{d}{dt}\mathcal F(x(t))&=
    \llangle d\mathcal F_{x(t)},\dot x(t)\rrangle
=-g_{x(t)}(\dot x(t),\dot x(t))\le 0,
\end{align}
where for the last equality we have used $d\mathcal F_{x(t)}=-\mathbb G_{x(t)}(\dot x(t)),$ which follows from \eqref{eq:grad-flow}. Therefore, $\mathcal F(x(t))$ is non-increasing in time. 

We have seen how the gradient flow formulation is determined by:
(i) a state space $\mathcal M$,
(ii) a dissipation mechanism encoded by $\mathbb K$ (or equivalently $\mathbb G$), and (iii) a free energy functional $\mathcal F$.
We next summarize how the gradient--flow structures specialize across three paradigms, highlighting the respective manifolds, Onsager operators, and free energy functionals. In doing so we review previous results on classical (continuous) optimal transport \cite{JKO1998}, discrete optimal transport~\cite{maas2011gradient,mielke2011gradient,chow2012fokker}, and quantum (non-commutative) optimal transport~\cite{carlen2017gradient,chen2017matrix,mittnenzweig2017entropic}. For a concise summary, see Table \ref{tab:fullwidth}. From now on, the time dependence of the elements in $\mathcal M$ and $T_x\mathcal M$ will be left implicit unless necessary.

\renewcommand{\arraystretch}{1.8}

\begin{table*}[tb]
\centering
\small
\begin{threeparttable}
\caption{Summary of the gradient-flow structure and geometry of entropy production across the continuous, discrete, and quantum optimal transport paradigms for purely dissipative systems.}
\label{tab:fullwidth}
  \begin{tabular}{cccc}

\rowcolor{DeepRed}
\textcolor{white}{\bfseries OT paradigm}
&
\textcolor{white}{\bfseries Continuous}
\textcolor{white}{$\;(\mu=\rho)$}
&
\textcolor{white}{\bfseries Discrete}
\textcolor{white}{$\;(\mu=p)$}
&
\textcolor{white}{\bfseries Quantum}
\textcolor{white}{$\;(\mu=\varrho)$}
\\
\midrule

\rowcolor{SoftRed}
\rowlabel{Manifold $\mathcal M$}
&
$\mathcal{P}_*^2(\mathbb{R}^d)$
&
$\mathcal{P}_{\ast}(\mathcal X)$
&
$\mathcal{P}_{\ast}(\mathbb C,d)$
\\

\rowlabel{System}
&
Overdamped Langevin
&
Finite Markov chain
&
Dissipative Lindblad
\\

\rowcolor{VerySoftRed}
\rowlabel{Dynamics}
&
{\footnotesize $\ \  \displaystyle
\dot{\rho}
=
\nabla\!\cdot\!\big(\rho D\nabla(\beta H+\ln\rho)\big)
\ \  $}
&
{\footnotesize $\displaystyle
\dot p_n
=
\sum_m\big(R_{nm}p_m-R_{mn}p_n\big)
$}
&
{\footnotesize $\displaystyle
\dot\varrho
=
\sum_k \gamma_k
\Big(
L_k\varrho L_k^\dagger
-\frac12\{L_k^\dagger L_k,\varrho\}
\Big)
$}
\\

\rowlabel{Detailed balance}
&
$\displaystyle
\rho_{\rm eq}\nabla(\beta H+\ln\rho_{\rm eq})=0
$
&
$\displaystyle
R_{nm}p_m^{\rm eq}
=
R_{mn}p_n^{\rm eq}
$
&
$\displaystyle
\gamma_k=\gamma_{-k}e^{-\beta\omega_k},$  $L_k^\dagger=L_{-k}$, $
[L_k,H]=-\omega_kL_k
$
\\

\midrule

\rowcolor{SoftRed}
\rowlabel{Mobility op. $M_\mu$\vspace{1.5mm}}
&
{\footnotesize $\displaystyle
\rho D(\cdot)
$}
&
{\footnotesize $\displaystyle\ \ 
\frac12 R_{nm}p_m^{\rm eq}\,
\theta\Big(
\frac{p_n}{p_n^{\rm eq}},
\frac{p_m}{p_m^{\rm eq}}
\Big)[\cdot]_{nm}
\ \ $}
&
{\footnotesize $\ \ \displaystyle
\frac{\gamma_k}{2}e^{\frac{\beta\omega_k}{2}}
\int_0^1
\big(e^{\frac{\beta\omega_k}{2}}\varrho\big)^s
(\cdot)
\big(e^{-\frac{\beta\omega_k}{2}}\varrho\big)^{1-s}
\,ds
\ \ $}
\\[-4pt]
\rowlabel{Onsager op. $\mathbb K_\mu$}
&
\multicolumn{3}{c}{
$\!\!\!\displaystyle
\mathbb K_\mu(\cdot)
=
-\nabla\!\cdot\!\big(M_\mu(\nabla(\cdot))\big)
$
}
\\
\rowcolor{SoftRed}
\rowlabel{\small Free energy\tnote{a}  $~\mathcal F$}
&
{\footnotesize $\displaystyle
\int_{\mathbb R^d}
\rho\ln\frac{\rho}{\rho_{\rm eq}}\,dx
$}
&
{\footnotesize $\displaystyle
\sum_{n\in\mathcal X}
p_n\ln\frac{p_n}{p_n^{\rm eq}}
$}
&
{$\displaystyle
\tr\{\varrho(\ln\varrho-\ln\varrho_{\rm eq})\}
$}
\\

\midrule

\rowcolor{SoftRed}
\rowlabel{Riemannian metric}
&
\multicolumn{3}{c}{
$\displaystyle
g_{\mu}(\cdot\,,\cdot)
=
\llangle \mathbb K_{\mu}^{-1}(\cdot),\cdot\rrangle
$
}\\

\rowlabel{Gradient flow}
&
\multicolumn{3}{c}{
$\displaystyle
\dot\mu=-\mathbb K_\mu(d\mathcal F_\mu)
$
}
\\


\rowcolor{VerySoftRed}
\rowlabel{$\ \ 2$-Wasserstein metric \ }
&
\multicolumn{3}{c}{
$\displaystyle
W_2(\mu_0,\mu_\tau)
=
\inf_{\{\mu(t)\}:\,\mu_0,\mu_\tau}\ell_\mu
$
}
\\

\rowlabel{Entropy prod. rate $\dot\Sigma$}
&
\multicolumn{3}{c}{
$\displaystyle
\dot\Sigma
=
g_\mu(\dot\mu,\dot\mu)
$
}
\\

\bottomrule
\end{tabular}

\begin{tablenotes}
\footnotesize
\item[a] The constant term $\ln Z$ is omitted. Thus, the entries in each column are $\mathcal F+\ln Z$, namely the corresponding relative entropy.
\end{tablenotes}

\end{threeparttable}
\end{table*}

\subsection{Continuous optimal transport}\label{sec:IIA}
Let $\mathcal M=\mathcal P_*^2(\mathbb R^d)$ be the space of strictly positive probability densities on $\mathbb R^d$, which we denote by $\rho$  \footnote{Specifically, $\mathcal P_*^2(\mathbb R^d)$ is the space of finite-second-moment probability measures on $\mathbb R^d$ that are absolutely continuous with respect to the Lebesgue measure and have a strictly positive density $\rho$ that satisfies a Poincaré inequality. 
 That is, 
 there exists a constant $C<\infty$ such that
$$
\int_{\mathbb R^d}\varphi^2\rho dx\leq C \int_{\mathbb R^d}\nabla\varphi^\top D\nabla\varphi\rho dx,
$$
for all sufficiently smooth functions $\varphi\in C^\infty_c(\mathbb R^d)$ with zero $\rho$-mean. Here, \(D\) is the fixed positive-definite matrix introduced in the main text.
}.  
Consider the Riemannian metric
$
    g_{\rho }(\dot \mu,\dot\nu):=\llangle\mathbb G_\rho(\dot\mu),\dot \nu\rrangle,
$
where $\dot\mu,\dot\nu\in T_\rho\mathcal M.$ Here, 
$\mathbb G_{\rho }:\dot\mu\mapsto \mathbb G_\rho(\dot\mu)=-\varphi\in T_\rho^*\mathcal M$, where $\varphi$ is the unique, zero-mean, weak solution to the Poisson equation~\footnote{On $\mathbb{R}^d$, the weighted Poisson equation $\nabla\cdot(D\,\rho\,\nabla\varphi)=\dot\mu$ has a (weak) solution, unique up to an additive constant, for admissible zero-mass tangent vectors \(\dot\mu\), provided that $\rho$ is strictly positive and satisfies a Poincaré inequality. A  sufficient condition is to assume a Gibbs density $\rho\propto e^{-\beta V}$ with $V$ uniformly convex outside a compact set. 
Since only $\nabla\varphi$ enters the dynamics, any additive constant is physically irrelevant.}
$$
\nabla\cdot(\rho D \nabla\varphi)=\dot\mu,
$$
with $D\in \mathbb R^{d\times d}$ a diagonal positive-definite  matrix,
and $\nabla$ the gradient operator in $\mathbb R^d$. Thus, the Onsager operator $\mathbb K_\rho:=\mathbb G_\rho^{-1}:\varphi\mapsto-\dot \mu$ is given by
\begin{equation}
    \label{eq:Krho}
    \mathbb K_{\rho }(\varphi):=-\nabla\cdot(M_{\rho }\nabla\varphi),
\end{equation}
where we have defined the mobility $M_{\rho }:=\rho D$. Alternatively, letting $\psi:=-\mathbb G_\rho(\dot\nu)$,
we can rewrite the metric as
$$
 g_{\rho }(\dot \mu,\dot\nu)=\llangle\mathbb G_{\rho }(\dot\mu),\dot \nu\rrangle=\llangle \varphi,\mathbb K_{\rho }(\psi)\rrangle=\langle \nabla\varphi, M_{\rho }\nabla\psi\rangle,
$$
where we have integrated by parts weakly, so no boundary terms are required, and $\langle\cdot,\cdot\rangle$ denotes the standard $L^2$ inner product. That is,
\begin{align}\label{eq:metric-cl}
    g_{\rho }(\dot \mu,\dot\nu)=\langle\nabla\varphi, M_{\rho }\nabla\psi \rangle:=\int_{\mathbb R^d} \nabla \varphi^\top M_{\rho }\nabla\psi dx,
\end{align}
with $^\top$  the transpose operator, and  $\dot\mu,\dot\nu\in T_{\rho }\mathcal M$ related to $\varphi,\psi\in T_{\rho }^*\mathcal M$ through
$
\dot\mu=-\mathbb K_{\rho }(\varphi) \mbox{ and }\dot\nu=-\mathbb K_{\rho }(\psi).
$

Next, consider the gradient flow
\begin{equation}
    \label{eq:grad-flow-cl}
    \dot \rho =-\mathbb K_{\rho }(d\mathcal F_{\rho }),
\end{equation}
where $\mathcal F$ is the free energy functional (taken here in units of entropy)
$$
\mathcal F(\rho ):=\int_{\mathbb R^d} ( \beta H +\ln\rho )\rho  dx,
$$
with $H:\mathbb R^d\to \mathbb R$  the internal energy of the system and $\beta>0$ the inverse temperature of the surrounding heat bath. Here and throughout, $k_B$ is taken to be equal to unity. 
By noting that, up to an irrelevant additive constant,
 $$
 d\mathcal F_{\rho }=\beta H+\ln\rho =:\phi,
 $$
 we see that equation \eqref{eq:grad-flow-cl} is nothing but the overdamped Fokker-Planck equation,
 \begin{align}
 \dot\rho&=\nabla\cdot(\rho \beta D\nabla H)+\nabla\cdot(D\nabla\rho),
 \end{align}
 that governs the evolution of the probability distribution associated with the stochastic process
\begin{equation}\label{eq:SDEover}
     dX_t=-\beta D\nabla Hdt+\sqrt{2}D^{1/2} dW_t,
\end{equation}
where $W_t$ is the standard Wiener process, and $D$ acts as the diffusion matrix associated with a single heat bath at inverse temperature $\beta$. 

The free energy functional $\mathcal{F}$ can be expressed in terms of the relative entropy with respect to the equilibrium state $\rho_{\rm eq}$,  
$$
\mathcal F(\rho )= \int_{\mathbb R^d}\rho \ln\frac{\rho }{\rho _{\rm eq}}dx -\ln Z,
$$
where $\rho _{\rm eq}:=e^{-\beta H}/Z$  with $Z:=\int_{\mathbb{R}^d} e^{-\beta H}dx$ the partition function, and the additive constant $-\ln Z$ plays no role. 
Further, it  satisfies
\begin{align}\label{eq:dotF-c}
    \frac{d}{dt}\mathcal F(\rho )&=\llangle d\mathcal F_{\rho },\dot\rho \rrangle=-g_{\rho }(\dot \rho ,\dot \rho )\leq 0.
\end{align}
Since the relative entropy is a positive functional, only zero when $\rho =\rho _{\rm eq}$, $\mathcal{F}$ (modulo $\ln Z$) acts as a Lyapunov function, ensuring stability and convergence to the equilibrium distribution. Therefore, the solution to \eqref{eq:grad-flow-cl} follows the steepest descent of the free energy until equilibrium is reached.

Note that, in view of~\eqref{eq:Krho}, eq. \eqref{eq:grad-flow-cl} can be understood as a continuity equation, where $-M_\rho(\nabla \phi)$ is the flux field. We observe that, for gradient flow dynamics, being at a steady state  ($\dot\rho=0$) is equivalent to having $\phi=0$ (up to a constant). That is, fluxes vanish in stationarity, implying that stationary states are equilibrium states. Systems whose stationary states have vanishing
probability fluxes are said to satisfy \emph{detailed balance}. 

We have shown that the ensemble evolution of a stochastic process following
\eqref{eq:SDEover} is the gradient flow of the free energy functional 
with respect to the Riemannian metric \eqref{eq:metric-cl}. This metric defines a geodesic distance on $\mathcal M$, known as the  Wasserstein-2 distance, through
\begin{align}\label{eq:cOT}
    W_2(\rho _0,\rho _\tau)^2&:=\inf_{\{\rho(t)\} :\rho _0,\rho _\tau}\tau\int_0^\tau g_{\rho }(\dot\rho ,\dot\rho ) dt,
\end{align} where we have used the shorthand $\{\rho(t)\} :\rho _0,\rho _\tau$ to signify that the optimization is over curves $\{\rho(t)\}_{t\in[0,\tau]}$
such that
$\rho(t)\in\mathcal{M}$ for all $t\in[0,\tau]$, $\rho (0)=\rho _0$ and $\rho (\tau)=\rho _\tau$.
The Cauchy-Schwarz inequality and the fact that equality can be achieved through constant arc-length parametrization lead to
$$
  W_2(\rho _0,\rho _\tau)^2=\inf_{\{\rho(t)\} :\rho _0,\rho _\tau}\ell_{\rho }^2,\nonumber
$$
where we have defined the Riemannian length of a curve
$$
\ell_{\rho }:=\int_0^\tau\sqrt{g_{\rho }(\dot\rho ,\dot\rho )}dt.
$$
That is, the Wasserstein-2 distance arises as the solution to problem \eqref{eq:cOT}, which is equivalent to finding the curve with minimum length that joins $\rho _0$ to $\rho _\tau$.
We may rewrite problem \eqref{eq:cOT} in terms of the Benamou-Brenier formulation of the optimal transport problem \cite{benamou2000computational}, namely, 
\begin{align}\label{eq:cost-OMT}
    W_2(\rho _0,\rho _\tau)^2=  &\inf_{V,\rho }\tau\int_0^\tau\|V\|^2_{M_{\rho }} dt \\& \mbox{s.t.} \ \dot\rho =\nabla\cdot(M_\rho  V),\ \rho (0)=\rho _0,\, \rho (\tau)=\rho _\tau,\nonumber
\end{align}
where $\|V\|^2_{M_\rho }=\langle V,M_\rho V\rangle. $  To see this, note that optimal velocities take the gradient form $V_{\rm opt}=\nabla\varphi_{\rm opt},$  and thus,
the cost function in~\eqref{eq:cost-OMT} can be taken to be the integral over time of $g_{\rho }(\dot\rho ,\dot\rho )$.

We refer the reader to \cite{ambrosio2005gradient} for a comprehensive development of the subject that has been succinctly summarized in this section.

\subsection{Discrete optimal transport}
\label{sec:discrete-optimal-transport}

Analogously to the development of the previous section, the ensemble evolution of discrete Markov processes may also be seen as the gradient flow of a free energy functional \cite{maas2011gradient,mielke2011gradient,chow2012fokker}. Specifically, 
consider a continuous-time Markov chain in a finite state space $\mathcal X$, and let $ p\in\mathcal M=\mathcal P_*(\mathcal X)$ denote its probability distribution, where $\mathcal{P}_*(\mathcal{X})$ is the space of strictly positive probability distributions on $\mathcal{X}$. The evolution of this distribution satisfies
\begin{equation}\label{eq:CTMC}
    \dot  p_n=\sum_{m\in\mathcal{X}}\big(R_{nm} p_{m}-R_{mn} p_n\big),~~n\in\mathcal{X},
\end{equation}
where $R_{mn}\geq 0$ for $m\neq n$ is the transition rate from state $n\in\mathcal X$ to $m\in\mathcal X\mathbin{\backslash} \{n\}$, while $R_{nn}:=-\sum_{m\in \mathcal X\mathbin{\backslash} \{n\}}R_{mn}$.
We assume that the Markov chain is irreducible (has no disconnected parts) and that the transition rates $R_{mn}$ satisfy detailed balance, that is,
\begin{align}\label{eq:detbal}
R_{nm}p_m^{\rm eq}=R_{mn}p_n^{\rm eq},
~~n,m\in\mathcal{X},
\end{align}
where $p_n^{\rm eq}:=e^{-\beta H_n}/Z$ with $Z:=\sum_{n\in\mathcal{X}} e^{-\beta H_n}$ is the equilibrium distribution. Here, $H:\mathcal{X}\rightarrow\mathbb{R}$ is the energy function (or Hamiltonian) and $\beta> 0$ is the inverse temperature of a single heat bath.
As will shortly become apparent, \eqref{eq:CTMC}  can be thought of as a discrete continuity equation with $R_{nm}p_m-R_{mn}p_n$ as the flux along the $m-n$ edge. Therefore, the detailed balance condition \eqref{eq:detbal} is equivalent to the fluxes vanishing in steady state.

We now introduce a discrete calculus on the graph induced by the Markov chain, mirroring the gradient–divergence structure of the continuous case.
 To this end, 
 let us define the
 inner product between real functions $\varphi$ and $\psi$ on $\mathcal X$ as
$$
\langle \varphi,\psi\rangle:=\sum_{n\in\mathcal X} \varphi_n\psi_n,
$$
and between real functions $U$ and $V$ on edges $\mathcal X\times\mathcal X$ as
$$
\langle U,V\rangle:=\sum_{n,m\in\mathcal X} U_{nm}V_{nm}.
$$
Moreover, define the discrete
gradient of a function $\varphi:\mathcal{X}\rightarrow\mathbb{R}$ on states as the edge map, $\nabla\varphi:\mathcal{X}\times \mathcal{X}\rightarrow \mathbb{R}$,
\begin{align*}
\nabla\varphi_{nm}:=\varphi_n-\varphi_m, ~~n,m\in\mathcal{X},
\end{align*}
and the discrete divergence of a function $U:\mathcal{X}\times\mathcal{X}\rightarrow\mathbb{R}$ on edges as the state map $\nabla\cdot U:\mathcal{X}\rightarrow \mathbb{R}$,
\begin{align*}
 [\nabla\cdot U]_n:=\sum_{m\in\mathcal X} (U_{mn}-U_{nm}), ~~ n\in\mathcal X.
\end{align*}
Then, it is easy to check that the discrete integration by parts formula,
\begin{align}\label{eq:IBPd}
\langle \nabla\varphi,U\rangle=-\langle \varphi,\nabla\cdot U\rangle,
\end{align}
holds.

Finally, define the mobility operator $ M_{ p}$ mapping an edge function $U$ to another edge function $M_{ p}(U)$, as  
\begin{equation}\label{eq:multiplic-d}
    [ M_{ p}( U)]_{nm}:=\frac12R_{nm}p_m^{\rm eq}\,\theta\bigg(\frac{ p_{n}}{p_n^{\rm eq}},\frac{ p_m}{p_m^{\rm eq}}\bigg)U_{nm},~n\neq m,
\end{equation}
while $ [ M_{ p}( U)]_{nn}:=0$. Here, the logarithmic mean $\theta(x,y)$ 
between two positive numbers $x$ and $y$ is given by
\begin{align}\label{logmean}
\theta(x,y)=\int_0^1x^{1-s}y^sds=\begin{cases}    \frac{x-y}{\ln(x)-\ln(y)},&\mbox{ if } x\neq y,\\x,&\mbox{ if }x=y.
\end{cases}
\end{align}
The mobility operator $M_{ p}$ can be alternatively understood as a weighted multiplication by $ p$. 
Note that if we are close to equilibrium, the mobility becomes a linear function of $p.$ Specifically, if $ p=p^{\rm eq}(1+\delta)$, then  $\theta\big(\frac{ p_{n}}{p_n^{\rm eq}},\frac{ p_m}{p_m^{\rm eq}}\big)=\frac12(\frac{ p_m}{p_m^{\rm eq}}+\frac{ p_n}{p_n^{\rm eq}})+O(\delta^2)$  and
\begin{align*}
    [ M_{ p}( U)]_{nm}=\frac14\big(R_{nm} p_m+R_{mn}p_n\big)U_{nm} +O(\delta^2),
\end{align*}
i.e., near equilibrium, the mobility is half of the edgewise dynamical activity $(R_{nm} p_m+R_{mn}p_n)/2$. 
In general, the edgewise mobility is bounded by the edgewise activity. In particular, using the fact that the logarithmic mean is upper-bounded by the arithmetic mean and lower-bounded by the geometric mean,
$$
A^f_{nm}\leq R_{nm}p_m^{\rm eq}\,\theta\bigg(\frac{ p_{n}}{p_n^{\rm eq}},\frac{ p_m}{p_m^{\rm eq}}\bigg)\leq A^d_{nm},
$$
where $A^f_{nm}:=\sqrt{R_{nm} p_mR_{mn} p_n}$ and $A_{nm}^d:=\frac{1}{2}(R_{nm} p_m+R_{mn} p_n)$ denote the edge contributions to the frenetic and dynamic activities, respectively. Thus, the mobility operator may also be understood as a measure of activity.

Moreover, in the appropriate continuum limit we recover the linear continuum mobility. Specifically, consider an underlying linear graph on the interval $[a,b]$ with $k+1$ nodes at a ``distance'' $\Delta x=(b-a)/k$ of each other. Let $H$ be a smooth function on $[a,b]$, $D\in(0,\infty)$, and consider the transition rates between neighboring nodes at positions $x_n$ and $x_{n+1}=x_n+\Delta x$, $R_{n+1,n}=\frac{D}{(\Delta x)^2}\exp\left[-\frac{\beta}{2}\left(H(x_{n+1})-H(x_n)\right)\right]$
and 
$R_{n,n+1}=\frac{D}{(\Delta x)^2}\exp\left[\frac{\beta}{2}\left(H(x_{n+1})-H(x_n)\right)\right]
$ for all $n\in\{1,\ldots,k\}$, and otherwise zero. Let $\rho$ and $\rho_{\rm eq} $ be fixed smooth densities on $[a,b]$, and consider their discrete counterparts $ p_n=\rho(x_n)\Delta x$ and $p^{\rm eq}_n=\rho_{\rm eq}(x_n)\Delta x$. Then, the mobility $  [ M_{ p}( U)]_{n+1n}$ is given by
\begin{align*}
\tfrac12\tfrac{D}{\Delta x}e^{-\frac{\beta}{2}\left(H(x_{n+1})-H(x_n)\right)}
\rho_{\rm eq}(x_n)\,\theta\big(\tfrac{\rho(x_{n})}{\rho_{\rm eq}(x_n)},\tfrac{\rho(x_{n+1})}{\rho_{\rm eq}(x_{n+1})}\big)U_{n+1 n}  ,
\end{align*}
for all $\Delta x>0.$ Since the ratio $\rho(x_{n+1})/\rho_{\rm eq}(x_{n+1})=\rho(x_{n})/\rho_{\rm eq}(x_{n})+O(\Delta x),$ the logarithmic mean to zeroth order in $\Delta x$ is $\rho(x_{n})/\rho_{\rm eq}(x_n)$. Therefore, as $\Delta x\to 0$ and we approach the continuum limit, we have
$$
 [ M_{ p}( U)]_{n+1n}=\frac{1}{2(\Delta x)^2}\Big(Dp_n + O\big((\Delta x)^2\big)\Big)U_{n+1n},
$$
thus recovering the linear in $ p_n$ relation of the mobility.


With these definitions, together with
$$
\phi_n:=\ln\frac{ p_n}{p_n^{\rm eq}}-\ln Z= \beta H_n+\ln  p_n,
$$ and the detailed balance condition \eqref{eq:detbal},
  the state dynamics 
\eqref{eq:CTMC} can be rewritten as
\begin{align}\label{eq:rhodotnew}
    \dot  p_n&=\sum_{m\in{\mathcal{X}}}R_{nm}p_m^{\rm eq}\bigg(\frac{ p_{m}}{p_m^{\rm eq}}-\frac{ p_n}{p_n^{\rm eq}}\bigg)
    \nonumber\\&=\sum_{m\in{\mathcal{X}}}R_{nm}p_m^{\rm eq}{\theta\bigg(\frac{ p_{m}}{p_m^{\rm eq}},\frac{ p_n}{p_n^{\rm eq}}\bigg)}(\phi_m-\phi_n)
      \nonumber\\&=2\sum_{m\in{\mathcal{X}}}[ M_{ p} (\nabla\phi)]_{mn}
      \nonumber\\&=\sum_{m\in{\mathcal{X}}}\big([ M_{ p} (\nabla\phi)]_{mn}-[ M_{ p}(\nabla\phi)]_{nm}\big)
       \nonumber\\&=\big[\nabla\cdot( M_{ p} (\nabla\phi))\big]_{n}.
\end{align}
The analogy to the continuous setting motivates us to consider the Riemannian metric
\begin{align}\label{eq:metric-d}
    g_{ p}(\dot \mu,\dot\nu):=\langle\nabla \varphi,  M_{ p} (\nabla\psi)\rangle,
\end{align}
where $\varphi,\psi\in T_{ p}^*\mathcal M$ are related to $\dot \mu,\dot\nu\in T_{ p}\mathcal M $ through the Onsager operator
\begin{align}\label{onsager}
\mathbb K_{ p}(\varphi):=-\nabla\cdot( M_{ p} (\nabla\varphi)),
\end{align}
as
$
\dot\mu=-\mathbb K_{ p}(\varphi) \mbox{ and }\dot\nu=-\mathbb K_{ p}(\psi).
$
The inverse of the Onsager operator is $\mathbb G_{ p}:\dot \mu\mapsto\mathbb G_{ p}(\dot\mu)=-\varphi$, where $\varphi$ is the unique (zero-mean) solution to
$
\nabla\cdot( M_{ p} ( \nabla\varphi))=\dot\mu
$
\footnote{Uniqueness up to an additive constant is ensured by the strict positivity of $ p~$\cite[Prop. 3.26]{maas2011gradient}.}.
Thus,  we can rewrite the metric $g$ using integration by parts from \eqref{eq:IBPd} as
$$
 g_{ p}(\dot \mu,\dot\nu)=\langle\nabla \varphi,  M_{ p} (\nabla\psi)\rangle=\llangle \varphi,\mathbb K_{ p}(\psi)\rrangle=\llangle\mathbb G_{ p}(\dot\mu),\dot \nu\rrangle.
$$

Using \eqref{eq:rhodotnew} and the definition of the Onsager operator in \eqref{onsager}, the evolution \eqref{eq:CTMC} can now be written as the gradient flow with respect to the metric \eqref{eq:metric-d} 
$$
\dot p_n=- [\mathbb K_{ p}(d\mathcal F_{ p})]_n,
$$
of the free energy functional
\begin{align}\label{eq:freedisc}
\mathcal F( p)=\sum_{n\in\mathcal X}(\beta H_n+ \ln  p_n)  p_n,
\end{align}
where
   $ [d\mathcal{F}_{p}]_n=\phi_n$, up to an irrelevant additive constant. As in the continuous setting, \eqref{eq:freedisc} can  be re-written in terms of the relative entropy
$$
\mathcal F( p)=\sum_{n\in\mathcal X}  p_n\ln \frac{ p_n}{p_n^{\rm eq}}-\ln Z,
$$
which implies that $\mathcal F( p)+\ln Z$ is positive, equal to zero only when ${ p_n=p_n^{\rm eq}}$. Moreover, it is non-increasing along solutions to \eqref{eq:CTMC}, since
\begin{align}\label{eq:dotF-d}
    \frac{d}{dt}\mathcal F( p)&=\llangle d\mathcal F_{ p},\dot p\rrangle=-g_{ p}(\dot p,\dot p)\leq 0.
\end{align}
Thus, $\mathcal{F}+\ln Z$ acts as a Lyapunov function, ensuring convergence to the equilibrium distribution.

Analogously to the continuous setting, this metric induces a discrete Wasserstein-2 distance,
\begin{align}\label{eq:discreteW2}
    W_2( p_0, p_\tau)^2&:=\inf_{ \{p(t)\}: p_0, p_\tau}\tau\int_0^\tau g_{ p}(\dot p,\dot p) dt\\&\phantom{:}=\inf_{ \{p(t)\}: p_0, p_\tau}\ell_{ p}^2,\nonumber
\end{align}
where the last equality is achieved for constant speed parametrization \cite[Lemma 3.9]{maas2011gradient}, and the Riemannian length of a curve is given by
$$
\ell_{ p}:=\int_0^\tau\sqrt{g_{ p}(\dot p,\dot p)}dt.
$$
Here, we have used the shorthand notation $\{p(t)\}: p_0,p_\tau$ to denote that the optimization is over the admissible trajectories $\{p(t)\}_{t\in[0,\tau]}$ that have $p_0$ and $p_\tau$ as endpoints. 
This geodesic distance can be seen as arising as the solution to the discrete optimal transport problem~\cite{maas2011gradient}
\begin{align}
\label{eq:discreteOMT}
  W_2( p_0, p_\tau)^2=&\inf_{V, p}\tau\int_0^\tau||V||^2_{M_{ p}} dt
\\    \nonumber& \mbox{s.t.} \ \dot  p=\nabla\cdot( M_{ p}(V)),\  p(0)= p_0,\, p(\tau)= p_\tau,
\end{align}
where $||V||^2_{M_{ p}}:=\langle V,  M_{ p}(V)\rangle$,
since optimal velocities are of the form $V_{\rm opt}=\nabla\varphi$ \cite[Lemma 3.6]{maas2011gradient}. 
Note that the metric depends on the mobility operator $M_p$, which needs to be fixed in advance as a function of $p$ and $V.$

\subsection{Quantum optimal transport}

Let us consider finite-dimensional quantum systems subject to purely dissipative Lindblad dynamics; we will again show that these dynamics can be recast as the gradient flow of a free energy functional \cite{carlen2017gradient,chen2017matrix,mittnenzweig2017entropic}. Let the state of the system be captured by the density matrix $\varrho \in\mathcal M:=\mathcal P_*(\mathbb C,d)$, where $\mathcal P_*(\mathbb C,d)$ is the smooth manifold of  $d\times d$ positive-definite Hermitian matrices with trace $1$. We consider the following Davies-type evolution~\cite{davies1974markovian,lindblad1976generators,gorini1976completely} 
\begin{equation}\label{eq:diss-lind}
    \dot\varrho =\sum_{k\in K}\gamma_k\Big(L_k\varrho  L_k^\dagger -\frac12\{L_k^\dagger L_k,\varrho \}\Big),
\end{equation}
which is termed ``purely dissipative'' for lacking a unitary part. Here, the symbol $\dagger$ denotes the Hermitian adjoint operation, the set of transition channels is
 $K:=\{-N,\ldots,-1,1,\ldots,N\}$, $L_k\in\mathbb C^{d\times d}$ are the jump operators, and $\gamma_k >0$ are the associated jump rates.
 
 The jump operators are assumed to occur in adjoint pairs $L_{-k}=L_k^\dagger$ and satisfy  $[H,L_k]=\omega_kL_k$,
 where $H=H^\dagger\in\mathbb{C}^{d\times d}$ is the Hamiltonian of the system. 
Here, $\omega_k$ is the Bohr frequency associated with channel $k$, and $\omega_{-k}=-\omega_k$
\footnote{
 Operators $L_k$ are called jump operators because they map an energy eigenstate with energy $H_m$ into another eigenstate with energy $H_m+\omega_k$. Indeed, if 
$H|m\rangle = H_m |m\rangle$, then $H\,L_k|m\rangle = (H_m + \omega_k)\,L_k|m\rangle.$ Thus the operators $L_k$ act as energy ladder (jump) operators implementing 
transitions between levels separated by the Bohr frequency $\omega_k$.
In the particular case of a rank-one channel $L_k\propto|n\rangle\langle m|$, channel $k$ represents the transition $m\to n$ and $\omega_k=H_n-H_m$. More generally, $L_k$ may combine several transitions sharing the same Bohr frequency $\omega_k$.}.
 The corresponding transition rates satisfy the thermal KMS relation \( \gamma_k=e^{-\beta\omega_k}\gamma_{-k}, \) where $\beta  > 0$ is the inverse temperature of a single heat bath. Together, these conditions imply GNS quantum detailed balance with respect to the Gibbs state $\varrho_{\rm eq}:=e^{-\beta H}/Z$ with \(Z:=\tr(e^{-\beta H})\)~\footnote{Here, GNS quantum detailed balance means that the corresponding dissipative Heisenberg-picture generator is self-adjoint with respect to the GNS inner product \[ \langle A,B\rangle_{\varrho_{\rm eq},{\rm GNS}} := \tr(\varrho_{\rm eq}A^\dagger B). \] This is the detailed-balance condition used in~\cite{carlen2017gradient}. It should be distinguished from KMS quantum detailed balance, which is defined using \[ \langle A,B\rangle_{\varrho_{\rm eq},{\rm KMS}} := \tr\!\left( \varrho_{\rm eq}^{1/2}A^\dagger \varrho_{\rm eq}^{1/2}B \right). \]}.
  Finally, we assume the system dynamics to be ergodic, i.e., the only operators $A$ such that $[L_k,A]=0$ for all $k\in K$ are scalar multiples of the identity operator~\footnote{Formally, let $\mathcal{A}$ be the $*$-algebra generated by the jump operators $\{L_k\}_{k\in K}$. 
The assumption that $[L_k,A]=0$ for all $k$ implies $A=c{\,\rm Id}$ means that the commutant 
$\mathcal{A}'=\{A:[A,X]=0\ \forall\,X\in\mathcal{A}\}$ is trivial, i.e., $\mathcal{A}'=\mathbb{C}\,{\rm Id}$. 
Equivalently, the representation of $\mathcal{A}$ on $\mathbb{C}^d$ is irreducible, in the sense of Schur's Lemma.}. 
The detailed-balance conditions ensure that the Gibbs state
\(\varrho_{\rm eq}\) 
is a stationary equilibrium state; the ergodicity assumption ensures that this
stationary state is unique.

Let $\mathfrak h$ denote the Hilbert space formed by equipping $\mathbb C^{d\times d}$ with the trace inner product $\langle A,B\rangle=\tr\{A^\dagger B\}.$ Consider the Hilbert space $\mathfrak h_K:=\oplus_{k\in K}\mathfrak h_k$, where each $\mathfrak h_k$ is a copy of $\mathfrak h.$  For $\mathbf V\in\mathfrak h_K$, let $V_k$ denote the component of $\mathbf V$ in $\mathfrak h_k$. We equip $\mathfrak h_K$ with the usual inner product
$$
\langle \mathbf U,\mathbf V\rangle=\sum_{k\in K}\tr\{U_k^\dagger V_k\}.
$$
Let us define the quantum partial derivatives $\partial_k:\mathfrak h\to\mathfrak h_k$
$$
\partial_k \varphi=[L_k,\varphi] \mbox{ so that } \partial_k^\dagger \varphi=[L_k^\dagger,\varphi],~~\varphi\in\mathfrak h,
$$
for all $k\in K.$
The associated gradient operator $\nabla:\mathfrak h\to\mathfrak h_{K}$ is defined as
$$
\nabla \varphi=\Big[\partial_{-N}\varphi,\cdots,\partial_{-1}\varphi,\partial_{1}\varphi,\ldots,\partial_N\varphi\Big]^\top,
$$
and the divergence operator $\nabla~\cdot:\mathfrak h_K\to\mathfrak h$ as
$$
\nabla \cdot \mathbf V=-\sum_{k\in K}\partial_k^\dagger V_{k}.
$$
By using $\tr([L_k,\varphi]^\dagger V_k)=\tr(\varphi^\dagger[L_k^\dagger,V_k])$, it is easy to check that the integration by parts formula holds, i.e.,
\begin{align}\label{eq:IBPquantum}
\langle\nabla\varphi ,\mathbf{V} \rangle=-\langle\varphi ,\nabla\cdot\mathbf{V}\rangle.
\end{align}

Moreover, let us define the quantum mobility operator  $ M_{\varrho }^k$  acting on 
$\mathfrak h_{k}
,~k\in K$, through a non-commutative (Bogoliubov-Kubo-Mori type) multiplication by $\varrho$, as
$$
 M_{\varrho }^k(V_k):=\frac{\gamma_k}{2}e^{\frac{\beta\omega_k}{2}}\int_0^1 \big(e^{\frac{\beta\omega_k}{2} }\varrho \big)^sV_k\big(e^{-\frac{\beta\omega_k}{2}}\varrho \big)^{1-s}ds,
$$
and the corresponding operator acting on $\mathfrak h_K$ as
\begin{equation}
    \label{eq:M-q}
    M_{\varrho }(\mathbf{V}):=\Big[M_{\varrho }^{-N}(V_{-N}),
 \ldots,
M_{\varrho }^{N}( V_{N})\Big]^\top.
\end{equation}
The operator $M^k_{\varrho }$ is the non-commutative analog of the edgewise mobility in the discrete case \eqref{eq:multiplic-d}. Indeed, let $H=\sum_{m=1}^dH_m|m\rangle\langle m|,$ where $|1\rangle,\cdots, |d\rangle$ is the energy eigenbasis.  If $\varrho$ is diagonal in the energy eigenbasis, $V_k=|n\rangle\langle m|$, with $n,m\in\{1,\ldots,d\}$, $n\neq m$, and \( H_n-H_m=\omega_k, \) then
\begin{align}\nonumber
     M_{\varrho }^k(V_k)&=\frac{\gamma_k}{2}e^{-\beta H_m}\int_0^1 \big(e^{\beta H_n}p_n\big)^s\big(e^{\beta H_m}p_m\big)^{1-s}ds~V_k\\
     &=\frac{1}{2} \gamma_k\,p_m^{\rm eq}\, \theta\bigg(\frac{p_{m}}{p_m^{\rm eq}},\frac{p_n}{p_n^{\rm eq}}\bigg)V_k,\label{eq:mobility-q2d}
 \end{align}
 where $p_{i}$ and $p^{\rm eq}_{i}$ denote the $i$-th eigenvalues of $\varrho $ and $\varrho _{\rm eq}$, respectively.
 This  mirrors the discrete mobility \eqref{eq:multiplic-d}.


With these definitions, we can
look back at the evolution equation \eqref{eq:diss-lind} and re-write it as
\begin{align}\label{eq:q-cont}
    \dot\varrho = \nabla\cdot( M_{\varrho }(\nabla\phi)),
\end{align}
where
$$
\phi:=\beta H+\ln\varrho =\ln\varrho -\ln{\varrho _{\rm eq}}-\ln Z.
$$
To prove \eqref{eq:q-cont}, first note that if we can show 
\begin{equation}
    \label{eq:miracle}
  M_{\varrho }^k(\partial_k\phi)=\frac{1}{2}(\gamma_kL_k\varrho  -\gamma_{-k}\varrho  L_k),
\end{equation}
which (for $\beta=0$) is the non-commutative analog of $\rho\nabla\ln\rho=\nabla\rho$~\footnote{For $\beta=0$, equation~\eqref{eq:miracle} reads
  $M^k_{\varrho }(\partial_k\!\ln\varrho )
   = \frac{\gamma_k}{2}\partial_k\varrho $.%
},
then rearranging the terms of the sum we obtain the desired result since 
\begin{align*}
 \nabla\cdot( M_{\varrho }(\nabla\phi)) &=-\sum_{k\in K}[L_k^\dagger  , M_{\varrho }^k(\partial_k\phi)]\\
 &=  \frac{1}{2}\sum_{k\in K}[\gamma_kL_k\varrho  -\gamma_{-k}\varrho  L_k,L_k^\dagger]\\
 &= \frac{1}{2}\sum_{k\in K}\Big([\gamma_kL_k\varrho ,L_k^\dagger] -[\gamma_{k}\varrho  L^\dagger_k,L_k]\Big)\\
 &=\sum_{k\in K}\gamma_k\Big(L_k\varrho  L_k^\dagger -\frac12\{L_k^\dagger L_k,\varrho \}\Big).
\end{align*}
To show \eqref{eq:miracle}, consider the matrix valued function $f(s)=\frac{\gamma_k}{2}e^{\frac{\beta\omega_k}{2}}(e^{\frac{\beta\omega_k}{2}}\varrho )^{1-s}L_k(e^{-\frac{\beta\omega_k}{2}}\varrho )^{s}$. Since using $[L_k,H]=-\omega_k L_k$ we have that
$
[L_k,\phi]=L_k\ln\big(e^{-\frac{\beta\omega_k}{2}}\varrho \big)-\ln\big(e^{\frac{\beta\omega_k}{2}}\varrho \big)L_k, 
$ the derivative of $f$ may be written as
\begin{align*}
    f'(s)&=\frac{\gamma_k}{2}e^{\frac{\beta\omega_k}{2}}(e^{\frac{\beta\omega_k}{2}}\varrho )^{1-s}[L_k,\phi](e^{-\frac{\beta\omega_k}{2}}\varrho )^{s}.
\end{align*}
Thus, $M_\varrho^k(\partial_k\phi)=\int_0^1f'(1-s)ds$, while the right hand side of 
 \eqref{eq:miracle} is $f(1)-f(0)$.
Therefore, applying the substitution $u=1-s$ and the fundamental theorem of calculus $\int_0^1f'(u)du=f(1)-f(0)$ yields \eqref{eq:miracle}  (see \cite[Lemma 5.5]{carlen2017gradient} for more details).

This motivates us to consider the following Riemannian metric \cite{carlen2017gradient}
\begin{align}\label{eq:metric-q}
    g_{\varrho }(\dot \mu,\dot\nu):=\langle \nabla\varphi, M_{\varrho }(\nabla \psi)\rangle,
\end{align}
 where $\varphi,\psi\in T_{\varrho }^*\mathcal M$ are related to $\dot \mu,\dot\nu\in T_{\varrho }\mathcal M $ 
 through the Onsager operator
\begin{equation}
\label{eq:q-onsager}
    \mathbb K_{\varrho }(\varphi):=-\nabla\cdot( M_{\varrho }(\nabla\varphi)),
\end{equation}
as
$
\dot\mu=-\mathbb K_{\varrho }(\varphi) \mbox{ and }\dot\nu=-\mathbb K_{\varrho }(\psi).
$
The inverse of the Onsager operator is $\mathbb G_{\varrho }:\dot\mu\to\mathbb G_\varrho(\dot\mu)=-\varphi$, where $\varphi$ is the unique traceless \cite[Theorem 7.3]{carlen2017gradient} solution to
$
\nabla\cdot( M_{\varrho }(\nabla\varphi))=\dot\mu.
$
Then, we can rewrite the metric as
$$
 g_{\varrho }(\dot \mu,\dot\nu)=\langle \nabla\varphi, M_{\varrho }(\nabla \psi)\rangle=\llangle \varphi,\mathbb K_{\varrho }(\psi)\rrangle=\llangle\mathbb G_{\varrho }(\dot\mu),\dot \nu\rrangle,
$$
where we have used integration by parts formula \eqref{eq:IBPquantum}.

Combining equations \eqref{eq:q-cont} and \eqref{eq:q-onsager}, the evolution \eqref{eq:diss-lind} can  be written as the gradient flow with respect to the metric \eqref{eq:metric-q} 
\begin{equation}\label{eq:q-grad}
  \dot\varrho = -\mathbb K_{\varrho }(d\mathcal F_{\varrho }),  
\end{equation}
of the free energy functional
\begin{align}
\mathcal F(\varrho )=\tr\{( \beta H+ \ln \varrho ) \varrho \},\label{freequanenergy}
\end{align}
where
   $ d\mathcal{F}_{\varrho }=\phi$, up to an additive multiple of the identity. As in the continuous and discrete settings, \eqref{freequanenergy} can be re-written in terms of the quantum relative entropy
$$
\mathcal F(\varrho )=\tr\{(\ln\varrho -\ln\varrho _{\rm eq})\varrho \}-\ln Z.
$$
 This implies that $\mathcal F(\varrho )+\ln Z$ is positive, equal to zero only when ${\varrho =\varrho _{\rm eq}}$. Moreover, it is non-increasing along solutions to~\eqref{eq:diss-lind}, since
\begin{align}\label{eq:dotF-q}
    \frac{d}{dt}\mathcal F(\varrho )&=\llangle d\mathcal F_{\varrho },\dot\varrho \rrangle=-g_{\varrho }(\dot\varrho ,\dot\varrho )\leq 0.
\end{align}
Thus, $\mathcal{F}$ (modulo $\ln Z$) acts as a Lyapunov function, ensuring convergence to the equilibrium state.

The metric \eqref{eq:metric-q} defines a quantum Wasserstein-2 geodesic distance,
\begin{align}\label{eq:qOT}
    W_2(\varrho _0,\varrho _\tau)^2:&=\inf_{\{\varrho(t)\} :\varrho _0,\varrho _\tau}\tau\int_0^\tau g_{\varrho }(\dot\varrho ,\dot\varrho ) dt\\&=\inf_{\{\varrho(t)\} :\varrho _0,\varrho _\tau}\ell_{\varrho }^2,\nonumber
\end{align}
where the Riemannian length of the curve is
$$
\ell_{\varrho }:=\int_0^\tau\sqrt{g_{\varrho }(\dot\varrho ,\dot\varrho )}dt,
$$
and we have used the shorthand $\{\varrho(t)\}: \varrho_0,\varrho_\tau$ to signify that we are searching for trajectories $\{\varrho(t)\}_{t\in[0,\tau]}$ with endpoints $\varrho(0)=\varrho_0$ and $\varrho(\tau)=\varrho_\tau.$
This distance squared is the solution to the
quantum optimal transport problem
\begin{align}
\label{eq:QOMT}
W_2(\varrho _0,\varrho _\tau)^2=&\inf_{\mathbf V,\varrho 
}\tau\int_0^\tau||
\mathbf V||^2_{M_{\varrho }} dt
\\& \mbox{s.t.} \ \dot \varrho =\nabla\cdot( M_{\varrho }(\mathbf{V})),\ \varrho (0)=\varrho _0,\,\varrho (\tau)=\varrho _\tau,     \nonumber
\end{align}
where $\|\mathbf V\|^2_{M_{\varrho }}:=\langle \mathbf{V}, M_{\varrho }(\mathbf{V})\rangle$. Once again, this is due to optimal velocities being of the form $\mathbf V_{\text{opt}}=\nabla\varphi_{\rm opt}$ \cite[Theorem 7.3]{carlen2017gradient}.
Note that the metric depends on the mobility operator $M_\varrho$ and the choice of jump operators $L_k$ through the definition of the gradient. 
Alternatively, we may define system-independent derivatives in analogy to the classical (continuous and discrete) settings, shifting all the system dependence to a weighted mobility operator $\hat M_\varrho$, see Appendix \ref{app:basisless} for more details. Then, we may think of the metric as depending only on $\hat M_\varrho$, which needs to be fixed in advance as a function of $\varrho$ and $\mathbf V$. Thus, in this sense, we will understand this geometry as depending uniquely on the mobility operator.

A couple of remarks are in order. Consider the optimal transport problem with fixed mobility operator prescribed by the dynamics of the form \eqref{eq:diss-lind} through \eqref{eq:M-q}, where the associated Hamiltonian $H$ is non-degenerate, and we have rank-one jump operators.
When the endpoints $\varrho_0$, $\varrho_\tau$ 
commute with the Hamiltonian $H$,
there exists an optimal solution to the quantum optimal transport problem $\varrho$ that commutes with $H$ at all intervening times (see Appendix \ref{app:q-d-W2}).  Consequently, the optimal dynamics may be written in the energy eigenbasis as a discrete Markov process in the form of \eqref{eq:CTMC}. In fact, these dynamics coincide with the optimal dynamics arising from the discrete optimal transport problem, with the mobility operator prescribed by $R_{nm}=\gamma_k$ where $k:m\to n$, between endpoints  $[p_0]_i$ and  $[p_\tau]_i$ given by the $i$-th eigenvalues of $\varrho_0$ and $\varrho_\tau$, respectively. Moreover, in this case,
the discrete and quantum distances coincide
\begin{equation}
    \label{eq:q-d-W2}
     W_2(\varrho _0,\varrho _\tau)= W_2(p_0,p_\tau).
\end{equation}
While it is known that the quantum metric, when restricted to the submanifold of diagonal states with respect to the $H$ basis, is equivalent to the discrete metric~\cite{carlen2020non}, this result is stronger.
It implies that the quantum optimization is not able to reduce the cost through coherences when the endpoints commute. 
We prove these statements in Appendix~\ref{app:q-d-W2}, where we derive the first-order conditions for optimality of the discrete and quantum problems. We show that, when the endpoints commute with the Hamiltonian, a minimizer of the quantum problem can be built through a minimizer of the discrete problem, leading to the same distance. This relies on the convexity of the optimization problem, which we now briefly discuss.

\subsection{Convexity of the optimal transport problems}
Indeed, the three optimal transport problems admit a formulation in flux (current) variables in which the minimization is convex. Concretely, let us introduce the current
\(\mathcal J := M_\mu(V)\), where $\mu$ is to be replaced by $\rho$ and $ p$ to obtain  the continuous and  discrete expressions, respectively, and $(\mu,V)$ by $(\varrho,\mathbf V)$ for the quantum setting. Here, the mobility operator $M_\mu$ is given by $\rho D,$ \eqref{eq:multiplic-d}, and \eqref{eq:M-q}, respectively. Then, we can rewrite each problem as
\begin{align}\label{eq:convexity}
    &\inf_{\mu,\,\mathcal J}\ \tau\int_0^\tau \langle \mathcal J,\, M_\mu^{-1}(\mathcal J)\rangle\,dt\\
    &\text{s.t.}\  \dot{\mu}= \nabla\!\cdot \mathcal J, \  \mu(0)=\mu_0,\ \mu(\tau)=\mu_\tau,\nonumber
\end{align}
where \(M_\mu^{-1}\) is understood on the appropriate space (range of $M_\mu$), so that the inverse is well-defined. In this flux formulation, the admissible set is affine (linear continuity equation with fixed endpoints), and the action density \((\mu,\mathcal J)\mapsto \langle \mathcal J, M_\mu^{-1}(\mathcal J)\rangle\) is jointly convex \cite[Prop. 9.6]{carlen2020non}. Hence, the three optimal transport problems are convex optimization problems in the flux variables, implying that any pair $(\mu,\mathcal J)$ that satisfies the first-order optimality conditions is a global minimizer.

\section{Geometry of entropy production in purely dissipative dynamics}
\label{sec:III}

As we have seen in Section \ref{sec:grad-flows}, a very similar gradient flow structure is present across three different  purely dissipative dynamics (see Table \ref{tab:fullwidth}). 
For these autonomous systems, i.e., systems whose dynamics are time-independent, detailed balance ensures convergence to equilibrium. Moreover, as the systems relax towards equilibrium, free energy decreases only as a result of dissipation.
Consequently, the rate at which free energy decreases is identified with the entropy production rate
\begin{equation}
    \label{eq:def-epr}
    \dot \Sigma:=-\llangle d\mathcal{F}_\mu,\dot\mu\rrangle=g_{\mu}(\dot\mu,\dot\mu)~ =-\dot{\mathcal{F}}, 
\end{equation}
where $d\mathcal{F}_\mu = \beta H+\ln \mu$. 
Alternatively, the quantity $\dot{\Sigma}$ can be understood as the sum of the entropy production rate in the system $-\langle \ln\mu,\dot\mu\rangle$ plus the entropy production rate in the environment $-\langle \beta H,\dot\mu\rangle$ \cite{sekimoto2010stochastic}. Here, and in what follows, $\mu$ is to be replaced by $\rho, p,$ or $\varrho,$ to obtain  the continuous, discrete, and quantum expressions, respectively.
Thus, the three optimal transport metrics  \eqref{eq:metric-cl}, \eqref{eq:metric-d}, and \eqref{eq:metric-q} introduced in Section \ref{sec:grad-flows}  quantify the total entropy production rate in their respective settings. Note that, since detailed balance is assumed, no housekeeping entropy is produced, and the entropy production rate corresponds to the excess entropy production rate.

The relationship between optimal transport and stochastic thermodynamics was first uncovered for continuous (overdamped) systems \cite{aurell2011optimal,aurell2012refined}, where entropy production was seen to be bounded by the (continuous) Wasserstein-2 distance. In this setting, the connection has been thoroughly studied~\cite{chen2019stochastic,nakazato2021geometrical,ito2024geometric}, and found applications in the derivation of speed limits and uncertainty relations~\cite{nakazato2021geometrical,otsubo2020estimating,ito2024geometric}, the design of thermodynamic engines~\cite{fu2021maximal,movilla2021energy,taghvaei2021relation,oikawa2025experimentally}, and
the decomposition of entropy production for systems that do not satisfy detailed balance~\cite{dechant2022geometric,dechant2022geometric2,miangolarra2024minimal}. 
Various 
extensions to underdamped~\cite{dechant2019thermodynamic,sabbagh2024wasserstein,watabe2025lower}, discrete~\cite{van2021geometrical,dechant2022minimum,van2023thermodynamic,yoshimura2023housekeeping,delvenne2024thermokinetic,nagayama2025geometric,kolchinsky2026generalized},  and quantum systems~\cite{van2021geometrical,van2023thermodynamic,yoshimura2025force}  have been developed since.  However, the way to extend the classical continuous result to discrete and quantum systems is not unique.
Several different results have arisen from diverse approaches that focus on different properties of continuum optimal transport.

Specifically, 
some works on discrete systems have found that analogs to the continuous Wasserstein-1 distance bound entropy production \cite{dechant2022minimum, van2023thermodynamic,delvenne2024thermokinetic,kolchinsky2026generalized} and provide speed limits \cite{van2023topological}. Other works have followed the approach that will be presented herein, and characterized entropy production by the discrete Wasserstein-2 distance \eqref{eq:discreteW2} arising from the gradient flow structure \cite{van2021geometrical,yoshimura2023housekeeping,nagayama2025geometric}.
While 
the quantum setting has been significantly less studied, both Wasserstein-1 \cite{van2023thermodynamic} and Wasserstein-2  \cite{van2021geometrical,yoshimura2025force}  approaches that bound entropy production have been proposed. 
We now follow an avenue similar to the latter.

As a result of \eqref{eq:def-epr}, if $\{\mu(t)\}_{t\in[0,\tau]}$ denotes a thermodynamic transition with endpoints $\mu_0$ and $\mu_\tau$ arising from detailed-balanced autonomous dynamics, and $\dot \Sigma$ denotes the  entropy production rate associated to such transition, then we have that
\begin{equation}
    \label{eq:W2-time-invar}
\int_0^\tau\dot\Sigma \, dt\geq \frac{1}{\tau}W_2(\mu_0,\mu_\tau)^2,
\end{equation}
in all three cases. This is clear since the trajectories $\{\mu(t)\}_{t\in[0,\tau]}$ and associated velocities are in the admissible set for the respective optimal transport problems (see \eqref{eq:cost-OMT},\eqref{eq:discreteOMT},\eqref{eq:QOMT}), and therefore the minimal cost (Wasserstein-2 distance) is not greater than the cost evaluated at   $\{\mu(t)\}_{t\in[0,\tau]}$ (left-hand side).  
In this way, the Wasserstein-2 distance between the specified endpoint states provides a finite-time correction to the second law, bounding total entropy production along a given thermodynamic transition. We may rearrange the terms of the inequality to view it as a \emph{thermodynamic speed limit} that bounds the minimum time $\tau$ required for a thermodynamic process ($\mu_0\to\mu_\tau$) with a given entropic budget $\int_0^\tau\dot\Sigma dt$. This result was first obtained in~\cite{aurell2011optimal,aurell2012refined} for the continuous setting and in~\cite{van2021geometrical}, for discrete and quantum settings. Furthermore, since by definition $g_\mu(\dot\mu,\dot\mu)$ is the square of the metric derivative, in view of \eqref{eq:def-epr} the entropy production rate can be written in terms of the $W_2$-speed of the thermodynamic transition through \begin{align*}
    \dot{\Sigma}(t)=\lim_{\delta\rightarrow 0^+}\cfrac{W_2(\mu(t),\mu(t+\delta))^2}{\delta^2},
\end{align*}
which holds for the continuous, discrete, and quantum settings. 



Thus far, we have assumed that the systems considered herein are autonomous, that is, the involved parameters are independent of time. If the systems are externally driven through a time-varying Hamiltonian instead, the results presented until now still hold, even if three important changes must be made.  Namely, in Section~\ref{sec:grad-flows}, (i) the free energy functional $\mathcal{F}$ should be defined using the \textit{instantaneous equilibrium distribution} $\mu^{\mathrm{eq}}(t)$ $\propto e^{-\beta H(t)}$, and as a consequence, (ii) the rate of change of the free energy  has an extra term (c.f. (\ref{eq:dotF-c},\,\ref{eq:dotF-d},\,\ref{eq:dotF-q})),
$$
\dot{\mathcal F}=\beta\langle\dot H(t),\mu\rangle-g_{\mu}(\dot\mu,\dot\mu),
$$
where the extra term $\beta\langle\dot H(t),\mu\rangle$ accounts for work put into the system.  Here, we have made explicit the time dependence to highlight that some maps (e.g. $H$), which were previously introduced as time-invariant, are now time-varying.
These two changes (i,ii) do not alter the gradient flow structure, which is understood as the steepest descent towards the instantaneous equilibrium at each instant in time. 
The entropy production rate is still given by the strictly negative, ``non-work'' contribution to $\dot{\mathcal F}$, $\dot\Sigma=g_{\mu}(\dot\mu,\dot\mu)$.

However, crucially, (iii) the discrete and quantum mobility operators introduced in Section \ref{sec:grad-flows} become time-varying. 
This implies that the related
Riemannian metrics are also time-varying, thus preventing us from defining a Wasserstein-2 distance as we understand it in Section \ref{sec:grad-flows}.
Still, one can define a pseudo-distance with a time-varying mobility operator and metric; this is the approach followed in \cite{van2021geometrical}.  Specifically,  for a thermodynamic transition $\{\mu(t)\}_{t\in[0,\tau]}$ with specified endpoints $\mu_0$ and $\mu_\tau$, arising from possibly time-varying detailed-balanced dynamics, we may use the fact that $\dot\Sigma=g_{t,\mu}(\dot\mu,\dot\mu)$ and the Cauchy-Schwarz inequality to bound entropy production as
\begin{equation}
    \label{eq:W2-time-var}
\tau\int_0^\tau\dot\Sigma \, dt\geq \bigg(\inf_{\mu :\mu_0,\mu_\tau}\int_0^\tau \sqrt{g_{t,\mu }(\dot\mu ,\dot\mu )}dt\bigg)^2,
\end{equation}
where $g_{t,\mu }(\dot\mu,\dot\mu)$ is given by the metric introduced in Section \ref{sec:grad-flows} with respect to the instantaneous mobility operator $M_{t,\mu}$, which is in general time-dependent in the discrete and quantum settings. 
   In the discrete and quantum settings, the square root of the right-hand side of \eqref{eq:W2-time-var} does not in general satisfy symmetry in $\mu_0,\mu_\tau$, nor the triangle inequality, and is thus termed a pseudo-distance. Indeed, because the metric varies in time, it cannot be interpreted as a Riemannian geodesic distance between endpoints. In particular,  constant arc-length reparametrization is not always possible, and therefore equality in \eqref{eq:W2-time-var} need not be achieved. Thus, this time-dependent formulation does not usually yield a sharp variational characterization of the minimum dissipation.


Alternatively to bounding entropy production through a time-varying mobility induced metric as in~\eqref{eq:W2-time-var}, we may consider physical systems with fixed mobilities $M_\mu$ as a function of $\mu$~\cite{yoshimura2023housekeeping,nagayama2025geometric}. That is, we may choose to modify kinetics in order to keep the mobility operator fixed, even while the Hamiltonian is changing. Then, the state space is endowed with a genuine Wasserstein-2 geometry, and we obtain a tighter notion of optimality.
Specifically, we obtain
\begin{align}\nonumber
\!\!\!\frac{1}{\tau}W_2(\mu_0,\mu_\tau)^2=\inf_{\phi,\mu }\bigg\{\int_0^\tau\dot\Sigma dt:\ \dot\mu &=\nabla\cdot(M_\mu(\nabla \phi)),\\\mu (0)&=\mu_0,\, \mu (\tau)=\mu_\tau
\bigg\}.\label{eq:W2-minE} 
\end{align}
 Since any potential $\phi$ can be obtained through $\phi=\beta \tilde H+\ln\mu$ by an appropriate Hamiltonian $\tilde H$, we may understand the square of the Wasserstein-2 distance as the minimum entropy production required to drive a system through Hamiltonian $\tilde H$ between two endpoints, for a fixed $M_\mu$. 
Fixing the mobility operator is natural in the classical setting, where the diffusion matrix $D$ is typically fixed. 
However, in the discrete and quantum settings, it is less intuitive.

We may motivate fixing the mobility operator in the discrete and quantum settings by noting that, with no other constraints, any thermodynamic process can be carried out in finite time with vanishing entropy production by suitably adjusting the rates/jump operators at each instant.  
Specifically, it is known that the rates $R_{nm}$ can be chosen to drive the discrete system through any trajectory $\{p(t)\}_{t\in[0,\tau]}$ with arbitrarily small entropy production \cite{dechant2022minimum}. A typical approach to overcome this unphysical result is to fix the activity~\cite{dechant2022minimum}. Moreover, in Appendix \ref{app:q-dechant} we show that an analogous statement is true in the quantum setting. That is, we show that there exist Lindblad dynamics (containing a unitary term) that drive the quantum system along any trajectory $\{\varrho(t)\}_{t\in[0,\tau]}$ with arbitrarily small entropy production. The proof of this statement is similar to that of the discrete result and likewise 
requires allowing the Hamiltonian, jump operators, and jump rates to be not necessarily detailed-balanced, time-varying, and to depend on the given trajectory (see Appendix~\ref{app:q-dechant}).
Thus, in the absence of kinetic constraints, any quantum trajectory in $\mathcal P_*(\mathbb C,d)$ can be implemented with arbitrarily small entropy production, rendering unconstrained finite-time entropy-production bounds trivial.
In this work, instead of fixing the activity or the averaged mobility~\cite{van2023thermodynamic}, we must fix the mobility operator $M_\mu$, leading to a fully specified dissipative structure $\mathbb K_\mu$. 

Physically interpreting such fixed mobility dynamics turns challenging in the discrete and quantum settings. One may enforce fixed mobilities in discrete systems by allowing the transition rates to depend on the state itself. Specifically, let $[M_{p}(V)]_{nm}=[M_p]_{nm} V_{nm}$ be given, then the transition rates
\begin{equation}\label{eq:d-R}
    \tilde R_{nm}(t)=\frac{2[M_p]_{nm}}{p^{\rm eq}_m(t)\theta\Big(\frac{ p_n}{p^{\rm eq}_n(t)},\frac{ p_m}{p_m^{\rm eq}(t)}\Big)},
\end{equation}
give rise to the desired mobility structure through~\eqref{eq:multiplic-d}.
 This construction, even if not unique, is detailed-balanced with respect to any chosen $p^{\rm eq}(t)\propto e^{-\beta \tilde H(t)}$, and preserves positivity and irreducibility. 

However, a similar construction does not seem in general possible in the quantum setting.  That is, it is not in general possible to find new $\tilde L_k$ and $\tilde \gamma_k$, detailed-balanced with respect to any Hamiltonian $\tilde H(t)$, that give rise to a given mobility operator. This can be traced to the fact that the quantum mobility operator depends on the frequencies $\omega_k$ in such a way that changing $\omega_k$ may change every element of the matrix $M_{\varrho}^k(V_k)$, not just by a scalar multiplicative factor. Since the jump operators $\tilde L_k$ must be detailed-balanced with respect to the Hamiltonian $\tilde H(t)$, their basis is essentially fixed. Therefore, the freedom in choosing $\tilde L_k$ and $\tilde \gamma_k$ is not enough to prescribe any fixed mobility operator with frequencies $\omega_k$ different from the Hamiltonian $\tilde H(t)$ frequencies.

In Section \ref{sec:LR}, we motivate fixing the mobility as a function of the state by noting that, in the linear-response regime, the mobility operators introduced in Section \ref{sec:grad-flows} depend only on the distribution, as required. Therefore, fixed mobility operators are seen to be meaningful in finding optimal counterdiabatic protocols.

In any case, whether we opt for a time-varying \eqref{eq:W2-time-var} or a time-invariant \eqref{eq:W2-minE} approach, the results presented until now are only valid for purely dissipative systems. Next, we go further and consider systems that evolve according to a mix of Hamiltonian and dissipative dynamics.

\section{Geometry of entropy production in Hamiltonian-dissipative dynamics}\label{sec:IV}
In the previous section, we considered purely dissipative detailed-balanced systems, whose dynamics are gradient flows. For detailed-balanced systems whose dynamics include non-dissipative (inertial or unitary) terms, these results are no longer directly applicable. 
In that case, it is necessary to redefine the metric to be able to quantify entropy production through a Wasserstein-like distance. In the classical setting, this leads to a new optimal transport distance that we introduce next. 

\subsection{Classical}
Consider a continuous system subject to both Hamiltonian and dissipative dynamics. In particular, let us consider a state $\rho\in\mathcal P_*^2(\mathbb R^{2d})$, a
probability density on $\mathbb R^{2d}$ of generalized positions $x\in\mathbb R^d$ and momenta $v\in\mathbb R^d$, that evolves according to
\begin{equation}\label{eq:FP-mixed}
    \dot\rho =\{H,\rho \}+ \nabla \cdot (\rho D  \nabla \phi),
\end{equation}
where $H(t,x,v)$ is the Hamiltonian, $\phi= \beta H+\ln\rho,$ $\beta=1/T$ the inverse temperature, and $D>0$ is a positive-definite $2d\times 2d$ diagonal diffusion matrix. Note that, $D$ being positive definite, these dynamics differ from the usual underdamped setting. Here, $\{f,g\}$ denotes the Poisson bracket, that is
$$
\{f,g\}=\nabla_x f^\top\nabla_v g-\nabla_vf^\top\nabla_x g.
$$ 
With no subindex, the $\nabla$ symbol denotes gradient with respect to both $x$ and $v$.

We may write the evolution explicitly as a mixture between Hamiltonian and gradient flows, namely,
$$
\dot\rho =-(\mathbb J_\rho +\mathbb K_{\rho })(\phi),
$$
where we have defined
\begin{equation}\label{eq:Jrho}
    \mathbb J_\rho(\phi):=T\{\rho,\phi \},
\end{equation} and $\mathbb K_{\rho }(\phi)=-\nabla \cdot (\rho D  \nabla \phi)$,   as before.
The Hamiltonian part does not contribute to the entropy production 
\begin{equation*}
    \dot \Sigma=-\llangle\phi,\dot\rho \rrangle=\langle\phi,\mathbb K_{\rho }(\phi)\rangle=\langle\nabla\phi, \rho D \nabla\phi\rangle,
\end{equation*}
since integrating by parts we have $\int_{\mathbb R^{2d}}\phi\{\phi,\rho \}dxdv=0.$ 

Our goal is again to use optimal transport to characterize entropy production for these dynamics. 
However, we see that problem \eqref{eq:cost-OMT} no longer quantifies  minimal entropy production for this system, since the dynamics do not include the Hamiltonian part. Specifically,
 we no longer have that
$
\dot \Sigma(t)$ is equal to  $\llangle\mathbb K_{\rho }^{-1}(\dot\rho ),\dot \rho \rrangle
$,
since it is no longer true that 
$
\dot\rho $ is given by $-\mathbb K_{\rho }(\phi).
$
To overcome this difficulty, we define a new metric that associates tangent elements to their duals in a way that incorporates the Hamiltonian part. Specifically, let us define the Riemannian metric
$$
h_{\rho }(\dot \mu,\dot\nu):=\langle \nabla \varphi,\rho D \nabla\psi\rangle,
    $$
    where now $\dot\mu$ and $\dot\nu$ are related to $\varphi$ and $\psi$ through
\begin{equation}\label{eq:one-to-one-c}
    \dot\mu=-(\mathbb J_\rho +\mathbb K_{\rho })(\varphi)\mbox{ and }\dot\nu=-(\mathbb J_\rho +\mathbb K_{\rho })(\psi).
\end{equation}
With this definition, we clearly have  that 
$$
\dot\Sigma=h_{\rho }(\dot\rho ,\dot\rho ).
$$
This is well-defined since, given $\dot \mu$, the equation 
$-(\mathbb J_\rho +\mathbb K_{\rho })(\varphi)=\dot\mu$ has a unique zero-mean solution for $\varphi$, provided $\rho\in\mathcal P_*^2(\mathbb R^{2d})$~\footnote{Specifically, $-(\mathbb J_\rho +\mathbb K_{\rho })(\varphi)=\dot\mu$ has a unique weak solution for $\varphi$ provided that $\int\dot\mu=0$ and $\rho$ is strictly positive and satisfies a Poincar\'{e}
inequality.}. Uniqueness of solutions critically depends on the strict positive-definiteness of $D.$ This is the reason why this setting does not apply directly to the standard underdamped regime where $D={\rm diag}(0_d,1_d)$~\footnote{Extending the present construction to that regime would require handling the degeneracy of the dissipation operator, since the uniqueness result used above relies on $D$ being strictly positive definite. A natural route would be to regularize $D$ as $D_\varepsilon = \mathrm{diag}(\varepsilon 1_d, 1_d)$ and study the singular limit $\varepsilon \to 0$, which may lead to a hypoelliptic or sub-Riemannian analog of the metric.}.

Alternatively, integrating by parts, we may rewrite the metric as
\begin{align}\nonumber
 h_{\rho }(\dot \mu,\dot\nu)&=-\frac12\Big(\langle\varphi, \nabla \cdot(\rho D \nabla\psi)\rangle+\langle\psi, \nabla \cdot(\rho D \nabla\varphi)\rangle\Big)
    \\&=\frac12\Big(\langle\varphi, (\mathbb J_\rho+\mathbb K_\rho)(\psi)\rangle+\langle\psi, (\mathbb J_\rho+\mathbb K_\rho)(\varphi)\rangle\Big),\label{eq:sym}
\end{align}
where for the second equality we have used  the fact that $\langle\varphi,\{\psi,\rho\}\rangle=-\langle \psi,\{\varphi,\rho\}\rangle$. Therefore, the metric reads
$$
h_{\rho }(\dot \mu,\dot\nu)=\frac12\Big(\langle(\mathbb J_\rho+\mathbb K_\rho)^{-1}(\dot\mu), \dot\nu\rangle+\langle (\mathbb J_\rho+\mathbb K_\rho)^{-1}(\dot\nu),\dot\mu\rangle\Big).
$$
We observe that since $(\mathbb J_\rho+\mathbb K_\rho)^{-1}$ is not self-adjoint, we must symmetrize it to obtain a symmetric bilinear form.

Let us now define the geodesic distance
\begin{align}\label{eq:Wh}
    W_{2,h}(\rho _0,\rho_\tau)^2:&=\inf_{\{\rho(t)\}:\rho_0,\rho_\tau}\tau\int_0^\tau h_{\rho}(\dot\rho,\dot\rho)dt\\&=\inf_{\{\rho(t)\}:\rho_0,\rho_\tau}{ l}_{\rho}^2,\nonumber
\end{align}
where we have defined the Riemannian length of a curve
$$
l_{\rho}:=\int_0^\tau\sqrt{h_{\rho}(\dot\rho,\dot\rho)}dt.
$$
The equality in the second line of \eqref{eq:Wh} follows from Cauchy-Schwarz inequality and the fact that  equality can be achieved through constant arc-length parametrization.
Thus,
 we again have
\begin{align}\label{eq:W2-minE-h} \nonumber
\!\!\!\!\frac{1}{\tau}W_{2,h}(\rho _0,\rho _\tau)^2=\inf_{\phi,\rho }\bigg\{\int_0^\tau\dot\Sigma\, dt:\, \dot\rho &=-(\mathbb J_\rho +\mathbb K_{\rho })(\phi),\\ \rho (0)&=\rho _0, \ \rho (\tau)=\rho _\tau\bigg\}.
\end{align}
 Since any admissible potential $\phi$ can be obtained through a time-varying Hamiltonian by setting $H=T(\phi-\ln\rho)$, the new optimal transport distance quantifies the minimum entropy produced through Hamiltonian driving between given endpoints. In particular, this result may be recast as an inequality,
 $$
 \int_0^\tau\dot\Sigma dt\geq \frac{1}{\tau}W_{2,h}(\rho_0,\rho_\tau)^2,
 $$
 that tightly bounds entropy production, providing a finite-time correction to the second law, and a new speed limit applicable for inertial systems.
The metric that enables this result has the same inner product structure between forces; what changes is only the way we associate elements in the tangent space to elements in the cotangent space.

We may compare the introduced distance \eqref{eq:Wh} to the classical Wasserstein-2 distance \eqref{eq:cOT}. Specifically, in Appendix \ref{app:equivalence} we show the following hierarchy
\begin{equation}
    \label{eq:hierarchy-app-main}
    \kappa({\rho })g_{\rm \rho }(\dot\rho ,\dot\rho )\leq h_{\rho }(\dot\rho ,\dot\rho )\leq g_{\rm \rho }(\dot\rho ,\dot\rho ),
\end{equation}
where we have defined 
$$
\kappa(\rho ):=\frac{1}{1+\|\mathbb K_{\rho }^{-1/2}
\mathbb J_{\rho }\mathbb K_{\rho }^{-1/2}\|_{\rm op}^2}
,$$
with $\|\cdot\|_{\rm op}$ denoting the operator norm.
Integrating over time and optimizing over trajectories, we obtain the following equivalence bound between metrics (see Appendix \ref{app:equivalence} for more details)
\begin{equation}\label{eq:hierarchy-c}
\kappa_{*}W_2(\rho _0,\rho _\tau)^2\leq  W_{2,h}(\rho _0,\rho _\tau)^2\leq W_2(\rho _0,\rho _\tau)^2 ,
\end{equation}
with $\kappa_*\in[0,1]$ given by $$\kappa_*:=\inf_{\rho \in\{{\rm \rho}(t)\}_{t\in[0,\tau]}}
\frac{1}{1+\|\mathbb K_{\rho }^{-1/2}
\mathbb J_{\rho }\mathbb K_{\rho }^{-1/2}\|_{\rm op}^2}.
$$ 
 In the last expression, the infimum is taken over states in the $W_{2,h}$-optimizing trajectory  $\{{\rm \rho }(t)\}_{t\in[0,\tau]}$, i.e., the one that minimizes $\int_0^\tau h_{\rho }(\dot\rho ,\dot\rho )dt$ between endpoints $\rho _0$ and $\rho _\tau$.
In words, bound \eqref{eq:hierarchy-c} implies that the minimum entropy production of an inertial system is upper bounded by the minimum entropy production of the corresponding overdamped system, and lower bounded by $\kappa_*$ times the overdamped minimum entropy production. 
That is, the Hamiltonian term can only lower the minimum entropy production, with the gain factor $\kappa_*$ quantifying the maximum factor by which it can be lowered.

 We may find an insightful explicit expression for a lower bound in terms of another constant $\kappa$ by slightly modifying the argument presented in Appendix \ref{app:equivalence} to the level of velocities.  We develop this argument here, since it provides insight into where the gap in the lower bound may come from.
Let us
define $\Gamma:=\beta D$ and the symplectic matrix \begin{equation}\label{eq:symplectic}
    J:=\left[\begin{array}{cc}
    0_{d\times d} & {\rm Id}_d \\
    -{\rm Id}_d & 0_{d\times d}
\end{array}\right].
\end{equation}
With these, we may rewrite the dynamics as
$$
\dot\rho =T\nabla\cdot\big((\Gamma-J)\rho  \nabla\phi\big)=T\nabla\cdot\big(\Gamma\rho  V\big),
$$
where $V=({\rm Id}-\Gamma^{-1}J)\nabla\phi.$
For such a $V$, by definition of $g_\rho$ \eqref{eq:metric-cl}, we have 
\begin{equation}
    \label{eq:vel}
     g_{\rho }(\dot\rho ,\dot\rho )\leq \langle V, \rho D V\rangle ,
\end{equation}
where the gap in the inequality comes from $V$ not being of gradient form. In fact, plugging in the expression for $V$ we obtain
\begin{align*}
\langle V, \rho D V\rangle
    &=\int_{\mathbb R^{2d}}\nabla\phi^\top({\rm Id}-\Gamma^{-1}J)^\top \rho D ({\rm Id}-\Gamma^{-1}J)\nabla\phi dxdv
    \\&= h_{\rho }(\dot\rho ,\dot\rho )+T\int_{\mathbb R^{2d}}\nabla\phi^\top J ^\top\Gamma^{-1}J\rho \nabla\phi dxdv.
\end{align*}
where we have used the fact that
$h_{\rm \rho }(\dot\rho ,\dot\rho )=T\langle  \nabla \phi, \Gamma \rho \nabla \phi\rangle$.
Letting $A=\Gamma^{-\frac12}J\Gamma^{-\frac12}$, the last term can be bounded as
\begin{equation}\label{eq:op-bound}
    T\int_{\mathbb R^{2d}}\nabla\phi^\top J^\top\Gamma^{-1}J\rho \nabla\phi dxdv\leq \|A\|_{\rm op}^2h_{\rm \rho }(\dot\rho ,\dot\rho )
\end{equation}
where $\|A\|_{\rm op}^2=\lambda_{\rm max}(A^\top A).$ This leads to the lower bound 
\begin{equation}
    \label{eq:lower-kappa-c}
    \kappa g_{\rho }(\dot\rho ,\dot\rho )\leq h_{\rho }(\dot\rho ,\dot\rho ),
\end{equation}
where
\begin{equation}
    \label{eq:kappa-c}\kappa:=\frac{1}{1+\|\Gamma^{-\frac12}J\Gamma^{-\frac12}\|_{\rm op}^2}.
\end{equation}
Integrating both sides over time and infimizing over trajectories with $\rho _0,\,\rho _\tau$ as endpoints, we obtain the lower bound in \eqref{eq:hierarchy-c} with $\kappa$ as in \eqref{eq:kappa-c}.

The lower bound in \eqref{eq:lower-kappa-c} is not generally attainable since the velocities in \eqref{eq:vel} are restricted in a suboptimal way. 
Note that equality in \eqref{eq:op-bound} is attainable, for example, if $\Gamma={\rm diag}(\gamma_x {\rm Id}_d,\gamma_p{\rm Id}_d),$ with $\gamma_x,\gamma_p>0$. Then, $J^\top\Gamma^{-1}J=\Gamma/(\gamma_x\gamma_p)$ and we can have equality  with $\|A\|^2_{\rm op}=1/\gamma_x\gamma_p$.
Therefore, in this case, we obtain
$
\kappa=\gamma_x\gamma_p/({\gamma_x\gamma_p+1}).
$
More generally, a simpler, looser bound can help us recover the overdamped (purely dissipative) limit. To do so, the operator norm $\|A\|^2_{\rm op}$ in \eqref{eq:op-bound} can be upper-bounded by $\|\Gamma^{-1}\|^2_{\rm op}\|J\|^2_{\rm op}$. Since $\|J\|^2_{\rm op}=1$, we obtain the bound~\eqref{eq:lower-kappa-c} with 
$$
\kappa=\frac{\lambda_{\rm min}(\Gamma)^2}{\lambda_{\rm min}(\Gamma)^2+1}.
$$
Clearly, as $\lambda_{\rm min}(\Gamma)\to 0$ we have that $\kappa\to 0,$ as is expected from purely Hamiltonian evolution. On the other hand, as $\lambda_{\rm min}(\Gamma)\to \infty,$ $\kappa\to 1$ and we recover the purely dissipative bound.

\renewcommand{\arraystretch}{1.65}

\begin{table*}[tb]
\centering
\small
\begin{threeparttable}

\caption{Summary of the geometry of entropy production in Hamiltonian-dissipative classical and quantum dynamics.}
\label{tab:conserv}

\begin{tabular}{@{}cccc@{}}
\rowcolor{DeepPurple}
\textcolor{white}{\bfseries OT paradigm}
&
\textcolor{white}{\bfseries Continuous}
\textcolor{white}{$\;(\mu=\rho)$}
&
\textcolor{white}{\bfseries Quantum}
\textcolor{white}{$\;(\mu=\varrho)$}
\\
\midrule

\rowcolor{VerySoftPurple}
\purplerowlabel{Manifold $\mathcal M$}
&
$\mathcal{P}^2_{\ast}(\mathbb{R}^{2d})$
&
$\mathcal{P}_{\ast}(\mathbb C,d)$
\\

\purplerowlabel{System}
&
Inertial Langevin
&
Full Lindblad
\\

\rowcolor{VerySoftPurple}
\purplerowlabel{Dynamics}
&
$\displaystyle \ \ 
\dot{\rho}
=
\{H,\rho\}
+
\nabla\!\cdot\!\big(\rho D\nabla(\beta H+\ln\rho)\big)
\ \ $
&
{\footnotesize
$\displaystyle \ \ 
\dot\varrho
=
-i[H,\varrho]
+
\sum_{k}\gamma_k
\Big(
L_k\varrho L_k^\dagger
-\frac12\{L_k^\dagger L_k,\varrho\}
\Big)
\ \ $}
\\

\midrule

\rowcolor{VerySoftPurple}
\purplerowlabel{Riemannian metric}
&
\multicolumn{2}{c}{
{\footnotesize
$\displaystyle
h_{\mu}(\dot\eta,\dot\nu)
=
\frac12
\Big(
\llangle
(\mathbb J_\mu+\mathbb K_{\mu})^{-1}(\dot\eta),
\dot\nu
\rrangle
+
\llangle
(\mathbb J_\mu+\mathbb K_{\mu})^{-1}(\dot\nu),
\dot\eta
\rrangle
\Big)
$}
}
\\

\purplerowlabel{ \ Hamiltonian-dissipative dynamics \ }
&
\multicolumn{2}{c}{
$\displaystyle
\dot{\mu}
=
-(\mathbb J_\mu+\mathbb K_\mu)(\phi)
$
}
\\

\rowcolor{VerySoftPurple}
\purplerowlabel{$2$-Wasserstein metric $W_{2,h}$}
&
\multicolumn{2}{c}{
$\displaystyle
W_{2,h}(\mu_0,\mu_\tau)
=
\inf_{\{\mu(t)\}:\,\mu_0,\mu_\tau}
l_\mu
$
}
\\

\purplerowlabel{Entropy production rate $\dot\Sigma$}
&
\multicolumn{2}{c}{
$\displaystyle
\dot\Sigma
=
h_\mu(\dot\mu,\dot\mu)
$
}
\\

\rowcolor{VerySoftPurple}
\purplerowlabel{Comparison with dissipative $W_2$}
&
\multicolumn{2}{c}{
{\centering $\displaystyle
\kappa_{\ast}W_2(\mu_0,\mu_\tau)^2
\leq
W_{2,h}(\mu_0,\mu_\tau)^2
\leq
W_2(\mu_0,\mu_\tau)^2
$
}}
\\

\bottomrule
\end{tabular}

\end{threeparttable}
\end{table*}

\subsection{Quantum}
Within the discrete stochastic framework considered in Section~\ref{sec:discrete-optimal-transport}, the dynamics are purely dissipative.
Unlike the continuum and quantum settings, this bare probability-simplex description does not come equipped with a canonical conservative term analogous to the Poisson bracket or the commutator. 
Although Hamiltonian-type dynamics on simplices can exist, they rely on additional noncanonical structure or a larger state space, falling outside the present geometric framework. We therefore turn to the quantum setting, where a natural Hamiltonian contribution is available on the same state space as the dissipative dynamics.

Indeed, Markovian open quantum dynamics are not typically purely dissipative, unlike those in \eqref{eq:diss-lind}. They normally have a non-negligible 
unitary term, leading to the form    \begin{align}\label{eq:full-lind}
    \dot\varrho =&-i[H,\varrho ]+\sum_{k\in K}\gamma_k\Big(L_k\varrho  L_k^\dagger -\frac12\{L_k^\dagger L_k,\varrho \}\Big),
\end{align}
where $L_k,\gamma_k$ satisfy the detailed balance conditions $[L_k,H]=-\omega_kL_k$ and $\gamma_k=\gamma_{-k} e^{-\beta \omega_k}$ for all $k\in K$, respectively, with $\omega_k$ defined as before. This assumption for all times (with a possibly time-varying Hamiltonian) is sometimes called the adiabatic assumption and requires the Hamiltonian to change slowly compared to heat bath dynamics. 

We may rewrite these dynamics as
\begin{equation}\label{eq:lind-grad-ham}
    \dot\varrho =-(\mathbb J_\varrho+\mathbb K_{\varrho})(\phi),
\end{equation}
where we have defined $$ \mathbb J_\varrho(\phi):=-iT[\varrho,\phi ],$$
 $\mathbb K_{\varrho}$ is as in \eqref{eq:q-onsager}, and  $\phi=\beta H+\ln\varrho $. Clearly, we see that the evolution is a mixture between a Hamiltonian and a gradient flow. The gradient flow part is the only part of the dynamics that contributes to entropy production. Specifically, at each instant of time we have
\begin{equation}\label{eq:q-entr-prod}
    \dot \Sigma=-\tr\{\phi\dot\varrho \}=\tr\{\phi\mathbb K_{\varrho}(\phi)\},
\end{equation}
since $\tr\{\phi[\phi,\varrho ]\}=0.$ Therefore, we still have that 
$
 \dot \Sigma=\langle \phi,\mathbb K_{\varrho}\big(\phi\big)\rangle\,.
$
However, in general, we no longer have that
$
\dot \Sigma$ is equal to  $\llangle\mathbb K^{-1}_{\varrho}(\dot\varrho ),\dot \varrho \rrangle
$,
since it is no longer true that 
$
\dot\varrho $ is given by $-\mathbb K_{\varrho}\big(\phi\big).
$
Therefore, metric \eqref{eq:metric-q} no longer characterizes entropy production for this system, since the dynamics do not include the Hamiltonian part.


As in the classical setting, we may circumvent this by defining a new metric that includes the Hamiltonian part in the dynamics; this is the approach taken in \cite{van2021geometrical}. Specifically, we may define a new Riemannian metric $h_{\varrho}$, similarly to before, as
\begin{equation}
    \label{eq:vanvu-metric}
    h_{\varrho }(\dot \mu,\dot\nu):=\langle \nabla\varphi,M_{\varrho }(\nabla\psi)\rangle, 
\end{equation}
 where $M_\varrho$ has the structure \eqref{eq:M-q}, while now $\dot\mu$ and $\dot\nu$ are related to $\varphi$ and $\psi$ through
\begin{equation}\label{eq:one-to-one-q}
    \dot\mu=-(\mathbb J_\varrho+\mathbb K_{\varrho })(\varphi)\mbox{ and }\dot\nu=-(\mathbb J_\varrho+\mathbb K_{\varrho })(\psi).
\end{equation}
This is well-defined since for each element of the tangent space $\dot\mu$ we can uniquely associate a dual $\varphi$ with zero trace that solves the above equation \cite{van2021geometrical}. 
With this definition, we clearly have that
$$
\dot\Sigma=-\llangle\phi,\dot\varrho \rrangle=\langle\phi,\mathbb K_{\varrho}(\phi)\rangle=h_{\varrho}(\dot \varrho ,\dot\varrho ).
$$

Alternatively, integrating by parts,
we may rewrite the metric as
\begin{align}\nonumber
 h_{\varrho }(\dot \mu,\dot\nu)&=-\frac12\Big(\langle\varphi, \nabla \cdot(M_\varrho  (\nabla\psi))\rangle+\langle\psi, \nabla \cdot(M_\varrho (\nabla\varphi))\rangle\Big)
    \\&=\frac12\Big(\langle\varphi, (\mathbb J_\varrho+\mathbb K_\varrho)(\psi)\rangle+\langle\psi, (\mathbb J_\varrho+\mathbb K_\varrho)(\varphi)\rangle\Big),\label{eq:sym-q}
\end{align}
where for the second equality we have used the fact that $\langle\varphi,[\psi,\varrho]\rangle=-\langle \psi,[\varphi,\varrho]\rangle$. Therefore, the metric alternatively reads
$$
h_{\varrho }(\dot \mu,\dot\nu)=\frac12\Big(\langle(\mathbb J_\varrho+\mathbb K_\varrho)^{-1}(\dot\mu), \dot\nu\rangle+\langle (\mathbb J_\varrho+\mathbb K_\varrho)^{-1}(\dot\nu),\dot\mu\rangle\Big).
$$

For a fixed thermodynamic trajectory $\{\varrho(t)\}_{t\in[0,\tau]}$ evolving from $\varrho_0$ to $\varrho_\tau$ according to~\eqref{eq:full-lind}, we may use this metric to lower bound entropy production in terms of a pseudo-distance as~\cite{van2021geometrical}
\begin{equation}
    \label{eq:W2h-timevar}
\int_0^\tau\dot\Sigma \, dt\geq \frac{1}{\tau}\bigg(\inf_{\{\varrho(t)\}:\varrho_0,\varrho_\tau}\int_0^\tau \sqrt{h_{t,\varrho }(\dot\varrho ,\dot\varrho )}dt\bigg)^2,
\end{equation}
where $h_{t,\varrho }(\dot\varrho,\dot\varrho)$ is characterized by the mobility operator \eqref{eq:M-q} with respect to the instantaneous dynamics at each instant of time.  Alternatively, 
we may consider the mobility operator as being fixed, and define the geodesic distance
\begin{align}
    \label{eq:Wh2-q}
W_{2,h}(\varrho_0,\varrho_\tau)^2:&=\inf_{\{\varrho(t)\}:\varrho_0,\varrho_\tau}\tau\int_0^\tau h_{\varrho}(\dot\varrho,\dot\varrho)dt\\&=\inf_{\{\varrho(t)\}:\varrho_0,\varrho_\tau}l_\varrho^2,\nonumber
\end{align}
where we have defined the length
$$
l_\varrho:=\int_0^\tau\sqrt{h_{\varrho}(\dot\varrho,\dot\varrho)}dt.
$$
Then, for systems with a fixed mobility, we obtain the stronger result
\begin{align}\label{eq:W2-minE-h-q}\nonumber
\frac{1}{\tau}W_{2,h}(\varrho _0,\varrho _\tau)^2=\inf_{\phi,\varrho }\bigg\{\int_0^\tau&\dot\Sigma\, dt:\, \dot\varrho =-(\mathbb J_\varrho+\mathbb K_{\varrho })(\phi),\\&  \varrho (0)=\varrho _0, \ \varrho (\tau)=\varrho _\tau\bigg\}. 
\end{align}
Thus, we again have results akin to \eqref{eq:W2-time-var} and \eqref{eq:W2-minE}, albeit for quantum systems that may have both Hamiltonian and gradient flows.


 As in the classical setting, we may compare the Hamiltonian-dissipative distance to the purely dissipative one. 
 Indeed, the same argument follows (see Appendix \ref{app:equivalence}), and we obtain,
\begin{equation}\label{eq:hierarchy-q}
    \kappa_*W_{2}(\varrho _0,\varrho _\tau)^2\leq W_{2,h}(\varrho _0,\varrho _\tau)^2\leq W_{2}(\varrho _0,\varrho _\tau)^2,
\end{equation}
with
$$
\kappa_*:=\inf_{\varrho \in\{{\rm \varrho}(t)\}_{t\in[0,\tau]}}\frac{1}{1+\|\mathbb K_{\varrho }^{-1/2}
\mathbb J_{\varrho }\mathbb K_{\varrho }^{-1/2}\|_{\rm op}^2},$$
where the infimum is taken over states in the trajectory that minimizes $\int_0^\tau h_{\varrho }(\dot\varrho ,\dot\varrho )dt$ between endpoints $\varrho _0$ and $\varrho _\tau$.
Thus, we have obtained that the distance-squared for a Hamiltonian-dissipative system is always smaller than that of a purely dissipative system, but no smaller than by a factor of $\kappa_*$. For physical systems with fixed mobilities, this implies the intuitive statement that a Hamiltonian term in the dynamics can only reduce the minimum entropy production. The lower bound quantifies how much the entropy production may be reduced through Hamiltonian evolution, mediated by coherences in the Hamiltonian basis, with the gain factor $\kappa_*.$
Note that, if all jump rates $\gamma_k$ are rescaled by a multiplicative constant $\gamma$, then $\kappa_*\to1$ as $\gamma\to\infty$, that is,  we recover the standard quantum Wasserstein-2 distance in the purely dissipative limit.

In the case of a time-varying metric, we obtain similar results. Indeed, for a particular instant of time, following the same argument, we can show (c.f. \eqref{eq:hierarchy-app-main})
\begin{equation}
    \label{eq:hierarchy}
    \kappa(t,{\varrho })g_{t,\rm \varrho }(\dot\varrho ,\dot\varrho )\leq h_{t,\varrho }(\dot\varrho ,\dot\varrho )\leq g_{t,\rm \varrho }(\dot\varrho ,\dot\varrho ),
\end{equation}
with
$
\kappa(t, \varrho ):=1/(1+\|\mathbb K_{t,\varrho }^{-1/2}
\mathbb J_{\varrho }\mathbb K_{t,\varrho }^{-1/2}\|_{\rm op}^2),$
where $\mathbb K_{t,\varrho }$ is the Onsager operator with respect to the instantaneous mobility.
Then, integrating over time and infimizing over trajectories, we obtain
\begin{align*}
\kappa_*\!\inf\!\!\int_0^{\tau}\!\!\!\!g_{t,\rm \varrho }(\dot\varrho ,\dot\varrho )dt&\leq \inf\!\!\int_0^{\tau}\!\!\!\!h_{t,\varrho }(\dot\varrho ,\dot\varrho )dt\leq \inf\!\!\int_0^{\tau}\!\!\!\!g_{t,\rm \varrho }(\dot\varrho ,\dot\varrho )dt,
\end{align*}
 where all infima are taken over trajectories $\{\varrho(t)\}_{t\in[0,\tau]}$ with $\varrho_0$ and $\varrho_\tau$ as endpoints. Here, $\kappa_*=\inf_{t\in[0,\tau]}\kappa(t,\varrho(t)),$ with $\{\varrho(t)\}_{t\in[0,\tau]}$ the $h_{t,\varrho}$-optimal trajectory.  Therefore, we obtain a hierarchy of inequalities that capture dissipation and the extent to which Hamiltonian motion can lower its minimal cost.

\section{Linear-response regime
and counterdiabatic driving}
\label{sec:LR}

In the near-equilibrium regime, the assumption of a fixed mobility operator is most natural. In this setting, we show that restricting the optimal transport metrics to equilibrium trajectories recovers the well-established thermodynamic length of linear response. 
While this equivalence had already been established for the overdamped~\cite{chennakesavalu2023unified,zhong2024beyond} and discrete geometries~\cite{sawchuk2026thermodynamic}, we show that it also holds for inertial and quantum systems, where the quantum thermodynamic length is recovered~\cite{scandi2019thermodynamic}.
As a consequence, we may interpret optimal transport protocols as counterdiabatic protocols \cite{guery2023driving} that enforce the system to follow linear-response geodesics in finite-time, thereby providing a natural extension of the linear-response Riemannian structure further from equilibrium.

Specifically, let $\lambda(t)\in\mathbb R^r$ denote control parameters of a detailed-balanced thermodynamic system, and $\mathbb L_\lambda$ denote the backward Markov generator dictating its dynamics. Here and in what follows, we drop the time dependence of $\lambda$ for simplicity of notation.
Define the set of observables
\begin{equation}\label{eq:deltaX}
    \delta X_i:=\partial_{\lambda_i}\ln\mu^\lambda,
\mbox{ for }i=1,\ldots, r,
\end{equation}
where 
$\mu^\lambda\propto e^{-\beta^\lambda H^\lambda}$ denotes the equilibrium distribution prescribed by $\lambda$, in the three settings $\mu\in\{\rho,p,\varrho\}$. These observables quantify the deviation of the ``conjugate force'' $-\partial_{\lambda_i} (\beta^\lambda H^\lambda)$ from its average.
Let the corresponding time evolved observables be denoted by
$$
\delta X_i(t):=e^{t\mathbb L_\lambda}(\delta X_i).
$$

It was shown in~\cite{sivak2012thermodynamic}
that the friction metric
\begin{equation}
    \label{eq:zeta2}
    \zeta_{ij}(\lambda):=\int_0^{\infty}\langle\delta X_i(0),\delta X_j(s)\rangle_{\mu^\lambda} ds,
\end{equation}
 quantifies dissipation in the linear-response regime through
 \begin{equation}
     \label{eq:EP-LR}
     \dot 
 \Sigma
 =\sum_{i,j=1}^r\dot\lambda_i\zeta_{ij}(\lambda)\dot\lambda_j.
 \end{equation}
 In \eqref{eq:zeta2}, $\langle A,B\rangle_{\mu^\lambda}$ denotes a $\mu^\lambda$-weighted inner product, which in the continuous and discrete settings is given by $\langle A,B\rangle_{\mu^\lambda}=\langle A,\mu^\lambda B\rangle$, while its quantum counterpart reads~\cite{scandi2019thermodynamic}
 $$
 \langle A,B\rangle_{\varrho^\lambda}=\langle A,\tilde M_{\varrho^\lambda}(B)\rangle, \mbox{ with }\tilde M_{\varrho}(A):=\int_0^1\varrho^s A \varrho^{1-s}ds.
 $$
Using the resolvent identity of Markov generators,  $(-\mathbb L_{\lambda})^{-1}=\int_0^\infty e^{s\mathbb L_{\lambda}}ds$, an equivalent expression for  $\zeta_{ij}$ may be found in terms of the generator, namely,~\cite{mandal2016analysis,scandi2019thermodynamic}
\begin{equation}
    \label{eq:zeta}
    \zeta_{ij}(\lambda)=\langle\delta X_i,(-\mathbb L_{\lambda})^{-1}(\delta X_j)\rangle_{\mu^\lambda} .
\end{equation}

Note that, for $\zeta_{ij}$ to define a metric, it must be symmetric. Indeed, the quadratic form \eqref{eq:EP-LR} is determined by the symmetric part of $\zeta_{ij}$. 
Therefore, if the backward generator $\mathbb L_\lambda$ is not self-adjoint with respect to the $\mu$-weighted inner product, then $\zeta_{ij}$ must be symmetrized. That is, in general, 
\begin{equation}
    \label{eq:zeta-sym}
    \zeta_{ij}(\lambda)=\frac12\langle\delta X_i,\big((-\mathbb L_{\lambda})^{-1}+(-\mathbb L_{\lambda})^{-\ddagger}\big)(\delta X_j)\rangle_{\mu^\lambda} ,
\end{equation}
where $\ddagger$ denotes adjoint with respect to the $\mu^\lambda$-weighted inner product, and $-\ddagger$ denotes the composition of inverse and adjoint, i.e., $\mathbb L_\lambda^{-\ddagger}=(\mathbb L_\lambda^\ddagger)^{-1}$.
In particular, this symmetrization is necessary for mixed Hamiltonian-dissipative dynamics.

We now show that this linear-response metric coincides with the corresponding optimal transport metric restricted to equilibrium distributions. This result will apply to continuous, discrete, and quantum settings, both purely dissipative and mixed.  

\subsection{Continuum}
We first consider a classical mixed Hamiltonian-dissipative system; the purely dissipative result will follow as a special case.
Let equilibrium states be parametrized by $\lambda(t)\in\mathbb R^r$ as $\rho^\lambda=e^{-\beta^\lambda H^\lambda}/Z^\lambda \in \mathcal P_*^2(\mathbb R^{2d}),$ where we have dropped the time dependence of $\lambda$ for simplicity of notation. 
Consider the restriction of the metric $h_{\rho }$ to the equilibrium submanifold; this restriction can be thought of as a pullback of the metric onto the space of parameters $\lambda$. Specifically, since $\dot\rho^\lambda=\sum_{i=1}^r \dot\lambda_i\partial_{\lambda_i}\rho^\lambda$, on the equilibrium submanifold we have
$$
h_{\rho^\lambda}(\dot\rho^\lambda,\dot\rho^\lambda)=\sum_{i,j=1}^r\dot\lambda_ih_{\rho^\lambda}(\partial_{\lambda_i}\rho^\lambda,\partial_{\lambda_j}\rho^\lambda)\dot\lambda_j.
$$
We will now show that $h_{\rho^\lambda}(\partial_{\lambda_i}\rho^\lambda,\partial_{\lambda_j}\rho^\lambda)$ is nothing but the symmetrized version of the friction tensor $\zeta_{ij}(\lambda).$

By definition, 
\begin{align*}h_{\rho^\lambda}(\partial_{\lambda_i}\rho^\lambda,\partial_{\lambda_j}\rho^\lambda)
    =\int_{\mathbb R^{2d}}\nabla\psi_i \rho^\lambda D^\lambda\nabla\psi_jdxdv,
\end{align*}
where $\psi_k$ satisfy 
\begin{equation}
        \label{eq:psi-j}  \partial_{\lambda_k}\rho^\lambda=\frac{1}{\beta^\lambda}\{\psi_k,\rho^\lambda\}+\nabla\cdot( \rho^\lambda D^\lambda\nabla\psi_k), \mbox{ for } k\in\{i,j\}.
\end{equation}
From \eqref{eq:sym}, 
we may rewrite $h_{\rho^\lambda}(\partial_{\lambda_i}\rho^\lambda,\partial_{\lambda_j}\rho^\lambda)$ as
\begin{align}
   -\frac12\bigg(\int_{\mathbb R^{2d}}\psi_i \partial_{\lambda_j}\rho^\lambda dxdv+\int_{\mathbb R^{2d}}\psi_j\partial_{\lambda_i}\rho^\lambda dxdv\bigg).\label{eq:h-c}
\end{align}

Given the definition of $\delta X_i$ \eqref{eq:deltaX}, we have that
$
\partial_{\lambda_i}\rho^\lambda=\rho^\lambda\delta X_i.
$
Then, we may solve for $\psi_k$ in \eqref{eq:psi-j} to obtain
\begin{equation}\label{eq:psik}
    \psi_k=\mathbb L_{\lambda}^{-\ddagger}(\delta X_k),  \mbox{ where } \mathbb L^{\ddagger}_{\lambda}(\cdot):=-\frac{1}{\rho^\lambda}(\mathbb J_{\rho^\lambda}+\mathbb K_{\rho^\lambda})(\cdot)\,.
\end{equation}
Note that $\mathbb L^{\ddagger}$ is invertible on the mean-zero subspace of observables like $\delta X_k$. Using the definitions of $\mathbb J_\rho$ and $\mathbb K_\rho$, \eqref{eq:Jrho} and \eqref{eq:Krho}, together with the fact that $\nabla\rho^\lambda=-\beta^\lambda \nabla H^\lambda\rho^\lambda$, we obtain
$$
\mathbb A_\lambda^{\ddagger}(A):=-\frac{1}{\rho^\lambda}\mathbb J_{\rho^\lambda}(A)=\{H^\lambda,A\},
$$
and 
$$
\mathbb S_\lambda(A):=-\frac{1}{\rho^\lambda}\mathbb K_{\rho^\lambda}(A)= -\beta^\lambda (\nabla H^{\lambda})^\top D^\lambda\nabla A+\nabla\cdot(D^\lambda\nabla A),
$$
where one can check that $\mathbb A_\lambda$ is anti-symmetric with respect to the $\rho^\lambda$-weighted inner product, while $\mathbb S_\lambda$ is symmetric. Then,
\begin{equation}
    \label{eq:L}
    \mathbb L_{\lambda}(\cdot)=
-\{H^\lambda,\cdot\}
-\beta^\lambda (\nabla H^{\lambda})^\top D^\lambda\nabla(\cdot)+\nabla\cdot(D^\lambda\nabla(\cdot)),
\end{equation}
that is,
$\mathbb L_{\lambda}$ is the backward generator of the Hamiltonian-dissipative evolution.

Therefore, using the expression for $\psi_k$, \eqref{eq:psik}, in \eqref{eq:h-c} and the fact that $
\partial_{\lambda_i}\rho^\lambda=\rho^\lambda\delta X_i
$, we obtain
$$
\frac12\Big(
   \langle\delta X_i,(-\mathbb L_\lambda)^{-\ddagger}(\delta X_j) \rangle_{\rho^\lambda}+\langle\delta X_j,(-\mathbb L_\lambda)^{-\ddagger}(\delta X_i) \rangle_{\rho^\lambda}\Big),
$$
where $\langle A,B\rangle_{\rho^\lambda}=\int_{\mathbb R^{2d}}A B\rho^\lambda dxdv.$
Rewriting the second term in terms of its adjoint leads to
\begin{align}
   h_{\rho^\lambda}(\partial_{\lambda_i}\rho^\lambda,\partial_{\lambda_j}\rho^\lambda)&=\frac12\langle
   \delta X_i,\big((\!-\mathbb L_\lambda)^{-1}\!\!+\!(\!-\mathbb L_\lambda)^{-\ddagger}\big)(\delta X_j) \rangle_{\rho^\lambda},\label{eq:ot-LR}
\end{align}
which is the symmetrized version of the friction tensor $\zeta_{ij}(\lambda),$ \eqref{eq:zeta-sym}. 
 Therefore, we have shown that the Hamiltonian-dissipative distance introduced herein ---when restricted to the equilibrium submanifold and pulled back to the space of control parameters $\lambda$--- gives the linear-response thermodynamic length.

The relationship between the purely dissipative  optimal transport metric and the linear-response regime metric has been previously laid out in the overdamped setting~\cite{chennakesavalu2023unified,zhong2024beyond}. 
In fact, an overdamped version of equation~\eqref{eq:ot-LR} is obtained in~\cite{zhong2024beyond}. To obtain that result, we may simply take the overdamped limit by setting $\mathbb J_{\rho^\lambda}$ in $\mathbb L_\lambda$ to zero. This leads to a self-adjoint $\mathbb L_\lambda$, so $\zeta_{ij}$ does not need to be symmetrized. Interestingly, Ref.~\cite{zhong2024beyond} uses this relationship to unify two approaches: a geometric one where entropy production is minimized subject to constraints on the endpoint probability densities, and an optimal-control one where work is minimized subject to constraints on the endpoint control parameters~\cite{schmiedl2007optimal,Proesmans2020Landauer,loos2024universal}. These different endpoint constraints give rise to characteristic discontinuities of the optimal-control protocol that are not observed in the present geometric setting.

Furthermore, the fact that the metrics coincide at equilibrium is used in~\cite{zhong2024beyond} to interpret the optimal transport solution as a counterdiabatic protocol that enforces the system to follow linear-response-metric geodesics. Specifically,
 if the space of $\lambda$-parametrized distributions is large enough \footnote{This may require the parametrization ($\lambda$-space) to be infinite-dimensional in general.  However, a finite-dimensional example is the Gaussian submanifold, where any two points are connected by $W_2$ geodesics that stay within the submanifold.}, then the optimal transport and the linear-response geodesics coincide, by virtue of their metrics coinciding~\eqref{eq:ot-LR}.
 The optimal Hamiltonian arising from the optimal transport problem has the form 
$
    \beta H_{\rm opt}=-\ln\rho^\lambda_{{\rm opt}}+\phi_{\rm opt},
$
where the first term of the right-hand side is prescribed by the linear-response geodesic, while the second term may be understood as the counterdiabatic term that ensures the system follows said geodesic 
(see \cite{zhong2024beyond} for more details of the overdamped setting). 
 Note that following such optimal protocols in the inertial setting requires full phase-space Hamiltonian control, which may be technically unfeasible.
We will see how an analogous perspective can be fruitful in the discrete setting.

\subsection{Discrete}
Let us now turn to the discrete setting, where equilibrium states are parametrized by $\lambda(t)\in\mathbb R^r$ as $p^\lambda_n=e^{-\beta^\lambda H_n^\lambda}/Z^\lambda,$ and the dynamics are purely dissipative. The transition rates $R_{nm}^\lambda$ are parametrized by $\lambda$ (or equivalently $p^\lambda$) through the detailed balance condition~\eqref{eq:detbal}, and so is the mobility operator $M_{p^\lambda}$ through~\eqref{eq:multiplic-d}.   Once again, $\dot p^\lambda_n=\sum_{i=1}^r \dot\lambda_i\partial_{\lambda_i}p^\lambda_n$,  and on the equilibrium submanifold we have
$$
g_{p^\lambda}(\dot p^\lambda,\dot p^\lambda)=\sum_{i,j}\dot\lambda_ig_{p^\lambda}(\partial_{\lambda_i}p^\lambda,\partial_{\lambda_j}p^\lambda)\dot\lambda_j,
$$
where $g_{p^\lambda}$ is prescribed by the fixed mobility operator $M_{p^\lambda}$.
Using integration by parts, we may rewrite the metric \eqref{eq:metric-d} on the space of parameters as
\begin{align}\label{eq:g-d}
    g_{p^\lambda}(\partial_{\lambda_i}p^\lambda,\partial_{\lambda_j}p^\lambda)=-\sum_n\partial_{\lambda_i}p^\lambda_n[\psi_j]_n,
\end{align}
with $\psi_j$ satisfying, for all $n$,
\begin{align} \label{eq:psi-j-d}
\partial_{\lambda_j}p^\lambda_n
    &=[\nabla\cdot(M_{p^\lambda}(\nabla\psi_j))]_n\\&=\sum_{m}R^\lambda_{nm}p^\lambda_m([\psi_j]_m-[\psi_j]_n),
    \nonumber
\end{align}
where 
we have used the fact that $[M_{p^\lambda}(\nabla \psi)]_{nm}=\frac12R^\lambda_{nm}p^\lambda_m[\nabla\psi]_{nm},$ which can be obtained from \eqref{eq:multiplic-d}.

By definition of $\delta X_j,$
$\partial_{\lambda_j}p^\lambda_n= p^\lambda_n[\delta X_j]_n.$ Therefore 
we may solve for $\psi_j$ in \eqref{eq:psi-j-d} as
$$
\psi_j=\mathbb L_{\lambda}^{-1}(\delta X_j),  \mbox{ with } \mathbb L_{\lambda}(\cdot):=-{\rm diag}(p^\lambda)^{-1}\mathbb K_{p^\lambda}(\cdot).
$$
 Note that $[\mathbb L_\lambda(\cdot)]_n=\sum_{m}R^\lambda_{mn}[\nabla(\cdot)]_{mn}$, that is, $\mathbb L_\lambda(\cdot)$ is the backward generator of the detailed-balanced Markov chain, which is self-adjoint with respect to the $p^\lambda$-weighted inner product. Using this expression for $\psi_j$ in \eqref{eq:g-d}, we obtain
\begin{align*}
    g_{p^\lambda}(\partial_{\lambda_i}p^\lambda,\partial_{\lambda_j}p^\lambda)&=\sum_n p^\lambda_n[\delta X_i]_n[(-\mathbb L_{\lambda})^{-1}(\delta X_j)]_n,
\end{align*}
yielding the usual friction tensor \eqref{eq:zeta}.

In analogy to the continuous setting, the coincidence between the optimal-transport metric restricted to the equilibrium submanifold and the linear-response friction tensor provides a constructive route to thermodynamically optimal counterdiabatic driving in discrete Markov systems. 
Rather than searching over arbitrary nonequilibrium paths, one may search over equilibrium-parametrized paths, for which the dissipation predicted by linear response is governed by the pullback metric $\zeta(\lambda)$. 
When the control of the Hamiltonian is full, any strictly positive distribution can be represented as an equilibrium state for some $H^\lambda$, since $p^\lambda \propto e^{-\beta^\lambda H^\lambda}$ (equivalently, $H^\lambda = -\ln p^\lambda/\beta^{\lambda}$ up to an additive constant). 
However, a finite-time protocol $\lambda(t)$ will not in general keep the system exactly on the instantaneous equilibrium curve $p^{\lambda}(t)$ under the bare detailed-balanced dynamics; this “lag” is precisely what generates dissipation beyond the quasistatic limit. 
Counterdiabatic control remedies this by modifying the kinetics so that the actual state tracks the target curve $p^\lambda(t)$ exactly in finite time. Through a conveniently chosen optimal transport problem, we may simultaneously optimize the linear-response path and find the corresponding counterdiabatic force that makes entropy production minimal.

Specifically, consider the discrete optimal transport problem with mobility operator $[M_{p^\lambda}(\nabla \psi)]_{nm}=\frac12R^\lambda_{nm}p^\lambda_m[\nabla\psi]_{nm}$, which is fixed as a function of $\lambda$, or equivalently, $p^\lambda$. 
That is, consider the problem 
\begin{align}\label{eq:discreteOT-LR}
&\inf_{\psi, p^\lambda}\tau\frac12\sum_{m,n}\int_0^\tau R^\lambda_{nm}p^\lambda_m[\nabla\psi]^2_{nm} dt
\\& \mbox{s.t.} \ \dot p^\lambda_n
    =\sum_{m}R^\lambda_{mn}p^\lambda_n(\psi_m-\psi_n),\  p^\lambda(0)= p^\lambda_0,\, p^\lambda(\tau)= p^\lambda_\tau.\nonumber
\end{align}
Note that prescribing the rates $R_{nm}^\lambda$ as a function of $\lambda$ fixes the mobility along the equilibrium submanifold. This dependence on $\lambda$ may come in different forms as long as detailed balance is satisfied with respect to $\beta^\lambda H^\lambda.$ For example, we may have the backward rates fixed, while the forward rates are enforced by detailed balance (see example in Section~\ref{sec:ex_d}), or we may have the symmetric part of the transition rates fixed, i.e., $R^\lambda_{nm}=d_{nm}A^\lambda_{nm}$, where $d_{nm}=d_{mn}$ are fixed and $A^\lambda_{nm}$ depend on $\lambda$ through $A^\lambda_{nm}/A^\lambda_{mn}=e^{-\beta^\lambda H^\lambda_n}/e^{-\beta^\lambda H^\lambda_m}$. 

With the solution to this discrete optimal transport problem \eqref{eq:discreteOT-LR}, we may uniquely determine detailed-balanced (with respect to a new Hamiltonian) rates $\tilde R_{nm}$ that give rise to the fixed mobility operator. Specifically, let $p^\lambda$ and $\psi$ denote the optimal solution to \eqref{eq:discreteOT-LR}. For $\tilde R_{nm}$ to give rise to the fixed mobility operator, it must satisfy
\begin{align*}
    \frac{\tilde R_{nm} p^\lambda_m - \tilde R_{mn} p^\lambda_n}{\ln \tilde R_{nm} p^\lambda_m - \ln \tilde R_{mn} p^\lambda_n} = R^\lambda_{nm} p_m^\lambda.
\end{align*}
Moreover, we require the rates to be detailed-balanced with respect to the tilted distribution $\propto {p^\lambda}e^{-\psi}$, which is the equilibrium distribution with $(\psi-\ln p^\lambda)/\beta^\lambda$ as the Hamiltonian. That is, $\tilde R_{nm}$ satisfy
$$
\ln \tilde R_{nm} p^\lambda_m - \ln \tilde R_{mn} p^\lambda_n=\psi_m-\psi_n.
$$
With these restrictions, we obtain that
\begin{equation}\label{eq:tilde-Rnm}
    \tilde R_{nm} = R^\lambda_{nm} \chi(\psi_m-\psi_n),
\end{equation}
with $\chi(x)=x/(1-  e^{-x})> 0,$  are non-negative rates that give rise to the optimal dynamics following $p^\lambda$.
Note that $\tilde R_{nm}\to R_{nm}^\lambda$ in the limit of slow driving (as $\nabla\psi\to 0$).

Analogously to the continuous setting, we may understand problem \eqref{eq:discreteOT-LR} as that of finding the linear-response geodesic, together with the counterdiabatic rates \eqref{eq:tilde-Rnm} that realize said geodesic in an arbitrary time. Both the geodesic and the counterdiabatic term are optimal in the sense that they minimize entropy production along the trajectory, for a fixed mobility operator.  Counterdiabatic protocols for discrete Markov processes are of interest in the context of biological networks~\cite{ilker2022shortcuts}, and
 have been suggested to control evolutionary processes~\cite{iram2021controlling}.
  Our construction provides a principled geometric way to identify a counterdiabatic protocol that is thermodynamically optimal.


\subsection{Quantum: purely dissipative and mixed}
\label{sec:q-LR}

Let us now move to the quantum setting and consider mixed Hamiltonian-dissipative dynamics; the purely dissipative result will follow as a special case. Let equilibrium states be parametrized by $\lambda(t)\in\mathbb R^r$ as $\varrho^\lambda=e^{-\beta^\lambda H^\lambda}/ Z^\lambda$, where $H^\lambda=(H^\lambda)^\dagger\in \mathbb C^{d\times d}.$ Equilibrium states change in time only through the parameters $\lambda$, and thus $\dot\varrho^\lambda=\sum_{i=1}^r \dot\lambda_i\partial_{\lambda_i}\varrho^\lambda$. Then, the restriction of the metric $h_{\varrho }$ to the equilibrium submanifold may be written as
$$
h_{\varrho^\lambda}(\dot\varrho^\lambda,\dot\varrho^\lambda)=\sum_{i,j}\dot\lambda_ih_{\varrho^\lambda}(\partial_{\lambda_i}\varrho^\lambda,\partial_{\lambda_j}\varrho^\lambda)\dot\lambda_j.
$$
Here,  $h_{\varrho^\lambda}$ is taken with respect to the fixed mobility operator $M_{\varrho^\lambda}$ (fixed as a function of $\lambda$, or equivalently of $\varrho^\lambda$) given by~\eqref{eq:M-q} with the jump rates $\gamma_k^\lambda$ and jump operators $L_k^\lambda$ detailed-balanced with respect to $H^\lambda$, i.e., at each value of $\lambda$ we have $[L^\lambda_k,H^\lambda]=-\omega^\lambda_kL^\lambda_k$  and $\gamma^\lambda_k=\gamma^\lambda_{-k} e^{-\beta^\lambda \omega^\lambda_k}$ for all $k\in K$, where $\omega^\lambda_k$ is the Bohr frequency associated with channel $k$ and parameter $\lambda$.

By definition,
\begin{align*}
    &h_{\varrho^\lambda}(\partial_{\lambda_i}\varrho^\lambda,\partial_{\lambda_j}\varrho^\lambda)
    =\tr\{\nabla\psi_i^\dagger M_{\varrho^\lambda}(\nabla\psi_j)\},
\end{align*}
where $\psi_k$ satisfies 
\begin{equation}
    \label{eq:dyn-LR-q}  \partial_{\lambda_k}\varrho^\lambda=-\frac{i}{\beta^\lambda}[\psi_k,\varrho^\lambda]+\nabla\cdot( M_{\varrho^\lambda}\nabla\psi_k), \mbox{ for } k\in\{i,j\}.
\end{equation}
From \eqref{eq:sym-q}, 
we may rewrite the metric as
\begin{align}
  h_{\varrho^\lambda}(\partial_{\lambda_i}\varrho^\lambda,\partial_{\lambda_j}\varrho^\lambda)= -\frac12\tr\big\{\psi_i \partial_{\lambda_j}\varrho^\lambda+\psi_j\partial_{\lambda_i}\varrho^\lambda \big\}.\label{eq:h-q}
\end{align}

Due to the identity
$$
\partial_{\lambda_i} e^{-\beta^\lambda H^{\lambda}}=-\int_0^1 e^{-s\beta^\lambda H^{\lambda}}\partial_{\lambda_i}(\beta^\lambda H^\lambda)  e^{-(1-s)\beta^\lambda H^{\lambda}}ds,
$$
we may write
the derivative of the equilibrium states as
$$
\partial_{\lambda_{i}} \varrho^\lambda=\tilde M_{\varrho^\lambda}(\delta X_i),
$$
where we recall that 
$$
\tilde M_{\varrho^\lambda}(A)=\int_0^1(\varrho^\lambda)^s A (\varrho^\lambda)^{1-s}ds,
$$ and
$
\delta X_i=-\big( \partial_{\lambda_i}(\beta^\lambda H^\lambda)  -\tr\{ \partial_{\lambda_i}(\beta^\lambda H^\lambda)  \varrho^{\lambda}\}{\rm Id}\big).
$ 
Therefore,
we may solve for $\psi_k$ in \eqref{eq:dyn-LR-q} as
$$
\psi_k=\mathbb L_{\lambda}^{-\ddagger}(\delta X_k),  \mbox{ where } \mathbb L^\ddagger_{\lambda}(\cdot):=-\tilde M^{-1}_{\varrho^\lambda}( \mathbb J_{\varrho^\lambda}(\cdot)+\mathbb K_{\varrho^\lambda}(\cdot)).
$$
In Appendix \ref{app:identity-LR}, we show that its $\tilde M_{\varrho^\lambda}$-weighted adjoint reads
$$
\mathbb L_\lambda(\cdot)=i[H^\lambda,\cdot]+\sum_k \gamma^\lambda_k\big((L^\lambda_k)^\dagger (\cdot) L^\lambda_k -\frac12\{ (L^\lambda_k)^\dagger  L^\lambda_k, \cdot\}\big),
$$
that is,  $\mathbb L_\lambda$ is the backward generator of the  Lindblad dynamics.

Using this expression for $\psi_k$ in \eqref{eq:h-q}, we obtain
$$
\frac12\tr\Big\{\tilde M_{\varrho^\lambda}(\delta X_j)  (-\mathbb L_{\lambda})^{-\ddagger}(\delta X_i)+\tilde M_{\varrho^\lambda}(\delta X_i) (-\mathbb L_{\lambda})^{-\ddagger}(\delta X_j)\Big\}.
$$
Rewriting the second term in terms of its adjoint yields the symmetrized friction tensor obtained in~\cite{scandi2019thermodynamic},
\begin{align*}
   h_{\varrho^\lambda}\!(\partial_{\lambda_i}\varrho^\lambda\!,\partial_{\lambda_j}\varrho^\lambda)&\!=\!\frac12\langle \delta X_i,\big((-\mathbb L_{\lambda})^{-1}\!+\!(-\mathbb L_{\lambda})^{-\ddagger}\big) (\delta X_j)\rangle_{\varrho^\lambda}.
\end{align*}
Therefore, we have shown that the Hamiltonian-dissipative quantum optimal transport distance  ---when restricted to the equilibrium submanifold and pulled back to the space of control parameters $\lambda$--- gives the linear-response regime thermodynamic length~\cite{scandi2019thermodynamic}. The purely dissipative counterpart to this result can be obtained by setting $\mathbb J_{\varrho^\lambda}$ to zero in the definition of $\mathbb L_\lambda,$
leading to a symmetric generator.

As in the discrete setting,  full control of $H^\lambda$ implies that any positive density matrix can be realized as an equilibrium density matrix with respect to a particular $\lambda$. Consider the  quantum optimal transport problem with the fixed mobility operator
$M_{\varrho^\lambda}$
and the corresponding jump operators $L^\lambda$ and jump rates $\gamma_k^\lambda$.  The solution $\varrho^\lambda$ to this problem corresponds to the linear-response geodesic in the limit of slow driving. 
 However, in contrast to the discrete setting, it is not in general possible to find new $\tilde L_k$ and $\tilde \gamma_k$, that are both detailed-balanced with respect to the new Hamiltonian $\tilde H=(\psi-\ln\varrho^\lambda)/\beta^\lambda$, and give rise to the given mobility operator. The reason for this is discussed in Section \ref{sec:III} (below Eq.~\eqref{eq:d-R}). Thus, we may not \emph{uniquely} associate a meaningful counterdiabatic driving in the quantum setting.

\section{Examples}
To make the abstract geometric constructions above more concrete, we now present a small set of representative examples. We first consider a continuous classical example with inertia, where explicit solutions can be obtained in the linear–Gaussian setting. We then illustrate the discrete framework by deriving optimal counterdiabatic protocols for a simple gene-regulation network. Finally, we turn to a qubit example to bound entropy production and highlight how coherent (Hamiltonian) terms can reduce dissipation.

\subsection{RLC circuit with two dissipative channels}

\begin{figure}
    \centering
\includegraphics[width=\linewidth, trim=0 0 0 1.525cm,clip]{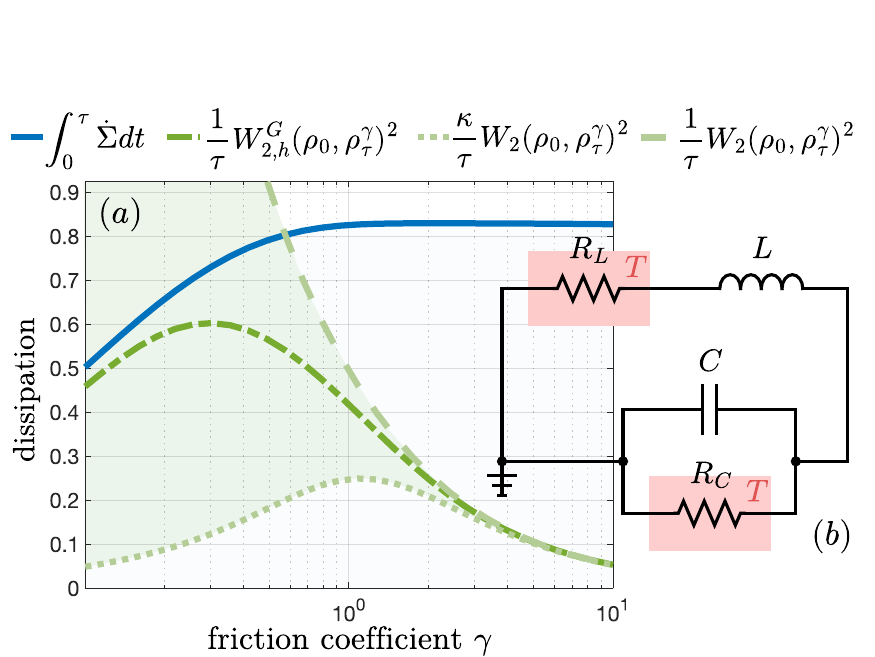}
\caption{(a) Entropy produced with the constant Hamiltonian, $\int_0^\tau\dot\Sigma dt$, and with the optimal quadratic Hamiltonian, $W_{2,h}^G(\rho_0,\rho_\tau^\gamma)^2/\tau$, between endpoints $\rho_0,\rho_\tau^\gamma$ , as a function of the friction coefficient $\gamma=R_L=1/R_C$. The shaded area illustrates the upper and lower bounds in terms of the purely dissipative metric. Constants $\tau, T, C,$ and $ L$ are set to $1.$  (b)  RLC circuit with resistors at temperature $T$. }
    \label{fig:under}
\end{figure}

Consider an RLC circuit with two resistors at temperature $T$, as depicted in Figure~\ref{fig:under}~(b).
Its equations of motion, expressed in terms of the flux at the inductor $p$ and the charge at the capacitor $q$, read
\begin{align*}
dq&=\left(\frac{p}{L}-\frac{1}{R_C}\frac{q}{C}\right)dt+\sqrt{\frac{2T}{R_C}}dW_C
\\
    dp&=-\left(\frac{q}{C}+R_L\frac{p}{L}\right)dt+\sqrt{2TR_L}dW_L,
\end{align*}
where $L$ is the inductance, $C$ the capacitance, $R_L$ and $R_C$ the series and parallel resistances, and $\{W_C\}$, $\{W_L\}$ are independent Brownian motions arising from the resistors being at a finite temperature $T$.
Equivalently, the system follows the ensemble equation \eqref{eq:FP-mixed} with
$H(q,p)=\frac{q^2}{2C}+\frac{p^2}{2L}$ and $D={\rm{diag}}(T/R_C, TR_L)$.

Since the system of stochastic equations is linear, if the initial state is Gaussian, it will continue to be Gaussian at all times. Thus, the state of the system is fully characterized by its mean $m\in\mathbb R^2$ and covariance matrix $P\in\mathbb R^{2\times 2}$.
Its mean evolves as
\begin{subequations}\label{eq:H-c-ev}
  \begin{equation}
\dot m=(J-\beta D)\mathsf H m,    
  \end{equation}
where $\mathsf H={\rm diag}(1/C, 1/L)$ is the Hessian of $H(q,p)$ and $J$ is the symplectic matrix \eqref{eq:symplectic}. The covariance matrix 
evolves as
\begin{equation}
    \label{eq:covar}
    \dot P= (J-\beta D)\mathsf{H}P+P\mathsf{H}(-J-\beta D)+2D.
\end{equation}
\end{subequations}
Since $\nabla\phi=(\beta \mathsf H -P^{-1})x+P^{-1}m$ with $x=(q,p)^\top,$ the entropy production rate at the resistors
may be written for this linear system as
\begin{align}\displaystyle\nonumber
\dot\Sigma=&\int_{\mathbb R^2}x^\top \Phi D\Phi x \rho dx+2m^\top \Phi DP^{-1}m+ m^\top P^{-1}DP^{-1}m\\
    =&\tr(\Phi D\Phi P)+\beta^2 m^\top\mathsf H D \mathsf H m,
\label{eq:entropy-rlc}
\end{align}
where we have defined $\Phi=(\beta \mathsf{H}-P^{-1})$, and for the second equality we have used the fact that $\int_{\mathbb R^2}x^\top \Phi D\Phi x \rho dx=\tr(\Phi D\Phi P)+m^\top \Phi D \Phi m$.

We now consider the problem of optimally choosing the Hamiltonian so as to minimize entropy production between two given Gaussian endpoints. Motivated by the fact that linear dynamics preserve Gaussianity, we will restrict ourselves to minimizers within the Gaussian/quadratic class.
In particular, we consider quadratic Hamiltonians leading to linear dynamics, i.e., we take the ansatz $\phi=x^\top A x/2+b^\top x,$ where $A$ is symmetric. Then, the system dynamics may be written as
\begin{equation}\label{eq:mean}
        \dot m=(TJ- D) (Am +b),
    \end{equation}
    and
    \begin{equation*}
       \dot P= (TJ-D)AP+PA(-TJ-D). 
    \end{equation*}
Since $\nabla\phi=Ax+b$,  following the same steps as earlier, the corresponding entropy production may be written as
\begin{align*}
\nonumber
&\int_0^\tau\Big(\int_{\mathbb R^2}x^\top A DAx \rho dx+b^\top D Am +m^\top ADb+ b^\top Db\Big)dt\\
   & =\int_0^\tau\Big(\tr(A DA P)+(b^\top+m^\top A)D (Am+b)\Big)dt.
\end{align*}
Note that, using \eqref{eq:mean}, the second term is simply
$$
\int_0^\tau \dot m^\top Q\dot mdt, 
 \mbox{ with } Q:= (TJ-D)^{-\top}D(TJ-D)^{-1},$$
 where $(TJ-D)$ is invertible by virtue of $D$ being positive definite and $J$ skew-symmetric (so every eigenvalue of $TJ-D$ has strictly negative real part).

Therefore, if we restrict the search for optimizers of the optimal transport problem with Gaussian endpoints to quadratic functions $\phi$, we obtain a natural finite-dimensional reduction of the problem on the Gaussian/quadratic subclass:
\begin{align}\label{eq:Gaus-ot}   W_{2,h}^G(\rho_0,&\rho_\tau)^2\!:=\tau\!\!\inf_{m,P,A} \int_0^\tau\!\!\Big(\tr(A DA P)+\dot m^\top \!Q\dot m\Big)dt,\!
\\&\qquad\ \  \mbox{s.t. } \dot P= (TJ-D)AP+PA(-TJ-D), \nonumber\\&\ \ P(0)=P_0,\, P(\tau)=P_\tau,\, m(0)=m_0,\, m(\tau)=m_\tau,\nonumber
\end{align}
where  $\rho_0\sim\mathcal N(m_0,P_0)$ and $\rho_\tau\sim\mathcal N(m_\tau,P_\tau)$ are Gaussians with mean $m_0,\, m_\tau$ and covariance matrix $P_0,\, P_\tau$, respectively. Note that, since this is a restricted optimal transport problem, in general  $W_{2,h}^G(\rho_0,\rho_{\tau})\geq W_{2,h}(\rho_0,\rho_{\tau})$.
 \begin{figure*}[tb]
    \centering
\includegraphics[width=0.88\textwidth,trim= 0cm 7.7cm 0cm 0cm,clip]{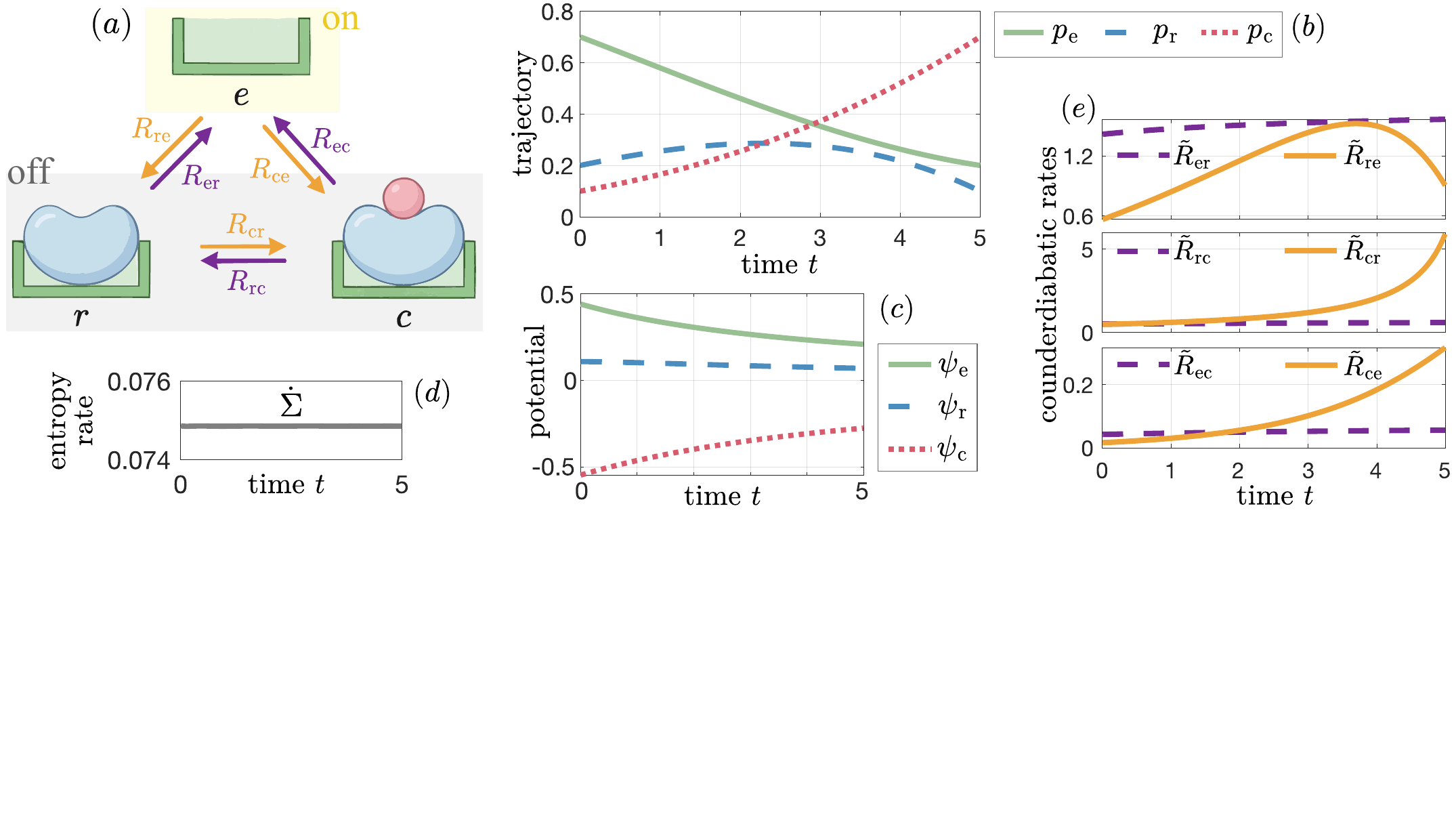}
    \caption{Minimum entropy counterdiabatic switching for gene regulation. (a) Illustration of the discrete Markov process, the operator site is depicted in green, the repressor in blue, and the corepressor in red.   Portrayal of (b) optimal trajectories, (c) optimal potentials, (d) resulting entropy production rate, and (e) detailed-balanced counterdiabatic rates that realize the optimal trajectory in finite time. We have set $\tau=5$ and $T=1.$}
    \label{fig:d}
\end{figure*}
By inspecting \eqref{eq:Gaus-ot} we realize that,
as in the case of standard optimal transport, the problem uncouples into two independent ones. The first one is that of optimally driving the mean between the given endpoints, so as to minimize the second term in the cost~\eqref{eq:Gaus-ot}. The second problem is that of optimally driving the covariance matrix through $A$ so that the first term in~\eqref{eq:Gaus-ot} is minimized. Therefore, we may build a solution to \eqref{eq:Gaus-ot} by independently solving: (i) the problem of driving the mean between endpoints with identical covariance, and (ii) the problem of driving the covariance matrix between endpoints with zero mean.

The solution to the first problem is rather straightforward.
Indeed, the optimal mean that minimizes entropy production between Gaussian endpoints with means $m_0$ and $m_\tau$ has constant velocity and is given by the linear interpolation
$$
m(t)=\bigg(1-\frac{t}{\tau}\bigg)m_0+\frac{t}{\tau} m_\tau.
$$
This leads to the minimum entropy production to drive the mean
$$
\frac{1}{\tau} (m_\tau-m_0)^\top Q(m_\tau-m_0).
$$
In other words, let $\rho_0\sim\mathcal N(m_0,P_0)$  and $\rho_\tau\sim\mathcal N(m_\tau,P_0).$ The endpoint covariances being identical, the choice $P=P_0$ and $A=0$ is admissible, making the first term of the cost~\eqref{eq:Gaus-ot} minimal and equal to zero.
 Then, we find that 
\begin{equation}\label{eq:W2-m}
    W^G_{2,h}(\rho_0,\rho_\tau)^2=(m_\tau-m_0)^\top Q(m_\tau-m_0).
\end{equation}
It is worth highlighting that, while the value of the minimum entropy produced does depend on the inertia 
of the system through $Q$, the optimal trajectory $m(t)$ does not; remarkably, it is the same as in the overdamped setting. 

In this case, we can actually show that 
$$
W^G_{2,h}(\rho_0,\rho_\tau)^2=W_{2,h}(\rho_0,\rho_\tau)^2,
$$
implying that linear Gaussian dynamics are optimal. To do so, we prove the inequality $h_{\rho}(\dot\rho,\dot\rho)\geq \dot m^\top Q\dot m$, leading to $ W_{2,h}(\rho_0,\rho_\tau)^2\geq W^G_{2,h}(\rho_0,\rho_\tau)^2$, and thus both distances must be equal. Indeed, 
 $$
\dot m=\int_{\mathbb R^2} x\dot \rho dx=(TJ-D)\mathbb E[\nabla\phi],
 $$
 where $\mathbb E[\nabla\phi]=\int_{\mathbb R^2}\nabla\phi\rho dx.$
 By Jensen's inequality
 $$
 \mathbb E[\nabla\phi^\top D\nabla \phi]\geq \mathbb E[\nabla\phi]^\top D \mathbb E[\nabla\phi],
 $$
 thus, 
 $$
 h_\rho(\dot\rho,\dot\rho)\geq \dot m ^\top(TJ-D)^{-\top} D(TJ-D)^{-1}\dot m,
 $$
establishing the desired inequality.

Next, we consider the case where the initial and final endpoints have zero mean  and different covariances.
Since this problem does not have a straightforward closed-form solution, 
we solve it numerically. 
In Figure \ref{fig:under}a, we display the value of the entropy produced $\int_0^\tau\dot\Sigma dt$, computed using~\eqref{eq:entropy-rlc}, through the constant Hamiltonian evolution \eqref{eq:H-c-ev} starting from $\rho_0\sim\mathcal N(0,P_0)$, for different values of $\gamma$, where $\gamma=R_L=1/R_C$. We compare this to the minimum entropy production $ W^G_{2,h}(\rho _0,\rho _\tau^\gamma)^2/\tau$, which is
obtained by solving the restricted optimal transport problem between endpoints $\rho_0\sim \mathcal N(0,P_0)$ and $\rho_\tau^\gamma \sim \mathcal N(0,P^\gamma_\tau)$, where $P_\tau^\gamma$ is the covariance evolved for time $\tau$ from \eqref{eq:covar} with a given $\gamma$, starting at $P_0.$ We display this distance together with the upper and lower bounds in \eqref{eq:hierarchy-c}. Indeed, that the same bounds hold for the Gaussian restricted case follows from integrating \eqref{eq:hierarchy-app-main} and \eqref{eq:lower-kappa-c}, optimizing only over Gaussian trajectories, and noting that Gaussian trajectories are optimal in the purely dissipative setting. Thus, the green shaded area corresponds to the envelope provided by $\kappa W_{2}(\rho _0,\rho _\tau^\gamma)^2/\tau$ and $ W_{2}(\rho _0,\rho _\tau^\gamma)^2/\tau$. 
We observe that the envelope shrinks 
as $\gamma\to\infty$ and we enter the overdamped regime, where $\kappa\to 1$ and dissipative and mixed distances agree. On the other hand, when $\gamma$ is small, the purely dissipative $W_2$ distance fails to bound entropy production since Hamiltonian dynamics dominate.

\subsection{Optimal counterdiabatic switching in a gene regulatory network}
\label{sec:ex_d}

Let us illustrate the applicability of geometrically optimal counterdiabatic protocols in the discrete setting through a gene regulation example~\cite{ilker2022shortcuts}. A common form of gene regulation in bacteria is enacted through repressor proteins that have the ability to bind to an operator site on DNA, preventing the transcription of the genes associated to that site. Moreover, some weakly binding repressor proteins may additionally require the presence of a small corepressor molecule to effectively prevent transcription.
The collective dynamics act as a genetic switch between two positions:  gene expression is ``on'' when the operator site is empty, while an occupied operator site (whether with a repressor protein or both a repressor and corepressor) describes the ``off'' position.

These dynamics can be modeled by a Markov process on three states: empty operator site (e), repressor bound to operator site (r), and repressor and corepressor bound to operator site (c) (see Figure \ref{fig:d}a). The unbinding rates $R_{\rm er},\, R_{\rm rc}, R_{\rm ec}$ are typically prescribed by the unbinding reactions. On the other hand, the binding rates $R_{\rm re},\, R_{\rm cr}, R_{\rm ce}$
 can be externally modulated by the concentrations of the bare repressors, corepressors and repressor-corepressor complexes~\cite{ilker2022shortcuts}. In this example, we use unbinding rates taken from \emph{in vitro} measurements of the purine repressor system for \emph{E. coli}~\cite{schumacher1995mechanism,xu1998kinetic}. We are interested in switching protocols that have minimal thermodynamic cost.

To this end,
we compute optimal counterdiabatic protocols and trajectories by solving \eqref{eq:discreteOT-LR}. We consider endpoints $(p_{\rm e}(0),p_{\rm r}(0),p_{\rm c}(0))=(0.7,0.2,0.1)$ and $(p_{\rm e}(\tau),p_{\rm r}(\tau),p_{\rm c}(\tau))=(0.2,0.1,0.7)$, corresponding to a switch from the approximately ``on'' to the ``off'' position. In this model, the mobility operator is fixed to $[M_{p^\lambda}(\cdot)]_{nm}=\frac12R^\lambda_{nm}p^\lambda_m[\cdot]_{nm}$,  where the unbinding rates are given by
$R_{\rm er}=1.68
,$ $ R_{\rm rc}=0.72
,$ $R_{\rm ec}=0.072
$~\cite{schumacher1995mechanism,xu1998kinetic}, while the binding rates depend on $\lambda,$ and satisfy detailed balance: $R^{\lambda}_{\rm re}=R_{\rm er}p^{\lambda}_{\rm r}/p^{\lambda}_{\rm e},$ $ R^{\lambda}_{\rm cr}=R_{\rm rc}p_{\rm c}^\lambda/p_{\rm r}^\lambda, $ $ R^{\lambda}_{\rm ce}=R_{\rm ec}p_{\rm c}^\lambda/p_{\rm e}^\lambda.$
In Figure \ref{fig:d}, we plot the optimal trajectories (b), the optimal potentials $\psi$ that lead to those trajectories (c), the associated entropy production rate (d) and the corresponding counterdiabatic rates, detailed-balanced  with respect to the tilted distribution $\propto p^\lambda e^{-\psi}$, given by \eqref{eq:tilde-Rnm} (e).
 
 This example is based on the model studied in Section IV A in~\cite{ilker2022shortcuts}. In that paper,   different counterdiabatic protocols that lead to a prescribed trajectory are studied. In contrast, here we simultaneously optimize over the trajectory and find the counterdiabatic detailed-balanced rates that lead to such trajectory.
Specifically, we illustrate the fact that optimizing over the trajectory may lead to non-monotonic behavior of the elements of the distribution (see $p_{\rm r}$ in Fig.~\ref{fig:d}b). Moreover, we obtain constant entropy production rate (see Fig.~\ref{fig:d}d), as expected from the Riemannian framework.

\subsection{Quantum two-level system 
}
\begin{figure}
    \centering
\includegraphics[width=1\linewidth,trim= 0.2cm 0.2cm 2.25cm 0cm,clip]{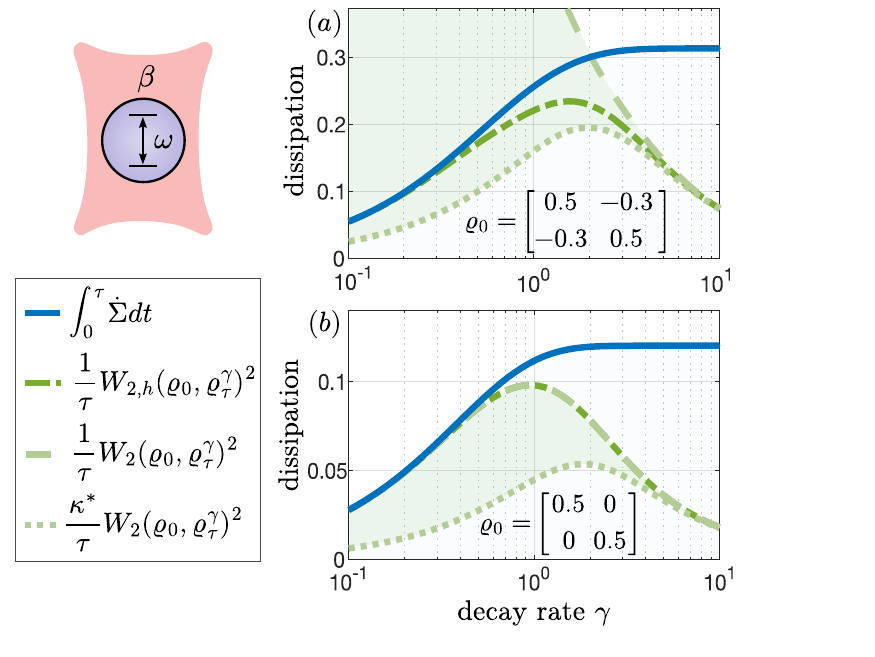}
    \caption{Comparison between entropy produced and the derived bounds for different values of the decay rate $\gamma$; $\beta,\omega,\tau$ are set to $1$. (a) The initial state is not diagonal, and the Hamiltonian term plays a role. (b) The initial state is diagonal, and the Hamiltonian term plays no role in the dynamics. }
    \label{fig:qbit}
\end{figure}

For our last example, we consider a two-level system satisfying the Lindblad equation \eqref{eq:full-lind} with Hamiltonian
$$
H=\frac{\omega}{2}\left[\begin{array}{cc}
   -1  & 0 \\
   0 &  1
\end{array}\right].
$$
We consider two possible transitions $K=\{+,-\}$, with jump operators
$$L_+=\left[\begin{array}{cc}
   0  & 0 \\
   1  &  0
\end{array}\right],\  L_-=\left[\begin{array}{cc}
 0  & 1 \\
   0  &  0
\end{array}\right],
$$
and jump rates $\gamma_-=\gamma$ and $\gamma_+=\gamma e^{-\beta\omega}$. Starting from an initial state $\varrho _0$, we may compute the entropy produced through the constant Hamiltonian dynamics up to some time $\tau$, $\int_0^\tau\dot\Sigma dt$, as shown in Figure \ref{fig:qbit} for different values of $\gamma$ and initial states. 

We compare this entropy production with the bound provided by the Hamiltonian-dissipative Wasserstein-2 distance measured between the same endpoints, namely, $\varrho_0$ and $\varrho_\tau^\gamma$, where $\varrho_\tau^\gamma$ is the state evolved from $\varrho_0$ according to the constant Hamiltonian up to time $\tau$ with the given $\gamma$. Moreover, we provide upper and lower bounds of $W_{2,h}^2/\tau$ through $\kappa_*$ and the dissipative Wasserstein-2 distance between the corresponding endpoints. 
As is expected, when the initial density matrix has coherences, the dissipative Wasserstein-2 distance fails to lower-bound entropy production for small $\gamma$ (see Figure \ref{fig:qbit}a).
This is due to the dissipative Wasserstein-2 distance ignoring the Hamiltonian evolution that is dominant when $\gamma$ is small and coherences are present, leading to an overestimation of the entropy produced along the trajectory whose dynamics are mostly unitary. As $\gamma$ becomes large, the upper ($W_2^2/\tau$), lower ($\kappa_* W_2^2/\tau$) and Hamiltonian-dissipative ($W_{2,h}^2/\tau$) curves coalesce.

Figure \ref{fig:qbit}b shows the case where $[H,\varrho_0 ]=0$, and therefore, the case where $H$ and $\varrho^\gamma_\tau$ commute too.
An exact agreement of the Hamiltonian-dissipative and purely dissipative distances is observed, where the Hamiltonian term is seen to play no role, and we effectively have discrete (classical) purely dissipative dynamics. 
This suggests equality of the Hamiltonian-dissipative, purely dissipative and discrete distances, when the endpoints commute. 
From result \eqref{eq:q-d-W2}, we know that the purely dissipative and discrete distances are equal in this case.
However, we were not able to rigorously prove this statement for the Hamiltonian-dissipative distance. Even if one can show that the purely dissipative solution provides an extremum for the Hamiltonian-dissipative problem, convexity may not be guaranteed, and therefore we cannot conclude that it is in general a minimum.


\section{Conclusions}
 In this work, we provide a unified perspective on how Wasserstein-2 geometries can quantify and bound entropy production for overdamped, inertial, discrete, and quantum systems alike. We have seen how relaxation towards equilibrium can be expressed in terms of a gradient flow of free energy with respect to the purely dissipative Riemannian metric, while conservative dynamics provide an additional Hamiltonian flow. We have compared mixed and purely dissipative geometries and quantified the effects of inertia and coherent dynamics via an equivalence of metrics. Finally, we have shown that the restriction of these metrics to equilibrium states gives rise to linear-response-regime thermodynamic lengths, and how this may be exploited to design counterdiabatic protocols that minimize entropy production.

Building on this, several results readily follow. 
First, the introduced metrics induce a decomposition of entropy production into housekeeping and excess terms, for systems that do not satisfy detailed balance. Specifically, a general irreversible force $F$, giving rise to a dissipative current $M_\mu(F)$, may be split into gradient $G$ and divergence-free $\chi$ terms with respect to the inner product defined by $M_\mu$, so that
$$
\langle F,M_\mu(F)\rangle=\langle G,M_\mu(G)\rangle+\langle \chi,M_\mu(\chi)\rangle,
$$
thus decomposing into two non-negative terms, the excess and housekeeping terms, respectively. This has already been explored for overdamped~\cite{dechant2022geometric,miangolarra2024minimal}, discrete~\cite{yoshimura2023housekeeping} and quantum settings~\cite{yoshimura2025force}.

Second,  thermodynamic uncertainty relations (TUR) may be derived from these geometries. In particular, we may easily derive a short-time TUR for both purely dissipative and mixed systems. To this end, let $\mu$ evolve according to mixed conservative-dissipative dynamics, $\dot\mu=-(\mathbb J_\mu+\mathbb K_\mu)(\phi)$. Consider a time-invariant observable $\psi$ and let $\bar\psi$ denote its ensemble average with respect to $\mu$. Then, its time derivative satisfies
$$
\dot{\bar\psi}=\llangle\psi,\dot\mu\rrangle=-\langle\psi,(\mathbb J_\mu+\mathbb K_\mu)(\phi)\rangle=\langle \mathbb K^{-\frac12}_\mu(\mathbb J_\mu-\mathbb K_\mu)\psi,\mathbb K^{\frac12}_\mu\phi\rangle,
$$
where in the last equality we have used that $\mathbb K_\mu$ is symmetric and $\mathbb J_\mu$ skew-symmetric.
Taking the square of this expression, we have
$$
\dot{\bar\psi}^2=\langle \mathbb K^{-\frac12}_\mu(\mathbb J_\mu-\mathbb K_\mu)\psi,\mathbb K^{\frac12}_\mu\phi\rangle^2\leq \langle \psi,\mathbb Q_\mu\psi\rangle\langle\phi,\mathbb K_\mu(\phi)\rangle,
$$
where we have used Cauchy-Schwarz inequality and defined $\mathbb Q_\mu:=(\mathbb J_\mu-\mathbb K_\mu)^\dagger\mathbb K^{-1}_\mu (\mathbb J_\mu-\mathbb K_\mu).$
Therefore, we obtain, for the continuous, discrete, and quantum cases, the short-time TUR 
$$
\frac{\dot{\bar\psi}^2}{\langle\psi,\mathbb Q_\mu(\psi)\rangle}\leq\dot\Sigma,
$$
where entropy production is lower bounded by the rate of change of the average and the $\mathbb Q_\mu$-weighted ``variance'' of the observable. The corresponding purely dissipative result is obtained by replacing $\mathbb Q_\mu$ with $\mathbb K_\mu,$ since setting $\mathbb J_\mu$ to zero in $\mathbb Q_\mu$ leads to $\mathbb Q_\mu=\mathbb K_\mu$. In its present form, this variance may not be experimentally accessible, and thus more tractable expressions of these short-time TURs are of interest~\cite{yoshimura2023housekeeping}.

Other directions may be pursued.  Indeed, it is possible to extend this $W_2$ optimal transport perspective to other systems. For instance, deterministic chemical reaction networks with nonlinear dynamics can also be framed within this setting~\cite{yoshimura2023housekeeping}.
Furthermore, optimal transport bounds that account for the possibility of measurement and feedback are of interest. These have been developed for overdamped \cite{taghvaei2021relation,kamijima2025optimal} and discrete systems \cite{nagase2024thermodynamically}, but remain otherwise largely unexplored. Moreover, the quantum framework developed here is restricted to Davies-type generators whose jump operators are exact eigenoperators of the Hamiltonian commutator. This structure is essential for the identity~\eqref{eq:miracle}, and hence for the gradient-flow representation~\eqref{eq:q-grad}. More general quantum master equations satisfying KMS detailed balance need not admit this Bohr-frequency-resolved representation~\cite{chen2025efficient}. Extending the optimal-transport geometry to general KMS-detailed-balanced Lindbladians is an interesting direction for future work.

Furthermore, in applications, full control of the Hamiltonian is not typically available. Instead, a finite number of parameters can be manipulated, leading to only a subset of the space of thermodynamic states being reachable. Thus,  accounting for limited control in building optimal trajectories is of practical interest. A possible route is paved by the provided link between optimal transport and linear-response geodesics. As shown in \cite{zhong2024beyond} for the overdamped setting,
it is possible to find linear-response geodesics and a corresponding counterdiabatic driving that are accessible through limited control by loosening the endpoint condition.

In addition, in  the overdamped setting, optimal transport geometry  has proven useful in the design of optimal thermodynamic engines~\cite{movilla2021energy,zhong2025optimal}. Specifically, work output over a cycle can be characterized as the addition of a dissipative term, captured by the optimal transport length, and a quasi-static term that can be cast as a line integral with respect to the optimal transport geometry~\cite{movilla2021energy,zhong2025optimal}. Consequently, optimal cycles balance these terms through isoperimetric or isoholonomic problems~\cite{movilla2021energy,huang2016optimal,huang2020sub}, and isoperimetric inequalities can bound entropy production in 2-dimensional submanifolds~\cite{movilla2021energy,frim2022geometric}. Similar results may extend to inertial, discrete, and quantum settings, and would yield fundamental geometric bounds on power and efficiency in these regimes.

Finally, it is of central relevance to better understand the connection between Wasserstein-2 and Wasserstein-1 distances. 
With Wasserstein-1 distances, the powerful Riemannian structure is lost. Moreover, they generally do not recover the overdamped result or the linear-response metric. On the other hand, Wasserstein-1 distances are typically defined in terms of optimal control problems under scalar constraints on average kinetic activity (e.g., dynamical activity, mobility, frenesy, etc.)~\cite{dechant2022minimum,van2023thermodynamic,nagayama2025infinite,kolchinsky2026generalized}. These constraints are significantly less restrictive than fixing the full mobility operator, as is required in the presented Wasserstein-2 approach. This is particularly taxing in the quantum setting, where arbitrary mobility operators may not be obtained for fixed Hamiltonians. The difference in what is being constrained precludes direct comparisons between the corresponding thermodynamic bounds, complicating their connection.

This raises the question of whether Riemannian geometries are natural structures in the far-from-equilibrium regime. A way to understand this is by noting that the presented Riemannian metrics may be seen as arising from a large deviation principle with a quadratic 
functional \cite{adams2011large}. 
However, large deviation functionals of interest need not be quadratic arbitrarily far away from equilibrium, yielding non-Riemannian geometries~\cite{mielke2014relation}. Whereas diffusion processes have quadratic large deviation functions, Poisson processes, for instance, do not ---unless close to equilibrium~\cite{mielke2017non,kolchinsky2026generalized}. 
Through the presented framework, we have pushed the Riemannian endeavor further away from equilibrium, being able to give meaningful bounds on entropy production and optimal finite-time counterdiabatic protocols. The question is now whether there are other geometries that recover the presented Wasserstein-2 structures and the linear-response-regime geometry when near equilibrium, but that far-from-equilibrium, outside the diffusive setting, take non-Riemannian form.
\\

\noindent
\emph{Acknowledgments:} OMM would like to thank Tryphon T. Georgiou and Luis A. Correa for their guidance and support.
OMM was supported by the European Union's Horizon 2020 research and innovation
programme under the Marie Skłodowska-Curie Grant Agreement No. 101151140, and acknowledges support from Ministerio de Ciencia e Innovaci\'on and European Union (FEDER) (PID2022-138269NB-I00). RS is supported by the National Science Foundation under grant ECCS-2347357, the Air Force Office of Scientific Research under FA9550-24-1-0278, and the Army Research Office under W911NF-22-1-0292. AK is partly supported by John
Templeton Foundation (Grant No. 62828) and by the
European Union’s Horizon 2020 research and innovation
programme under the Marie Skłodowska-Curie Grant
Agreement No. 101068029.

\bibliography{main}

\appendix


\section{System-independent quantum formulation}\label{app:basisless}
Here we show that the optimal transport problem \eqref{eq:QOMT} can be written through system-independent derivatives in analogy to the classical (continuous and discrete) settings. To do so, we will shift the system dependence of the derivatives with respect to $L_k$ to a weighted mobility operator $\hat M_\varrho$.

To this end, consider the Hermitian, traceless, orthonormal operator basis $\{E_\alpha\}_{\alpha\in A}$, with $A=\{1,\ldots,d^2-1\},$ satisfying $E_\alpha=E^\dagger_\alpha,$ $\tr \{E_\alpha\}=0,$ $\tr\{E_\alpha E_\beta\}=\delta_{\alpha\beta}$.
Define the Hilbert space $\mathfrak h_A:=\oplus_{\alpha\in A}\mathfrak h_\alpha$, where each $\mathfrak h_\alpha$ is a copy of $\mathfrak h.$  For $\mathbf V_A\in\mathfrak h_A$, let $V_\alpha$ denote the component of $\mathbf V_A$ in $\mathfrak{h}_{\alpha}$. We equip $\mathfrak h_A$ with the inner product
$$
\langle \mathbf U_A,\mathbf V_A\rangle_A=\sum_{\alpha\in A}\tr\{U_\alpha^\dagger V_\alpha\},
$$
and define the quantum partial derivatives $\partial_\alpha:\mathfrak h\to\mathfrak h_\alpha$
$$
\partial_\alpha \varphi=[E_\alpha,\varphi] =[E_\alpha^\dagger,\varphi]=\partial_\alpha^\dagger \varphi,~~\varphi\in\mathfrak h,
$$
for all $\alpha\in A.$
The associated gradient operator $\nabla_A:\mathfrak h\to\mathfrak h_{A}$ is defined as
$$
\nabla_A \varphi=\Big[\partial_{1}\varphi,\cdots,\partial_{d^2-1}\varphi\Big]^\top,
$$
 and the divergence operator $\nabla_A~\cdot:\mathfrak h_A\to\mathfrak h$ as
$$
\nabla_A \cdot \mathbf V_A=-\sum_{\alpha\in A}\partial_\alpha^\dagger V_{\alpha}.
$$
Again, it is easy to check that the integration by parts formula holds for the new derivative and inner product.

Let us write jump operators $L_k$ in terms of the introduced basis as $L_k=\sum_{\alpha\in A}c_{k\alpha}E_\alpha$, with $c_{k\alpha}\in\mathbb C$. We define the $2N\times (d^2-1)$ complex-valued matrix $C$ with elements $c_{k\alpha}$. That is, $C:\mathfrak h_A\to\mathfrak h_K$ such that 
$$
[C\mathbf V_A]_k=\sum_{\alpha\in A} c_{k\alpha}V_\alpha
.$$
Consider $\hat M_\varrho:\mathfrak h_A\to\mathfrak h_A$ given by
\begin{equation}\label{eq:basisless-m}
  \hat M_\varrho(\mathbf V_A):= C^* M_\varrho(C\mathbf V_A),  
\end{equation}
where $
C^*:\mathfrak h_K\to\mathfrak h_A$ is given by 
$$
[C^*\mathbf V]_\alpha=\sum_{k\in K}c^*_{k\alpha}V_k.
$$
With these definitions, we have
$$\nabla=C\nabla_A\mbox{ and }\nabla\cdot(\mathbf V)=\nabla_A\cdot(C^*\mathbf V),
$$
leading to
$$
    \dot\varrho = \nabla_A\cdot( \hat M_{\varrho }(\nabla_A\phi)),
$$
and 
$$
  g_{\varrho }(\dot \mu,\dot\nu):=\langle \nabla_A\varphi, \hat M_{\varrho }(\nabla_A \psi)\rangle_A.
$$
Therefore, we may write~\eqref{eq:qOT} in terms of system-independent derivatives, where all the model dependence is contained in the mobility operator $\hat M_\varrho$.

 \section{Quantum and discrete \texorpdfstring{$W_2$}{W2} distances coincide when \texorpdfstring{$[H,\varrho_0]=[H,\varrho_\tau]=0$}{[H,varrho0]=[H,varrhotau]=0}}
\label{app:q-d-W2}

 Consider a quantum system following dynamics~\eqref{eq:diss-lind}, where the jump operators are assumed to be rank one, and the corresponding Hamiltonian $H$ is non-degenerate.
Let $\varrho_0$, $\varrho_\tau$ and $H$ be simultaneously diagonalizable, i.e. $[H,\varrho_0]=[H,\varrho_\tau]=0$. We would like to show that, in this case, the quantum Wasserstein distance $W_2(\varrho_0,\varrho_\tau)$, with the fixed mobility prescribed by $\gamma_k, L_k$, simplifies to the discrete classical one $W_2(p_0,p_\tau)$ between the eigenvalue probabilities, with the fixed mobility prescribed by $R_{nm}=\gamma_k$ for $k:m\to n$
. To show this, we will first provide the first-order optimality conditions for the discrete and quantum problems in the convex flux formulation. Then, we show that if there is a solution to the first-order optimality conditions of the discrete problem~\eqref{eq:discreteOMT}, then the same solution (stacked as diagonal matrices) satisfies the necessary conditions for optimality of the quantum problem. By convexity of the optimization problems, it follows that those solutions are global minimizers. Finally, we show that both distances are equal since both cost functions are identical when evaluated at the respective solutions.

\subsection{Optimality conditions for the discrete problem}
Consider the discrete optimal transport problem \eqref{eq:discreteOMT} in the convex flux formulation, namely,
\begin{align}\label{eq:discreteOMT-app} 
&\inf_{ p\in\mathcal P_*(\mathcal X),\mathcal J}\tau\sum_{n\neq m}\int_0^\tau \frac{\mathcal J_{nm}^2}{a_{nm}(p)}dt
\\& \nonumber \mbox{s.t.} \ \dot  p_n=\sum_{m\in\mathcal X\backslash\{n\}}(\mathcal J_{mn}-\mathcal J_{nm}),\ p(0)= p_0,\, p(\tau)= p_\tau,\nonumber
\end{align}
where 
$$
a_{nm}(p)=\frac12  R_{nm}p_m^{\rm eq}\,\theta\Big(\frac{ p_{n}}{p_n^{\rm eq}},\frac{ p_m}{p_m^{\rm eq}}\Big),
$$
with the rates $R_{nm}$ satisfying detailed balance with respect to $p_m^{\rm eq}\propto e^{-\beta H_m},$ and
\begin{align*}
\theta(x,y)=\int_0^1x^{1-s}y^sds=\begin{cases}    \frac{x-y}{\ln(x)-\ln(y)},&\mbox{ if } x\neq y,\\x,&\mbox{ if }x=y.
\end{cases}\end{align*}
To obtain the first-order necessary condition for optimality we define the Lagrangian
\begin{align*}
    \mathcal L=\!&\int_0^\tau \!\!\sum_{n\in\mathcal X}\!\bigg\{\tau+\lambda_{n}\dot p_n+\!\!\!\! \!\!\sum_{m\in\mathcal X\backslash\{n\}}\!\!\!\!\!\!-\lambda_{n}(\mathcal J_{mn}\!-\!\mathcal J_{nm})+\!\frac{\mathcal J_{nm}^2}{a_{nm}(p)} \bigg\}dt.
\end{align*}
Taking the first variation with respect to $\mathcal J_{kl}$ we obtain
\begin{align*}
    \delta_{\mathcal J_{kl}}\mathcal L=&\int_0^\tau\bigg(2\tau \frac{\mathcal J_{kl}}{a_{kl}(p)}+\lambda_k-\lambda_l\bigg)\delta_{\mathcal J_{kl}}dt.
\end{align*}
Setting the first variation to zero for all $\delta_{\mathcal J_{kl}}$ we obtain
\begin{equation}
    \label{eq:opt-V}
    \mathcal J_{kl}=-\frac{a_{kl}(p)}{2\tau}[\nabla\lambda]_{kl}.
\end{equation}
In particular, this implies that the optimal velocity $V=M_p^{-1}(\mathcal J)$ takes the gradient form
$$
V_{kl}=-\frac{1}{2\tau}[\nabla\lambda]_{kl}.
$$

Plugging in the optimality condition \eqref{eq:opt-V}, the Lagrangian reads
\begin{align*}
    \mathcal L\!=&\!\int_0^\tau \sum_{n\in\mathcal X}\bigg\{\lambda_{n}\dot p_n-\sum_{m\in\mathcal X}\frac{1}{4\tau}[\nabla\lambda]_{nm}^2 a_{nm}(p) \bigg\}dt
    \\=&\!\int_0^\tau \!\!\!\!\sum_{n\in\mathcal X}\!\bigg\{\lambda_{n}\dot p_n-\sum_{m\in\mathcal X}\frac{1}{8\tau} R_{nm}p_m^{\rm eq}\,\tfrac{\tfrac{p_{n}}{p_n^{\rm eq}}-\tfrac{p_m}{p_m^{\rm eq}}}{\ln\tfrac{p_{n}}{p_n^{\rm eq}}-\ln\tfrac{p_{m}}{p_m^{\rm eq}}}[\nabla\lambda]_{nm}^2 \bigg\}dt,
\end{align*}
where for the first equality we have used the fact that $\sum_{m,n\in\mathcal X}\lambda_n(\mathcal J_{mn}-\mathcal J_{nm})=\sum_{m,n\in\mathcal X}(\lambda_m-\lambda_n)\mathcal J_{nm}.$
The first variation with respect to $p_l$ yields 
\begin{align*}
    \delta_{p_l}\mathcal L=&\int_0^\tau \bigg\{ -\dot \lambda_{l}-\sum_{m\in \mathcal X}\frac{1}{4\tau}\frac{ R_{lm}p_m^{\rm eq}}{{\ln(\frac{p_{l}}{p_l^{\rm eq}})\!-\!\ln(\frac{p_{m}}{p_m^{\rm eq}})}}\times\\&\times\bigg(\frac{1}{p_l^{\rm eq}}-\frac{1}{p_l}\frac{p_{l}/p_l^{\rm eq}-p_m/p_m^{\rm eq}}{\ln(\frac{p_{l}}{p_l^{\rm eq}})\!-\!\ln(\frac{p_{m}}{p_m^{\rm eq}})}\bigg)[\nabla\lambda]_{lm}^2\bigg\}\delta_{p_l}dt.
\end{align*}
Setting it to zero for all $\delta_{p_l}$, we obtain
\begin{subequations}
\begin{equation}\label{eq:FONC-d}
    \dot\lambda_l=-\frac{1}{4\tau}\sum_{m\in\mathcal X}R_{ml}\frac{\frac{p_m}{p_m^{\rm eq}}\frac{p_l^{\rm eq}}{p_l}-\ln\big(\frac{p_m}{p_m^{\rm eq}}\frac{p_l^{\rm eq}}{p_l}\big)-1}{(\ln\tfrac{p_l}{p_l^{\rm eq}}-\ln\tfrac{p_m}{p_m^{\rm eq}})^2}[\nabla\lambda]_{lm}^2 .
\end{equation}
Therefore, to obtain the optimal solution for the discrete optimal transport problem \eqref{eq:discreteOMT-app}, we may solve for $\lambda$ and $p$ in \eqref{eq:FONC-d} together with
\begin{equation}
    \label{eq:FONC-d-p}
    \dot  p_n=\sum_{m\in\mathcal X}\frac{1}{2\tau} R_{nm}p_m^{\rm eq}\,\theta\Big(\frac{ p_{n}}{p_n^{\rm eq}},\frac{ p_m}{p_m^{\rm eq}}\Big)[\nabla\lambda]_{nm},
\end{equation}
\end{subequations}
and the boundary conditions $p(0)=p_0,\,p(\tau)=p_\tau.$

\subsection{Optimality conditions for the quantum problem}
Let us now write the optimality conditions for the convex formulation of the quantum problem.  
Specifically, we want to write the first-order optimality condition for the following  problem:
\begin{align}
&\inf_{\varrho\in\mathcal P_*(\mathbb C,d),\mathcal J
}\tau\int_0^\tau\sum_{k\in K}\tr\{\mathcal J_k^\dagger(M_\varrho^k)^{-1}(\mathcal J_k)\} dt\label{eq:QOMT-app} 
\\   \nonumber & \mbox{s.t.} \ \dot \varrho =-\sum_{k\in K}[L_k^\dagger,\mathcal J_k],\,\, \varrho (0)=\varrho _0,\,\varrho (\tau)=\varrho _\tau, 
\end{align}
where we recall that $\mathcal J_k=M_\varrho^k(V_k).$
To do so, we build the Lagrangian:
\begin{align*}
 \mathcal L=& \int_0^\tau\tr\bigg\{ \tau\sum_{k\in K}\mathcal J_k^\dagger(M_\varrho^k)^{-1}(\mathcal J_k)+\Lambda\Big(\dot\varrho+\sum_{k\in K}[L_k^\dagger,\mathcal J_k] \Big)
\bigg\}dt   .
\end{align*}
Its first variation with respect to $\mathcal J_k$ yields
\begin{align*}
 \delta_{\mathcal J_k}\mathcal L= \int_0^\tau\sum_{k\in K}&\tr\bigg\{ \tau\Delta_{\mathcal J_k}^\dagger(M_\varrho^k)^{-1}(\mathcal J_k)\\&+\tau\mathcal J_k^\dagger(M_\varrho^k)^{-1}(\Delta_{\mathcal J_k})+[\Lambda,L_k^\dagger]\Delta_{\mathcal J_k}
\bigg\}dt   .
\end{align*}
Noting that, by self-adjointness of $(M_\varrho^k)^{-1}$, we have $\langle \mathcal J_k,(M_\varrho^k)^{-1}(\Delta_{\mathcal J_k})\rangle=\langle(M_\varrho^k)^{-1}(\mathcal J_k),\Delta_{\mathcal J_k}\rangle$,  and that for any $A\in\mathfrak h$, $\tr\{A+A^\dagger\}=2{\rm Re}(\tr\{A\})$, we obtain
\begin{align*}
    \delta_{\mathcal J_k}\mathcal L= \!\int_0^\tau\!\!\!\sum_{k\in K}\!{\rm Re}\, \tr\Big\{& \Big(2\tau \big((M_\varrho^k)^{-1}(\mathcal J_k)\big)^\dagger\!\!+[\Lambda,L_k^\dagger]\Big)\Delta_{\mathcal J_k}
\Big\}dt,  
\end{align*}
where taking the real part in both terms is justified since the first variation of $\mathcal L$ must be real.
Setting the first variation to zero for all $\Delta_{\mathcal J_k}$ yields
\begin{equation}
    \label{eq:opt-V-q}
\mathcal J_k=-\frac{1}{2\tau }M_\varrho^k([L_k,\Lambda]).
\end{equation}
Therefore, optimal velocities take the gradient form
$$
    V_k=-\frac{1}{2\tau}[L_k,\Lambda].
$$

We may use the expression \eqref{eq:opt-V-q} to rewrite the Lagrangian as
\begin{align*}
 \!\mathcal L\!=& \!\int_0^\tau\!\!\tr\bigg\{ \Lambda\dot\varrho-\!\!\sum_{k\in K}\frac{\gamma_k}{8\tau}[L_k,\Lambda]^\dagger\!\!\int_0^1\!\!e^{\beta \omega_k s}\varrho^s[L_k,\Lambda]\varrho^{1-s}ds
\bigg\}dt   .\!
\end{align*}
To compute its first-order variation with respect to $\varrho$, note that~\cite{carlen2014analog}
$$
\delta_\varrho \varrho^\alpha=\int_0^1\int_0^\alpha\frac{\varrho^{\alpha-\zeta}}{(1-\sigma)I+\sigma\varrho}\Delta_\varrho \frac{\varrho^{\zeta}}{(1-\sigma)I+\sigma\varrho}d\zeta d\sigma.
$$
Then,
\begin{subequations}\label{eq:FONC-q}
\begin{widetext}
 \begin{align*}
 \delta_\varrho\mathcal L= \int_0^\tau\tr\bigg\{& -\sum_{k\in K}\frac{\gamma_k}{8\tau}[L_k,\Lambda]^\dagger\int_0^1e^{\beta \omega_k s}\int_0^1\int_0^s\frac{\varrho^{s-\zeta}}{(1-\sigma)I+\sigma\varrho}\Delta_\varrho \frac{\varrho^{\zeta}}{(1-\sigma)I+\sigma\varrho}d\zeta d\sigma [L_k,\Lambda] \varrho^{1-s} ds
 \\& -\sum_{k\in K}\frac{\gamma_k}{8\tau}[L_k,\Lambda]^\dagger\int_0^1e^{\beta \omega_k s}\varrho^s[L_k,\Lambda]\int_0^1\int_0^{1-s}\frac{\varrho^{1-s-\zeta}}{(1-\sigma)I+\sigma\varrho}\Delta_\varrho \frac{\varrho^{\zeta}}{(1-\sigma)I+\sigma\varrho}d\zeta d\sigma ds-\dot\Lambda\Delta_\varrho
\bigg\}dt   .
\end{align*} 
Rearranging terms we obtain
 \begin{align*}
 \delta_\varrho\mathcal L= \int_0^\tau\tr\bigg\{\bigg(& -\sum_{k\in K}\frac{\gamma_k}{8\tau}\int_0^1e^{\beta \omega_k s}\int_0^1\int_0^s\frac{\varrho^{\zeta}}{(1-\sigma)I+\sigma\varrho} [L_k,\Lambda] \varrho^{1-s} [L_k,\Lambda]^\dagger \frac{\varrho^{s-\zeta}}{(1-\sigma)I+\sigma\varrho} d\zeta d\sigma ds
 \\& -\sum_{k\in K}\frac{\gamma_k}{8\tau}\int_0^1e^{\beta \omega_k s}\int_0^1\int_0^{1-s}\frac{\varrho^{\zeta}}{(1-\sigma)I+\sigma\varrho}[L_k,\Lambda]^\dagger \varrho^s[L_k,\Lambda]\frac{\varrho^{1-s-\zeta}}{(1-\sigma)I+\sigma\varrho} d\zeta d\sigma ds-\dot\Lambda\bigg)\Delta_\varrho
\bigg\}dt   .
\end{align*} 
Setting this first variation to zero for all $\Delta_\varrho$ Hermitian with zero trace implies (up to an irrelevant term proportional to the identity matrix)
\begin{align}\nonumber
 \dot\Lambda= -\sum_{k\in K}\frac{\gamma_k}{8\tau}\int_0^1e^{\beta \omega_k s}\int_0^1&\bigg(\int_0^s\frac{\varrho^{\zeta}}{(1-\sigma)I+\sigma\varrho} [L_k,\Lambda] \varrho^{1-s} [L_k,\Lambda]^\dagger \frac{\varrho^{s-\zeta}}{(1-\sigma)I+\sigma\varrho} d\zeta 
\\&\!\!+\int_0^{1-s}
\frac{\varrho^{\zeta}}{(1-\sigma)I+\sigma\varrho}[L_k,\Lambda]^\dagger \varrho^s[L_k,\Lambda]\frac{\varrho^{1-s-\zeta}}{(1-\sigma)I+\sigma\varrho} d\zeta\bigg) d\sigma ds   .\label{eq:FONC-q-Lambda}
\end{align} 
\end{widetext}  
This equation, together with 
\begin{equation}\label{eq:opt-dyn-app}
    \dot\varrho=\frac{1}{2\tau}\sum_{k\in K}[L_k^\dagger,M_\varrho^k([L_k,\Lambda])]
\end{equation}
\end{subequations}
and the endpoint conditions $\varrho(0)=\varrho_0$ and $\varrho(\tau)=\varrho_\tau$, constitute the first-order optimality conditions.
If there is a solution to this set of equations with the required endpoints, then, by the convexity of the optimization problem, this solution minimizes our cost.

\subsection{Quantum solution from discrete solution}
We now explicitly assume that $H$ is non-degenerate, and that the jump operators take the form $L_k=|n\rangle\langle m|$, where $\{|n\rangle\}_{n\in\mathcal X}$ is the energy eigenbasis and $\mathcal X$ is the set of energy eigenstates.
To find a solution to the system of equations~\eqref{eq:FONC-q}, let us take the ansatz that $\Lambda$ and $\varrho$ are diagonal in the energy eigenbasis at all times, and let $\{\lambda_i\}$ and $\{p_i\}$ denote their eigenvalues. Then, 
$
[L_k,\Lambda]=-(\lambda_n-\lambda_m)L_k,
$ 
 where $k$ is associated with the transition $m\to n$. 
 Then, the optimal velocity is simply
\begin{equation}\label{eq:Vq2d}
    V_k=\frac{1}{2\tau}(\lambda_n-\lambda_m)L_k.
\end{equation}
With these assumptions, the mobility takes the form in~\eqref{eq:mobility-q2d}, and thus equation \eqref{eq:opt-V-q} reads
\begin{equation}\label{eq:Jq2d}
    \mathcal J_k=
\frac{\gamma_k}{4\tau}\,p_m^{\rm eq}\, \theta\bigg(\frac{p_{m}}{p_m^{\rm eq}},\frac{p_n}{p_n^{\rm eq}}\bigg)(\lambda_n-\lambda_m) L_k.
\end{equation}

On the other hand, using the fact that 
$$
L_k D=D_{m}L_k \ \mbox{ and }\ DL_k=D_{n}L_k,
$$
where $D$ is a diagonal matrix in the energy eigenbasis, we may rewrite \eqref{eq:FONC-q-Lambda} as
    \begin{align*}
 &\dot\Lambda= -\sum_{m,n\in \mathcal X}\frac{\gamma_k}{8\tau}(\lambda_n-\lambda_m)^2\int_0^1e^{\beta  (H_n-H_m) s}p_m^{1-s}p_n^{s}\int_0^1\times\\&\times\bigg(  \frac{s}{(1-\sigma+\sigma p_n)^2} |n\rangle\langle n| 
+ \frac{(1-s)}{(1-\sigma+\sigma p_m)^2} |m\rangle\langle m|\bigg) d\sigma ds    .
\end{align*} 
Note that with the chosen ansatz, the $\Lambda$ dynamics stay diagonal, as required for consistency.
Using the integral
$$
\int_0^1\frac{1}{(1+\sigma(a-1))^2}d\sigma=\frac{1}{a},
$$
and looking at the $l$-th eigenvalue of $\Lambda$, we obtain
\begin{align*}
 &\dot\lambda_l=-\sum_{m\in\mathcal X}\frac{R_{ml}}{8\tau}(\lambda_m-\lambda_l)^2\int_0^1\bigg(\frac{p_m}{p_m^{\rm eq}}\bigg)^{s}\bigg(\frac{p_l}{p_l^{\rm eq}}\bigg)^{-s} (1-s) ds
 \\&
   -\!\sum_{m\in\mathcal X}\!\frac{R_{lm}}{8\tau}(\lambda_l-\lambda_m)^2e^{\beta  (H_l-H_m)}\!\int_0^1\!\bigg(\frac{p_m}{p_m^{\rm eq}}\bigg)^{1-s}\bigg(\frac{p_l}{p_l^{\rm eq}}\bigg)^{s-1}\!\!\! s ds,
\end{align*} 
where we have used $p_n^{\rm eq}=\frac{e^{-\beta H_n}}{Z}$, and defined $R_{ml}=\gamma_k$ where $k:l\to m$.
We may change variables in the second integral $s\to 1-s$, to realize that the integrals of both terms have the form
$$
\int_0^1 (1-s)r^sds=\frac{r-\ln r-1}{(\ln r)^2},
$$
for $r=\frac{p_mp_l^{\rm eq}}{p_m^{\rm eq}p_l}.$
Therefore, we obtain
\begin{align*}
 &\dot\lambda_l=-\sum_{m\in\mathcal X}\frac{R_{ml}}{4\tau}\frac{\frac{p_mp_l^{\rm eq}}{p_m^{\rm eq}p_l}-\ln \frac{p_mp_l^{\rm eq}}{p_m^{\rm eq}p_l}-1}{(\ln \frac{p_mp_l^{\rm eq}}{p_m^{\rm eq}p_l})^2}(\lambda_m-\lambda_l)^2,
\end{align*} 
which exactly coincides with \eqref{eq:FONC-d}. 

With these conventions, the dynamics in \eqref{eq:opt-dyn-app} become diagonal, and can be written in terms of the $n$-th eigenvalue of $\varrho$ as
\begin{align*}
\dot p_n &=-\langle n|\sum_{k\in K}[L_k^\dagger,\mathcal J_k]|n\rangle\ \mbox{ with }k:m\to l,
\\&
=-\sum_{m,l\in\mathcal X}\frac{\gamma_k}{4\tau}\,p_m^{\rm eq}\, \theta\bigg(\frac{p_{m}}{p_m^{\rm eq}},\frac{p_l}{p_l^{\rm eq}}\bigg)(\lambda_l-\lambda_m)\langle n|[L_k^\dagger,L_k]|n\rangle
\\&
=\sum_{m\in\mathcal X}\frac{R_{nm}}{2\tau}\,p_m^{\rm eq}\, \theta\bigg(\frac{p_{m}}{p_m^{\rm eq}},\frac{p_n}{p_n^{\rm eq}}\bigg)(\lambda_n-\lambda_m),
\end{align*}
which exactly coincides with 
\eqref{eq:FONC-d-p}. Therefore, if we can solve for $\lambda$ and $p$ in the first-order optimality equations, \eqref{eq:FONC-d} and \eqref{eq:FONC-d-p}, for the discrete problem with the appropriate endpoints, then we can build a quantum solution to \eqref{eq:FONC-q} by setting $\Lambda=\sum_{n\in\mathcal X}\lambda_n|n\rangle\langle n|,$ and $\varrho=\sum_{n\in\mathcal X}p_n|n\rangle\langle n|$.

It only remains to show the equality of both optimization costs when evaluated at their respective solutions. Indeed, we can rewrite the cost in \eqref{eq:QOMT-app} as $\tau\int_0^\tau\sum_{k\in K}\tr\{\mathcal J_k^\dagger V_k\} dt$, which using \eqref{eq:Vq2d} and \eqref{eq:Jq2d} reads
\begin{align*}
&\int_0^\tau\sum_{k\in K}\frac{\gamma_k}{8\tau}\,p_m^{\rm eq}\, \theta\bigg(\frac{p_{m}}{p_m^{\rm eq}},\frac{p_n}{p_n^{\rm eq}}\bigg)(\lambda_n-\lambda_m)^2 \tr\{L_k^\dagger L_k\} dt
\\&=\int_0^\tau\sum_{n,m\in\mathcal X}\frac{R_{nm}}{8\tau}\,p_m^{\rm eq}\, \theta\bigg(\frac{p_{m}}{p_m^{\rm eq}},\frac{p_n}{p_n^{\rm eq}}\bigg)(\lambda_n-\lambda_m)^2dt
\end{align*}
which coincides with \eqref{eq:discreteOMT-app} evaluated at the optimal $(p,\mathcal J)$.

Therefore, putting everything together, we have shown that the diagonal pair $(\Lambda,\varrho)$ constructed above from a solution of the discrete first-order optimality equations is feasible for the quantum problem and satisfies the quantum stationarity conditions. 
Since the optimal transport problems in flux variables are convex, the first-order optimality conditions are not only necessary but also sufficient for global optimality. 
Their actions at these trajectories being equal implies that the discrete and quantum Wasserstein-2 distances coincide,  that is,
$$
W_2(\varrho_0,\varrho_\tau)=W_2(p_0,p_\tau),
$$
where $[p_0]_n$ and $[p_\tau]_n$ are the eigenvalues of $\varrho_0$ and $\varrho_\tau$, respectively. Here, the discrete optimal transport problem is with the fixed mobility prescribed through $R_{nm}=\gamma_k$, with $k:m\to n.$ 

\section{Vanishing entropy production for a given quantum trajectory}
\label{app:q-dechant}

We would like to show that one can choose a Hamiltonian, admissible jump operators, and jump rates,  such that entropy production can be made arbitrarily small for any given quantum trajectory in $\mathcal P_*(\mathbb C,d)$. The Hamiltonian, jump operators, and jump rates are not necessarily detailed-balanced, and are allowed to be time-varying and to depend on the given trajectory, as in the discrete counterpart of this result~\cite{dechant2022minimum}. The underlying idea of the proof is to split the quantum evolution into a populations part and a unitary part: the unitary can be generated by a Hamiltonian with no entropy produced, while the populations can be changed with arbitrarily low entropy production by using an argument analogous to the classical one \cite{dechant2022minimum}. 

Thus, as in the discrete setting, we must relax the detailed balance assumption to consider systems that may only satisfy local detailed balance. 
Specifically, we consider a quantum system evolving according to
\begin{equation}\label{eq:diss-lind-2}
    \dot\varrho =-i[H,\varrho]+\sum_{k\in K}\gamma_k\Big(L_k\varrho  L_k^\dagger -\frac12\{L_k^\dagger L_k,\varrho \}\Big),
\end{equation}
where $K:=\{-N,\ldots,-1,1,\ldots,N\}.$ 
Let us relax the detailed balance assumption and make it local by considering rates $\gamma_k$  such that $\gamma_k=\gamma_{-k}e^{s_k}$, with $s_k=-s_{-k}$ denoting the entropy change in the environment due to jump operator $L_k,$ whose only requirement is to be such that $L_k=L_{-k}^\dagger.$
Let $\varrho(t)=\sum_{n}p_{n}(t)|n(t)\rangle\langle n(t)|$, at each instant of time; we will drop the time dependence for simplicity of notation. Define the transition rates between each pair of eigenstates as $R^k_{nm}=\gamma_k|\langle n|L_k|m\rangle|^2$.  Note that $R_{nm}^k=e^{s_k}R_{mn}^{-k}$. 

The time derivative of the instantaneous populations, $p_n=\langle n|\varrho|n\rangle$, is $\dot p_n=\langle n|\dot\varrho|n\rangle+p_n(\langle \dot n|n\rangle+\langle n|\dot n\rangle)$.
Since $\tfrac{d}{dt}\langle n|n\rangle=0$, from \eqref{eq:diss-lind-2} we obtain,
\begin{equation}\label{eq:master-eq-q}
    \dot p_n=\sum_{k\in K}\sum_{m}(R^k_{nm}p_m-R^k_{mn}p_n),
\end{equation}
that is, we obtain a master equation for the population dynamics. Alternatively, we may define $J_{nm}=\sum_{k\in K}(R^k_{nm}p_m-R^{-k}_{mn}p_n)$, and write $\dot p_n=\sum_{m} J_{nm}$. Note that the Hamiltonian term in the Lindblad dynamics~\eqref{eq:diss-lind-2} does not affect \eqref{eq:master-eq-q}.

The entropy production rate in the system is given by 
$$
\dot \Sigma_{\rm sys}=-\tr\{\dot\varrho\ln\varrho\}=-\sum_{k\in K}\sum_{m,n}(R^k_{nm}p_m-R^k_{mn}p_n)\ln p_n,
$$
where we have used \eqref{eq:master-eq-q}.
Since $s_k$ is the entropy change in the environment due to a jump, on average the entropy production rate of the environment is
$$
\dot \Sigma_{\rm env}=\sum_{k\in K}\sum_{m,n}s_kR_{nm}^kp_m.
$$
Putting these together we obtain
\begin{align*}
  \dot\Sigma&=\sum_{k\in K}\sum_{m,n} \big(\ln(e^{s_k})+\ln(p_m/p_n)\big)R_{nm}^kp_m
  \\&=\sum_{k\in K}\sum_{m,n} R_{nm}^kp_m\ln\bigg(\frac{R_{nm}^kp_m}{R_{mn}^{-k}p_n}\bigg)
    \\&=\frac12\sum_{k\in K}\sum_{m,n} (R_{nm}^kp_m-R_{mn}^{-k}p_n)\ln\bigg(\frac{R_{nm}^kp_m}{R_{mn}^{-k}p_n}\bigg)
\end{align*}
where we have used the fact that $e^{s_k}=R_{nm}^k/R_{mn}^{-k}$.

Let $\{\varrho(t)\}_{t\in[0,\tau]}$ be a given smooth trajectory in $\mathcal P_*(\mathbb C,d)$, and consider its spectral decomposition $\varrho=\sum_np_n|n\rangle\langle n|$, with $p_n>0$, for all $t.$ We would now like to show that this trajectory can be achieved with arbitrarily low entropy production by a suitable choice of $H,\,\gamma_k,$ and $L_k.$
Indeed, its dynamics are characterized by 
$$
\dot \varrho=\sum_{n}(\dot p_n|n\rangle\langle n|+p_n|\dot n\rangle\langle n|+p_n|n\rangle\langle \dot n|),
$$
where the first term captures the population dynamics, while the last two terms capture the rotation of the state.

Let the population dynamics be governed by some antisymmetric current $J_{nm}$ such that
$
\dot p_n=\sum_{m}J_{nm}
$ and $J_{nn}=0$. 
Choose the jump operators $L_{xy}=|x\rangle \langle y|$ for all $|x\rangle\neq|y\rangle$ eigenstates of the instantaneous $\varrho(t),$ with corresponding rates $\gamma_{xy}=(\Lambda+ J_{xy})/(2{p_y})$ and $\gamma_{yx}=(\Lambda-J_{xy})/(2p_x)$, where $\Lambda$ is chosen large enough to make the rates positive. By construction, $L_{xy}^\dagger=L_{yx}$, and $\gamma_{xy}=\gamma_{yx}e^{s_{xy}}$ with $s_{xy}={\ln{\frac{(\Lambda+J_{xy})p_x}{(\Lambda-J_{xy})p_y}}}=-s_{yx}.$
These jump operators provide the required eigenvalue dynamics, since from \eqref{eq:master-eq-q},
\begin{align*}
\dot p_n&=\sum_{x,y\neq x}\sum_{m}(\gamma_{xy}|\langle n|L_{xy}|m\rangle|^2p_m-\gamma_{xy}|\langle m|L_{xy}|n\rangle|^2p_n)
\\&=\sum_{m\neq n}(\gamma_{nm}p_m-\gamma_{mn}p_n)
=\sum_{m} J_{nm}.
\end{align*}
 On the other hand, the rotation part of the dynamics, i.e., $\sum_np_n\big(|\dot n\rangle\langle n|+|n\rangle \langle \dot n|\big)$ can be generated through a Hermitian Hamiltonian  that does not generate any entropy production, e.g., $H=i\sum_{n}|\dot{n}\rangle \langle n|$.

Therefore, the entropy production is simply due to the change in the eigenvalue dynamics, and reads 
\begin{align*}
  \dot\Sigma&=\frac12\sum_{m,n} J_{nm}\ln\bigg(\frac{\Lambda+J_{nm}}{\Lambda-J_{nm}}\bigg).
\end{align*}
 Clearly, as $\Lambda\to \infty$,  $\dot \Sigma\to 0,$ and the entropy production vanishes, while the dynamics are kept unchanged, completing the proof. Note that to achieve this, the Hamiltonian, jump operators, and jump rates are allowed to be time varying and to depend on the given trajectory, as in the classical discrete setting \cite{dechant2022minimum}. Moreover, the Hamiltonian term is allowed to be chosen independently from the jump operators and rates, which only need to satisfy local detailed balance.

\section{Comparison between purely dissipative and mixed metrics}
 \label{app:equivalence}
We provide a proof to the equivalence bounds \eqref{eq:hierarchy-c} and~\eqref{eq:hierarchy-q} that is valid for both the classical and the quantum setting. 
Let $\psi$ and $\varphi$ be such that
$$
\dot\mu =-\mathbb K_{\mu }(\psi)=-(
\mathbb J_{\mu }+\mathbb K_{\mu })(\varphi),
$$
that is, $\psi$ and $\varphi$ are the potentials that generate the same velocity $\dot\mu $ in the purely dissipative and conservative-dissipative settings, respectively. Here, $\mu$ should be replaced by $\rho$ and $\varrho$ to obtain the classical and quantum expressions, respectively. 

Noting that $\psi=\mathbb K_{\mu }^{-1}((
 \mathbb J_{\mu }+\mathbb K_{\mu })(\varphi))$, we may write the classical $W_2$ metric as 
\begin{align}\label{eq:g-h}
   g_\mu (\dot\mu ,\dot\mu )&=\llangle\psi,\mathbb K_{\mu }(\psi)\rrangle \nonumber\\
   &=\langle
\mathbb K_{\mu }^{-1}((
\mathbb J_{\mu }+\mathbb K_{\mu })(\varphi)),
(\mathbb J_{\mu }+\mathbb K_{\mu })(\varphi)\rangle\nonumber\\
&=\langle\varphi,(
\mathbb J_{\mu }+\mathbb K_{\mu })^
\dagger\mathbb K_{\mu }^{-1}
(\mathbb J_{\mu }+\mathbb K_{\mu })(\varphi)\rangle\nonumber\\
&=\langle\varphi,\mathbb K_{\mu }(\varphi)\rangle+\langle\varphi,
\mathbb J_{\mu }^
\dagger\mathbb K_{\mu }^{-1}
\mathbb J_{\mu }(\varphi)\rangle,
\end{align}
where we have used the fact that $\langle\varphi,\mathbb J_\mu(\varphi)\rangle=0$.
Since $h_{\mu }(\dot\mu ,\dot\mu )=\langle\varphi,\mathbb K_{\mu }(\varphi)\rangle$ and the last term in \eqref{eq:g-h} is positive, we obtain
$
h_\mu (\dot\mu ,\dot\mu )\leq g_\mu (\dot\mu ,\dot\mu ).
$
Moreover, the last term can be bounded as
\begin{align*}
   \langle\varphi,
\mathbb J_{\mu }^
\dagger\mathbb K_{\mu }^{-1}
\mathbb J_{\mu }(\varphi)\rangle&= \langle\varphi,
\mathbb K_{\mu }^{1/2}\mathbb A_{\mu }^\dagger\mathbb A_{\mu }\mathbb K_{\mu }^{1/2}(\varphi)\rangle 
\\&\leq \|\mathbb A_{\mu }\|_{\rm op}^2\langle\varphi,
\mathbb K_{\mu }(\varphi)\rangle 
\end{align*}
where $\mathbb A_{\mu }:=\mathbb K_{\mu }^{-1/2}
\mathbb J_{\mu }\mathbb K_{\mu }^{-1/2}$ and $\|\mathbb A_{\mu }\|_{\rm op}$ is the operator norm. Thus, using this upper bound in \eqref{eq:g-h} and rearranging terms, we obtain 
$
 g_\mu (\dot\mu ,\dot\mu )\leq (1+\|\mathbb A_{\mu }\|_{\rm op}^2)h_\mu (\dot\mu ,\dot\mu ).
$ 

Putting all the ingredients together,
\begin{equation}
    \label{eq:hierarchy-app}
    \kappa({\mu })g_{\rm \mu }(\dot\mu ,\dot\mu )\leq h_{\mu }(\dot\mu ,\dot\mu )\leq g_{\rm \mu }(\dot\mu ,\dot\mu ),
\end{equation}
where we have defined 
$$
\kappa(\mu ):=\frac{1}{1+\|\mathbb K_{\mu }^{-1/2}
\mathbb J_{\mu }\mathbb K_{\mu }^{-1/2}\|_{\rm op}^2}
.$$
Integrating both sides of the first inequality over time and setting the trajectory $\{{\rm \mu }(t)\}_{t\in[0,\tau]}$ to be the minimizer of $\int_0^\tau h_{\mu }(\dot\mu ,\dot\mu )dt$ between endpoints $\mu _0$, $\mu _\tau$, we obtain
\begin{equation*}
    W_{2,h}(\mu _0,\mu _\tau)^2\geq \tau\int_0^\tau\kappa({\mu })g_{\rm \mu }(\dot\mu ,\dot\mu )dt\geq \kappa_{*}W_2(\mu _0,\mu _\tau)^2,
\end{equation*}
where $\kappa_*:=\inf_{\mu \in\{{\rm \mu}(t)\}_{t\in[0,\tau]}}\kappa(\mu ) $ is between 0 and 1. Note that if the minimizer does not exist, the inequality holds for $\kappa_*:=\lim\inf_{n\to \infty}\inf_{\mu \in\{{\rm \mu^{(n)}}(t)\}_{t\in[0,\tau]}}\kappa(\mu ),$  where $\{\mu^{(n)}\}_n$ is a minimizing sequence. Infimizing both sides of the rightmost inequality in \eqref{eq:hierarchy-app}  over paths with endpoints $\mu _0$ and $\mu _\tau$, we obtain the desired result
\begin{equation}
    \label{eq:hierW2c}\kappa_* W_{2}(\mu _0,\mu _\tau)^2\leq W_{2,h}(\mu _0,\mu _\tau)^2\leq W_{2}(\mu _0,\mu _\tau)^2.
\end{equation}

 \section{Proof that \texorpdfstring{$\mathbb L_{\lambda}$}{L-lambda} is the backward generator of the Lindblad dynamics}
\label{app:identity-LR}

Let $\mathbb L_{\lambda}^\ddagger$ be given by
$$
\mathbb L_{\lambda}^\ddagger(\cdot)=\mathbb S_{\lambda}(\cdot)+\mathbb A^\ddagger_{\lambda}(\cdot),
$$
where 
$$
\mathbb S_{\lambda}:=-\tilde M^{-1}_{\varrho^\lambda}(\mathbb K_{\varrho^\lambda}(\cdot)) \mbox{ and }\mathbb A^\ddagger_{\lambda}:=-\tilde M^{-1}_{\varrho^\lambda}(\mathbb J_{\varrho^\lambda}(\cdot)).
$$
We would like to show that its adjoint with respect to the $\tilde M_{\varrho^\lambda}$-weighted inner product is the backward generator of the Lindblad dynamics, that is,
\begin{equation}
    \label{eq:res-app}
    \mathbb L_\lambda=i[H^\lambda,\cdot]+\sum_{k\in K} \gamma^\lambda_k\big((L^\lambda_k)^\dagger (\cdot) L^\lambda_k -\frac12\{ (L^\lambda_k)^\dagger  L^\lambda_k, \cdot\}\big).
\end{equation}
To do so, we separately show that
\begin{align}\label{eq:diss-app}
 \mathbb S_{\lambda}&=\sum_{k\in K} \gamma^\lambda_k\big((L^\lambda_k)^\dagger (\cdot) L^\lambda_k -\frac12\{ (L^\lambda_k)^\dagger  L^\lambda_k, \cdot\}\big),
\end{align}
and
\begin{equation}\label{app:dissJ}
     \mathbb A^\ddagger_{\lambda}=-i[H^\lambda,\cdot].
\end{equation}
Note that $ \mathbb S_{\lambda}$ and $\mathbb A^\ddagger_{\lambda}$ are symmetric and antisymmetric with respect to the $\tilde M_{\varrho^\lambda}$-weighted inner product, respectively. Then, taking their adjoints, we obtain the desired result \eqref{eq:res-app}.

Let us first show \eqref{eq:diss-app}. To this end, note that we may swap $(L_k^\lambda)^\dagger$ and $\varrho^\lambda$ thanks to the identity 
$$
(L^\lambda_k)^\dagger (\varrho^\lambda)^\alpha=e^{-\beta^\lambda\omega^\lambda_k\alpha}(\varrho^\lambda)^\alpha (L^\lambda_k)^\dagger.
$$
To see this, let $f(t)=e^{tH^\lambda}(L_k^\lambda)^\dagger e^{-tH^\lambda},$ where $[H^\lambda,(L_k^\lambda)^\dagger]=-\omega^\lambda_k(L_k^\lambda)^\dagger$. Taking its time derivative we have
$$
\dot f(t)=e^{tH^\lambda}(H^\lambda(L_k^\lambda)^\dagger-(L_k^\lambda)^\dagger H^\lambda)e^{-tH^\lambda}=-\omega^\lambda_kf(t).
$$
Thus, $f(t)=e^{-\omega_k^\lambda t}(L_k^\lambda)^\dagger,$ and $f(\alpha\beta^\lambda)=e^{\alpha\beta^\lambda H^\lambda}(L_k^\lambda)^\dagger e^{-\alpha\beta^\lambda H^\lambda}=e^{-\alpha\beta^\lambda\omega^\lambda_k}(L_k^\lambda)^\dagger$, implying that  $(L_k^\lambda)^\dagger (\varrho^\lambda)^\alpha=e^{-\alpha\beta^\lambda\omega_k^\lambda}(\varrho^\lambda)^\alpha(L_k^\lambda)^\dagger.$ We may use this identity and~\eqref{eq:M-q} to write
\begin{align}\label{eq:Mtilde}\nonumber
&\!\mathbb K_{\varrho^\lambda}(A)
    =\sum_{k\in K}\big[(L_k^\lambda)^\dagger,\frac{\gamma_k^\lambda}{2}\int_0^1 e^{\beta^\lambda\omega^\lambda_k s}\big(\varrho^\lambda \big)^s\partial_k^\lambda A\big(\varrho^\lambda \big)^{1-s}ds\big]
     \\\nonumber &\!\!\!=\!\sum_{k\in K}\!\frac{\gamma_k^\lambda}{2}\!\int_0^1 \!\!\big(\varrho^\lambda \big)^s\Big((L_k^\lambda)^\dagger\partial_k^\lambda A-e^{\beta^\lambda\omega^\lambda_k }\partial_k^\lambda A(L_k^\lambda)^\dagger\Big)\big(\varrho^\lambda \big)^{1-s}ds 
    \\&\!\!\!=\frac12 \tilde M_{\varrho^\lambda}\Big(\sum_{k\in  K}\big(\gamma^\lambda_k (L^\lambda_k)^\dagger\partial^\lambda_kA-\gamma^\lambda_{-k}\partial^\lambda_kA (L^\lambda_k)^\dagger\big) \Big)\nonumber
     \\&\!\!\!=-\tilde M_{\varrho^\lambda}\Big(\sum_{k\in  K}\gamma^\lambda_k \big((L^\lambda_k)^\dagger A L_k^\lambda-\frac12\{(L^\lambda_k)^\dagger L_k^\lambda, A\}\big)\Big),
\end{align}
 where $\partial_k^\lambda A=[L_k^\lambda,A]$, and for the last equality we have rearranged the terms of the sum. Taking the inverse of $\tilde M_{\varrho^\lambda}$ on both sides we obtain \eqref{eq:diss-app}.

To show 
\eqref{app:dissJ}, we first use the definition of $\tilde M_\varrho$ and the fact that $\varrho^\lambda\propto e^{-\beta^\lambda H^\lambda},$ to write
\begin{align*}
 \tilde M_{\varrho^\lambda}(i[H^\lambda,A])&=
 i\int_0^1e^{-s\beta^\lambda H^\lambda}[H^\lambda,A]e^{s\beta^\lambda H^\lambda}ds\varrho^\lambda.
\end{align*}
Defining $f(s):=e^{-s\beta^\lambda H^\lambda}Ae^{s\beta^\lambda H^\lambda}$, and taking its derivative $f'(s)=\beta^\lambda e^{-s\beta^\lambda H^\lambda}[A,H^\lambda]e^{s\beta^\lambda H^\lambda}$, we see that
\begin{align*}
 \beta^\lambda\tilde M_{\varrho^\lambda}(i[H^\lambda,A]) &= -i\int_0^1f'(s)ds\varrho^\lambda
 \\&=i(A-e^{-\beta^\lambda H^\lambda}Ae^{\beta^\lambda H^\lambda})\varrho^\lambda,
\end{align*}
where we have used the fundamental theorem of calculus. Since we may write 
$$
\beta^\lambda \mathbb J_{\varrho^\lambda}(A)=i[A,\varrho^\lambda]=i(A-e^{-\beta^\lambda H^\lambda}Ae^{\beta^\lambda H^\lambda})\varrho^\lambda,
$$
we have that $\mathbb J_{\varrho^\lambda}(A)=\tilde M_{\varrho^\lambda}(i[H^\lambda,A]),$ and we obtain~\eqref{app:dissJ} by inverting $\tilde M_{\varrho^\lambda}.$

Then, noting that $\mathbb A_{\lambda}=-\mathbb A^\ddagger_{\lambda}$ and $\mathbb S_\lambda=\mathbb S_\lambda^\ddagger$,  together with $\mathbb L_{\lambda}(\cdot)=\mathbb S_{\lambda}(\cdot)+\mathbb A_{\lambda}(\cdot)$, we obtain the desired result~\eqref{eq:res-app}.

\end{document}